\catcode`\@=11

\newif\if@fewtab\@fewtabtrue

{\count255=\time\divide\count255 by 60
\xdef\hourmin{\number\count255}
\multiply\count255 by-60\advance\count255 by\time
\xdef\hourmin{\hourmin:\ifnum\count255<10 0\fi\the\count255}}
\def\ps@draft{\let\@mkboth\@gobbletwo
    \def\@oddhead{}
    \def\@oddfoot
       {\hbox to 7 cm{$\scriptstyle Draft\ version:\ \draftdate$
       \hfil}\hskip -7cm\hfil\rm\thepage \hfil}
    \def\@evenhead{}\let\@evenfoot\@oddfoot}

\def\ceqno{\global\@fewtabfalse
    \ifcase\@eqcnt \def\@tempa{& & &}\or \def\@tempa{& &}
      \or \def\@tempa{&}
      \or\def\@tempa{}\fi\@tempa
{\rm(\theequation)}}

\def\aeqno#1{\global\@fewtabfalse
    \ifcase\@eqcnt \def\@tempa{& & &}\or \def\@tempa{& &}
      \or \def\@tempa{&}
      \or\def\@tempa{}\fi\@tempa
{\rm(\theequation,#1)}}

\def\label#1{\ifnum\draftcontrol=1
 \global\def\draftnote{$\scriptstyle #1$}\fi
 \@bsphack\if@filesw {\let\thepage\relax
   \def\protect{\noexpand\noexpand\noexpand}%
\xdef\@gtempa{\write\@auxout{\string
      \newlabel{#1}{{\@currentlabel}{\thepage}}}}}\@gtempa
   \if@nobreak \ifvmode\nobreak\fi\fi\fi
  \@esphack}

\def\alabel#1#2{\label{#1}\global\@fewtabfalse
    \ifcase\@eqcnt \def\@tempa{& & &}\or \def\@tempa{& &}
      \or \def\@tempa{&}
      \or\def\@tempa{}\fi\@tempa
{\hbox to 3cm{\phantom{\rm(\theequation,#2)}
\draftnote \hfil}\hskip -3cm {\rm(\theequation,#2)}}}

\def\clabel#1{\label{#1}\global\@fewtabfalse
    \ifcase\@eqcnt \def\@tempa{& & &}\or \def\@tempa{& &}
      \or \def\@tempa{&}
      \or\def\@tempa{}\fi\@tempa
{\hbox to 3cm{\phantom{\rm(\theequation)}
\draftnote \hfil}\hskip -3cm{\rm(\theequation)}}}

\def\eqnarray{\def\draftnote{{}}\global\@fewtabtrue
\stepcounter{equation}\let\@currentlabel=\theequation
\global\@eqnswtrue
\global\@eqcnt\z@\tabskip\@centering\let\\=\@eqncr
$$\halign to \displaywidth\bgroup\@eqnsel\hskip\@centering\@eqcnt\z@
  $\displaystyle\tabskip\z@{##}$&\global\@eqcnt\@ne
  \hskip 1\arraycolsep \hfil${##}$\hfil
  &\global\@eqcnt\tw@ \hskip 1\arraycolsep
$\displaystyle\tabskip\z@{##}$
\hfil  \tabskip\@centering&\global\@eqcnt\thr@@\llap{##}\tabskip\z@
\cr}

\def\endeqnarray{\@@eqncr\egroup
      \global\advance\c@equation\m@ne$$\global\@ignoretrue}

\def\@eqnnum{\hbox to 3cm{\phantom{\rm(\theequation)} \draftnote
                         \hfil}\hskip -3cm {\rm(\theequation)}}

\def\@@eqncr{\let\@tempa\relax
    \ifcase\@eqcnt \def\@tempa{& & &}\or \def\@tempa{& &}
      \or \def\@tempa{&}
      \or\def\@tempa{}
\fi\@tempa
\if@eqnsw
\if@fewtab\@eqnnum\fi
\stepcounter{equation}\fi\global
\@eqnswtrue\global\@eqcnt\z@\global\@fewtabtrue\cr}

\def\draftcite#1{\ifnum\draftcontrol=1#1\else{}\fi}

\def\@lbibitem[#1]#2{\item{}\hskip -3cm \hbox to 2cm
{\hfil$\scriptstyle\draftcite{#2}$}\hskip
1cm[\@biblabel{#1}]\if@filesw
     {\def\protect##1{\string ##1\space}\immediate
      \write\@auxout{\string\bibcite{#2}{#1}}}\fi\ignorespaces}

\def\@bibitem#1{\item\hskip -3cm \hbox to 2cm
{\hfil $\scriptstyle\draftcite{#1}$}\hskip 1cm
\if@filesw \immediate\write\@auxout
       {\string\bibcite{#1}{\the\value{\@listctr}}}\fi\ignorespaces}

\def\nsection#1{\section{#1}\setcounter{equation}{0}}

\font\tendl=msbm10  scaled \magstep1%double line
\font\sevendl=msbm7 scaled \magstep1
\font\fivedl=msbm5 scaled \magstep1
\font\tengl=eufm10  scaled \magstep1% gothic letters
\font\sevengl=eufm7 scaled \magstep1
\font\fivegl=eufm5 scaled \magstep1

\newfam\dlfam  % \dl is double line
\textfont\dlfam=\tendl \scriptfont\dlfam=\sevendl
\scriptscriptfont\dlfam=\fivedl
\newfam\glfam  % \gl is gothic letters
\textfont\glfam=\tengl \scriptfont\glfam=\sevengl
\scriptscriptfont\glfam=\fivegl

\def\draftdate{\number\month/\number\day/\number\year\ \ \ \hourmin }

\global\def\draftcontrol{0}
\catcode`\@=12
\def\tilde{\widetilde}
\def\hat{\widehat}
\documentclass[12pt]{article}
\usepackage{epsfig}
\usepackage{bm}
\usepackage{amssymb}
\usepackage{comment}
\usepackage{amsfonts}\usepackage{color}
\renewcommand{\theequation}{\arabic{section}.\arabic{equation}}

\newcommand{\be}{\begin{eqnarray}}
\newcommand{\en}{\end{eqnarray}\vs 0.5 cm}
\newcommand{\non}{\nonumber}

\newcommand{\vs}{\vskip}

\newcommand{\Diff}{{\rm Di\hspace{-0.03cm}f\hspace{-0.07cm}f\hspace{-0.04cm}}}
\newcommand{\Vect}{{\rm V\hspace{-0.1cm}ect}}

\newcommand{\NR}{{{\mathbb R}}}%letra doble raya en modo matematico
\newcommand{\NH}{{{\mathbb H}}}%letra doble raya en modo matematico
\newcommand{\qq}{\begin{eqnarray}}

\newcommand{\da}{\partial}
\newcommand{\ee}{{\rm e}}

\newcommand{\qqq}{\end{eqnarray}}

\newcommand{\tr}{\hbox{tr}}

\newcommand{\CC}{{\cal C}}
\newcommand{\CD}{{\cal D}}
\newcommand{\CE}{{\cal E}}
\newcommand{\CF}{{\cal F}}
\newcommand{\CG}{{\cal G}}
\newcommand{\CH}{{\cal H}}

\newcommand{\CJ}{{\cal J}}

\newcommand{\CL}{{\cal L}}

\newcommand{\CO}{{\cal O}}
\newcommand{\CP}{{\cal P}}

\newcommand{\CX}{{\cal X}}

\newcommand{\s}{\hspace{0.05cm}}
\newcommand{\m}{\hspace{0.025cm}}

\begin{document}

\title{{\bf Turbulence on hyperbolic plane:\\
the fate of inverse cascade}}
\author{Gregory Falkovich$^{1}$ \,and \,Krzysztof Gaw\c{e}dzki$^{2}$ \\
%EndAName
\\
$^{1}$\small{Weizmann Institute of Science, Rehovot 76100, Israel}\cr
\small{Institute for Information Transmission Problems, Moscow, 
127994 Russia}\hfill\cr
$^{2}$\small{Laboratoire de Physique, C.N.R.S., ENS-Lyon,
Universit\'e de Lyon,}\cr
\small{46 All\'ee d'Italie, 69364 Lyon, France}\hfill}
\date{}
\maketitle

%\date{\quad version of Feb 2014}

\begin{abstract}
\noindent We describe ideal incompressible hydrodynamics on the hyperbolic
plane which is an infinite surface of constant negative curvature. We derive equations of motion, general symmetries and
conservation laws, and then
consider turbulence with the energy density linearly
increasing with time due to action of small-scale forcing. In a flat space, such energy growth is due to an inverse cascade, which builds a constant part
of the velocity autocorrelation function proportional to time and expanding
in scales, while the moments of the velocity difference saturate during
a time depending on the distance. For the curved space, we analyze the long-time long-distance scaling limit, that lives in
a degenerate conical geometry, and find that the energy-containing mode linearly growing with time is not constant in space. The shape of the velocity correlation function indicates that the energy builds up in vortical rings of arbitrary
diameter but of width comparable to the curvature radius of the hyperbolic
plane. The energy current across scales does not increase linearly
with the scale, as in a flat space, but reaches a maximum around the
curvature radius. That means that the energy flux through scales decreases at larger scales so that the energy is transferred in a non-cascade way, that is the inverse cascade
spills over to all larger scales where the energy pumped into the system
is cumulated in the rings. The time-saturated part
of the spectral density of velocity fluctuations contains a finite energy
per unit area, unlike in the flat space where the time-saturated
spectrum behaves as $\,k^{-5/3}$.
\end{abstract}
\vskip 0.5cm

\setcounter{section}{0}

\nsection{\bf Introduction}
\label{sec:intro}

Incompressible Navier-Stokes turbulence in two dimensions presents
a very different phenomenology from that in three dimensions
due to the presence of two different quadratic conserved quantities,
the energy and the enstrophy \cite{Batch,F,Frisch,Kr67}. It has been argued 
in \cite{Kr67,Batch}
that this leads to two coexisting cascades: the direct one towards short
distances for the enstrophy and the inverse one towards long distances
for the energy. The persistence of such two cascades requires, however,
special conditions of scale separation: the injection scale on which
forcing keeps operating must be much longer
than the dissipation scale on which viscous dissipation removes
enstrophy and much shorter than the integral scale (the size of the
container or friction scale) which does not allow the energy to 
cascade further. If
only the second condition is satisfied, no enstrophy cascade will
develop and a part of energy will be dissipated at short and intermediate
scales, but the remaining energy will still be transferred towards
longer distances in the inverse cascade. If there is no effective
friction mechanism removing energy at large scales, then, upon reaching
finite integral scale, the inverse cascade will start building energy
at that and intermediate distances developing large-scale coherent vortical
motions and, eventually, a stationary state with no cascades. Otherwise,
i.e. in the infinite space or in presence of a large scale friction,
the inverse cascade will persist for all times. Such a stationary inverse
energy cascade seems to be the simplest turbulent state as it exhibits
Kolmogorov scaling properties with no visible intermittency corrections
\cite{PT,BCV}, unlike the direct energy cascade in three dimensions
\cite{Aetall}. The absence of intermittency indicates possible existence
of the scaling limit where the injection scale is pushed to zero and the scale-invariance holds exactly. A theoretical understanding
of such a scaling state may be a challenge that is not completely out
of reach.  Although the existence of the scaling limit for the Navier-Stokes
inverse cascade is only a plausible assumption, it is reassuring to know
that there is a much simpler hydrodynamical system, the Kraichnan model
of passive scalar advection, that exhibits similar features at least partly
under analytic control: no intermittency in the inverse cascade and
intermittency in the direct one \cite{FGV}.
\vskip 0.2cm

To add to the puzzle about the Navier-Stokes inverse cascade, but possibly
providing an important clue, it has been observed \cite{BBCF,BBCF2} that
the zero-vorticity lines seem to behave as fractal
interfaces in some critical two-dimensional models of statistical
mechanics (e.g. the percolation), pointing to the possible presence
of a conformally invariant sector in the scaling limit of the theory
(such properties have no analogue for the passive scalar \cite{VFT}). Searching
for a possible source of this behavior was a part of our original
motivation to study the two-dimensional Navier-Stokes turbulence in
different background Riemannian geometries. As all such geometries
are locally conformally equivalent, comparing flows in different geometries
could permit to identify better the conformally invariant features
of the inverse cascade that should be present whenever the geometry
supports such a turbulent state. The latter condition eliminates
compact geometries (as that of a sphere), where the inverse cascade is eventually blocked. We also demand that the underlying geometry has
as many symmetries as the infinite flat space, admitting an analogue
of the homogeneous and isotropic turbulence. This leaves us with a single
choice: that of the hyperbolic plane with constant negative curvature.
Such geometry, that has the three-dimensional Lorentz group doubly covered
by $SL(2,\NR)$ as the isometry group, possesses many concrete realizations.
It may be viewed as a unit disc with the Poincar\'e metric, or as
the complex upper half plane with the hyperbolic metric, or as the upper
sheet of the two-sheeted hyperboloid in the three-dimensional Minkowski
space with the induced metric. Different presentations may be convenient
for the discussion of different aspects but are otherwise equivalent.
One of the basic features of the hyperbolic geometry is that on distances
longer than the curvature radius $\,R\,$ there is more room than in flat
space. Suppose that we start the inverse cascade in such geometry at
the forcing scale much shorter than $\,R$, \,where the dynamics of
the flow does not feel the curvature and thus should be similar to that
in the flat space. In principle, such a cascade should not be blocked
at scales longer than $\,R\,$ as it has more than enough room there to
evolve in. The main question we address in this paper is what happens
with the cascade in the hyperbolic plane at such long scales.
\vskip 0.2cm

\begin{figure}[h]
\begin{center}
\leavevmode
\includegraphics[width=7cm,height=7cm]{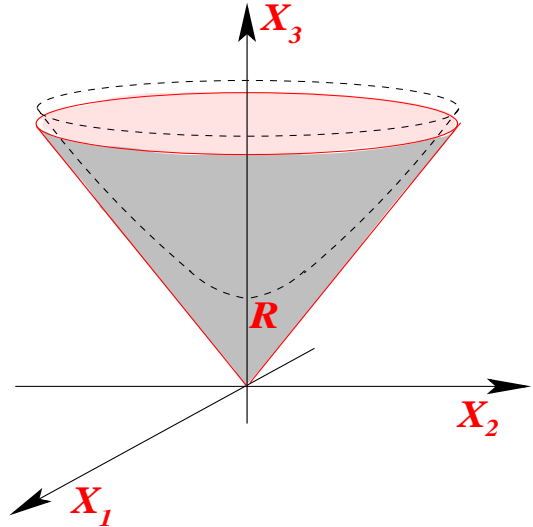}\\
\end{center}
\caption{Hyperboloid and cone}
\label{fig:lightcone}
\end{figure}

We do not have a firm answer to this question but we propose a scenario
describing the long-distance, long-time fate of the inverse cascade on
the hyperbolic plane that seems to pass several consistency checks. Assuming that the energy input is time-independent, we conclude that the energy density $\langle v^2\rangle$ grows linearly with time. The question is then: what is the part of the 2-point velocity autocorrelation function that grows linearly in time condensing energy in it? In a flat space, it may be inferred from simple scaling that this is a constant mode that extends to longer and
longer distances. Physically, it corresponds to flows that look locally as uniform jets getting longer and wider with time. On the hyperbolic plane, the exact form of such a
growing-in-time mode cannot be determined by scaling  because of
the presence of the additional distance scale $\,R$. One can only analyze its
asymptotic behavior at scales much shorter than $\,R\,$ (but longer
than the injection scale) or much longer than $\,R$. \,In particular,
such a mode should still be approximately constant at distances much
shorter than the curvature radius. What about its behavior on
distances much longer than $\,R$? \,We observe that on such distances
the geometry of a hyperboloid in the Minkowski space looks more and more
as that of the light-cone, see Figure \ref{fig:lightcone},
so the long distance asymptotics of the cascade might be realized
by a cascade on the Minkowski light-cone.

\vskip 0.2cm

\noindent The light-cone geometry is only
semi-definite being degenerate in the light-ray directions. It has the
infinite-dimensional group $\,\Diff\m S^1\,$
as the isometry group, which is also the group of conformal one-dimensional
symmetries. In particular, the Euler equation in the light-cone geometry
has this group as its symmetries and an infinity of related conserved
quantities. We set up a formal long-distance scaling limit of the randomly
forced Navier-Stokes  dynamics on the hyperboloid so that, if convergent,
it should determine the long-time long-distance asymptotics of the hyperbolic
plane inverse cascade. We note that the complete convergence cannot take
place because the scaling limit of the forcing covariance is partly
singular, but even a partial convergence should fix the long-distance
behavior of the energy-condensing mode, as well as that of the remaining
part of the velocity 2-point function if such a remainder stabilizes with
time, as in the flat space case. Both long distance behaviors are then
determined by scaling. This way we find the short-distance and long-distance 
asymptotic behaviors of the 2-point function mode growing linearly in time. 
Unlike in the flat space, where the constant
mode corresponds to the zero wavenumber, on the hyperbolic plane, the mode
growing linearly in time contains all wave numbers. This means that 
after reaching the wavenumber of order $\,R^{-1}$, \,the inverse energy 
cascade in the wavenumber space overspills to other, in particular
shorter, wavenumbers constantly feeding them with energy so that on such 
scales one cannot really speak of the wavenumber-space cascade in the usual 
sense. This is confirmed by the analysis of the flux relation
in the position space. In the flat space, this relation states that the
energy current across scales, expressed by a 2-point function built
of three velocities, grows linearly with the distance within the inverse
cascade regime, corresponding the constant energy flux through the scales.
We analyze an analogous flux relation on the hyperbolic plane and show
that the energy current reaches a maximum around the scale equal to the
curvature radius with the energy flux changing sign there. Moreover, a properly defined velocity autocorrelation function changes sign at distances exceeding $R$, which prompts the interpretation that the jets do not grow wider than $R$ and look like rings at larger scales, with circular motion providing velocity anti-correlation.
\vskip 0.2cm

A word of general warning to the reader is due. Despite of the use of
mathematical tools, this paper contains only a heuristic physical analysis of
the Navier-Stokes turbulence in the hyperbolic plane. Not much is
mathematically known about the existence and properties of a random process that solves
the stochastic Navier-Stokes equations on the hyperbolic plane of the type
considered here. Even in an infinite flat space, there are only weak results
about such general questions \cite{ViFu}, in contrast to the finite domains
where such problems were much more studied by the mathematical community
and the general questions (and more) are settled with full rigor \cite{KuSh}.
Infinite spaces present additional difficulty as they clearly require
appropriate conditions on the limits of flows at infinity for the evolution
equations to be well posed, see below.

\vskip 0.2cm
Let us briefly summarize the structure of the paper. In
Sec.\,\ref{sec:geodflow}, we recall the well known Lagrangian formulation
of the Euler equation on a general $d$-dimensional Riemannian manifold. 
This allows us
to discuss in Sec.\,\ref{sec:conslaws} the symmetries of such equation
within Noether's approach. Sec.\,\ref{sec:geomNS} recalls the formulation
of the Navier-Stokes equation in the geometric setup. In
Sec.\,\ref{sec:hyplane}, we discuss the particular case of the Euler
equation on the hyperbolic plane. Sec.\,\ref{sec:harman} collects
the basic facts about harmonic analysis in such a geometry related
to its $\,SL(2,\NR)\,$ isometry group representation theory.
The Navier-Stokes equation on the hyperbolic plane with a stochastic
forcing, the main object of interest in this paper, is introduced in
Sec.\,\ref{sec:NSonH}. Sec.\,\ref{sec:invcasc} recalls the inverse
cascade scenario in the flat space and proposes its extension to the case
of the hyperbolic plane. The global flux relation for the latter case is
derived in Sec.\,\ref{sec:GFR}. In Sec.\,\ref{sec:Scal}, we study
the behavior of the Navier-Stokes dynamics on the hyperbolic plane under
rescalings of time, distances and velocities. Sec.\,\ref{sec:R=0} discusses
the limiting large scale conical geometry and the Euler equation in its
background as well as its symmetry enhanced to the infinite-dimensional
group $\,\Diff\m S^1$. \,The long-time
long-distance scaling limit of the theory on the hyperbolic plane
is studied in Sec.\,\ref{sec:scalim}, first for the random Gaussian
white-in-time forcing and then for the velocity equal-time 2-point
function. In the latter case, we check the consistency of the assumption
that such a limit exists (up to logarithmic corrections for some
velocity components) and is given by the scaling solution obtained
directly in the conical geometry. This permits to fix the long-distance
behavior of the terms of velocity autocorrelation that are dominant within
the scenario, as well as the long-distance behavior of the energy
current across the scales. In Conclusions, we summarize our
main results about the ultimate fate of the inverse energy cascade on
the hyperbolic plane and speculate about their possible bearing on
the phenomenology of the two-dimensional turbulence. Six Appendices
collect a supplementary or even more technical material related to the main
body of the paper.
\vskip 0.3cm

\noindent{\bf Acknowledgements}. The work of G.F. is supported by the grants 
of BSF and Minerva Foundation. The work of K.G. was a part
of the STOSYMAP project ANR-11-BS01-015-02. Both authors are grateful
to their home institutions for the support of mutual visits. We thank 
R. Chetrite and A. B. Zamolodchikov for useful discussions and U. Frisch
for additional references and for pointing out a mistake in the original 
version of the paper.

\nsection{\bf Euler equation as a geodesic flow}
\label{sec:geodflow}

The Euler equation has an infinite-dimensional geometric
interpretation \cite{Arnold}: it describes the geodesic flow
on the group of volume preserving diffeomorphisms.
%\footnote{Recall that the Euler top
%describes the geodesic flow on the group $SO(3)$.}.
Let us briefly sketch this argument.  Take as the flow space an oriented
$\,d$-dimensional Riemannian manifold $\,(M,g)\,$ with metric volume form
$\chi$, see Appendix\,\ref{appx:Riemgeom} for the basic notions
of Riemannian geometry needed below. \,Let $\,\Diff_\chi\,$ be the group
of diffeomorphisms $\,\Phi\,$ of $\,M\,$ preserving the metric volume
form\footnote{More general volume forms could be also used \cite{Arnold}.}
$\,\chi$: $\,\Phi^*\chi=\chi$, \,where $\,\Phi^*\chi\,$ denotes the pullback
of form $\,\chi\,$ by diffeomerphism $\,\Phi$. \,The space
$\,\Vect_\chi\,$ of divergenceless vector fields $\,v\,$ may be considered
as the Lie algebra of $\,\Diff_\chi$\footnote{For non-compact $\,M\,$
we shall assume that $\,\Phi\,$ do not move points outside a compact
set and that $\,v\,$ have compact supports, although weaker assumptions
about the behavior at infinity may be more natural and/or more physiqcal.}.
\,Upon the identification of infinitesimal variations
of diffeomorphisms $\,\Phi\,$ with the vector fields $\,v\,$
by $\,\delta\Phi(x)=v(\Phi(x))$, \,the
scalar product of vector fields
\qq
\Vert v\Vert^2\s=\s\int\limits_M g(v,v)\s\m\chi\s\equiv\s
\int\limits_M g_{ij}\,v^i\s v^j\,\chi
\label{normsq}
\qqq
induces on $\Diff_\chi$ a right-invariant Riemannian
metric. This metric in not left-invariant.
\vskip 0.2cm

The geodesic flow with respect to a left-right-invariant
metric on a group is given by one-parameter subgroups, modulo
time-independent left and right translations. This is not the case if
the metric is only right-invariant. The geodesic motions
$\s t\mapsto\Phi_t\s$ on $\Diff_\chi$ corresponding to the right-invariant
Riemannian metric defined above are given by the extrema of the action
\qq
S_g(\Phi)\s=\s{_1\over^2}\int\Vert v(t)\Vert^2\s{\rm d}t\,,
\non
\qqq
for $\,\Phi(t,x)\equiv\Phi_t(x)$, \,where
\qq
\da_t\Phi(t,x)\s=\s v(t,\Phi(t,x))\,,
\non
\qqq
under the condition that $\,\Phi^*_t\chi=\chi\,$ for all $\,t$.
They satisfy the (generalized) Euler equation
\qq
\da_tv^i+v^k\nabla_k v^i = -g^{ik}\partial_kp
\label{Euler}
\qqq
together with the incompressibility condition
\qq
\nabla_kv^k=0\,,\label{incomp}
\qqq
see Appendix \ref{appx:varprinc}. The assignment $\,x\mapsto\Phi(t,x)\,$ 
is the Lagrangian map for the velocity field $\,v(t,x)$.
\,It is sometimes more convenient to rewrite the Euler equation and the
incompressibility condition in
terms of the 1-form $\,v^\flat\,$ with lower-index components $\,v_i\,$:
\qq
\da_tv_i+v^k\nabla_k v_i = -\partial_ip\,,\qquad g^{ij}\nabla_iv_k=0
\qqq
The latter condition may be also rewritten as $\,{\rm d}^\dagger v^\flat=0$, 
\,where
$\,{\rm d}^\dagger\,$ is the adjoint of the exterior derivative and 
$\,{\rm d}\,$ given
by Eqs.\,(\ref{extder}), (\ref{ddag}) and (\ref{ddag1}) of Appendix
\ref{appx:Riemgeom}.
On noncompact manifolds, one should demand that $\,v\,$ decays
sufficiently fast at infinity, see Appendix \ref{appx:varprinc}.
Pressure $p$ may be eliminated by introducing the vorticity $\omega$
\qq
\omega\,=\,*{\rm d}v^{\flat}\qquad{\rm or}\qquad\omega_{k_1\dots k_{d-2}}=\,
(\nabla^iv^j)\,\epsilon_{ijk_1\dots k_{d-2}}\,\sqrt{g}\,,
\label{vort}
\qqq
that is a $\,(d-2)$-form satisfying the relation $\,{\rm d}^\dagger\omega=0\,$
(trivial in $d=2$). Above, $\,*\,$ stands for the Hodge star, see
Eq.\,(\ref{H*}) of Appendix \ref{appx:Riemgeom} and
$\,\sqrt{g}\equiv\sqrt{\det(g_{ij})}$.
\,The Euler equation implies the vorticity evolution equation
\qq
\partial_t\omega_{j_1\dots j_{d-2}}+({\cal L}_v\omega)_{j_1\dots j_{d-2}}
\,=\,0
\label{Eulvort}
\qqq
from which the pressure has dropped out. $\,{\cal L}_v\,$
denotes the Lie derivative w.r.t. the vector field
$\,v$, \,see Eq.\,(\ref{LDf}) in Appendix \ref{appx:Riemgeom}. 
Eq.\,(\ref{Eulvort}) implies that the pullback of the time $t$ vorticity
$\,(d-2)$-form by the Lagrangian map $\,\Phi(t)\,$ is constant in time:
\qq
\partial_t\,\big(\Phi(t)^*\omega(t)\big)\,=\,0\,.
\label{vorttr}
\qqq 
For the exterior derivative of the vorticity form one obtains the relation
\qq
*^{-1}{\rm d}\omega\,=\,-\Delta v^{\flat}\,,
\label{Delv}
\qqq
where $\,\Delta\,$ is the Laplace-Beltrami operator of Eq.\,(\ref{LaplB})
of Appendix \ref{appx:Riemgeom}. This allows, in principle, to recalculate
the divergenceless velocity field from the vorticity as
\qq
v^{\flat}\ =\ -*^{-1}{\rm d}\hspace{0.03cm}\Delta^{-1}\omega
\label{vfromom}
\qqq
under condition of sufficient fast decrease of  $\,\omega\,$ at infinity.
For the fluid (kinetic) energy, one obtains then the expression:
\qq
E\,=\,\frac{_1}{^2}\int\Vert v\Vert^2\chi\ =\
\frac{_1}{^2}(v^{\flat},v^{\flat})\ =\ -\frac{_1}{^2}(\omega,\Delta^{-1}\omega)
\label{energy}
\qqq
in terms of the $L^2$ scalar products discussed in
Appendix \ref{appx:Riemgeom}.
\vskip 0.2cm

In two dimensions, where the vorticity is a scalar function,
Eqs.\,(\ref{vort}), (\ref{vfromom}) and (\ref{Eulvort}) reduce to
\qq
\omega\,=\,(\nabla^iv^j)\epsilon_{ij}\sqrt{g}\,=\,\Delta\psi
\label{omega2d}
\qqq
and
\qq
\partial_t\omega\,=\,-v^i\partial_i\omega\,=\,\frac{_1}{^{\sqrt{g}}}
\epsilon^{ij}(\partial_i\omega)(\partial_j\psi)
\label{evomega2d}
\qqq
where $\,\psi(t,x)\,$ is the ``stream function'' s.t.
\qq
v^i\,=\,\frac{1}{\sqrt{g}}\epsilon^{ji}\partial_j\psi\,.
\label{psi2d}
\qqq

\nsection{Symmetries and conservation laws}
\label{sec:conslaws}

It follows from the above discussion, see also Appendix \ref{appx:varprinc},
that the incompressible Euler equation
results from the variational principle for the field-theoretical action
functional
\qq
S(\Phi,\lambda)\ =\ \int\,L(x,\Phi^i(t,x),\partial_t\Phi^i(t,x),
\partial_j\Phi^i(t,x),\lambda(t,x))\,\chi(x)\,{\rm d}t
\label{faction}
\qqq
with the Lagrangian density
\qq
\!\!\!\!\!\!\!\!\!\!\!  &&L(x,\Phi^i,\partial_t\Phi^i,
\partial_j\Phi^i,\lambda)%\cr&&\qquad\qquad
= \frac{_1}{^2}\m g_{ij}(\Phi)\,(\partial_t\Phi^i)
(\partial_t\Phi^j) -
\lambda \Big[ 1-\sqrt{g(\Phi)}
 {\rm det}(\partial_i\Phi^j)/\sqrt{g(x)}\Big],\ \,
\qqq
where $\,\lambda\,$ is the Lagrange multiplier imposing the volume-preserving
condition related on the extremal field configurations to the pressure,
see Eq.\,(\ref{pressure}) of Appendix \ref{appx:varprinc}. \,The corresponding
field theory has multiple symmetries:
time translation, left composition of $\,\Phi\,$ with diffeomorphisms
preserving the metric and the orientation and the right composition
with diffeomorphisms preserving volume.
On the infinitesimal level those symmetries give rise by the Noether
theorem to conserved currents $\,(J^0(t,x),J^k(t,x))\,$ satisfying
the conservation laws
\qq
\partial_tJ^0+\nabla_kJ^k\,=\,0
\qqq
or, equivalently,
\qq
\partial_t(J^0\sqrt{g})+\partial_k(J^k\sqrt{g})\,=\,0\,.
\label{conslaw}
\qqq
In fact, due to the special feature of the theory whose fields
are maps of the same space, there are two different sets of
conserved Noether currents. To understand it, let us rewrite
the conservation law (\ref{conslaw}) as an identity
\qq
\int\big[J^0(t,x)\,\partial_tf(t,x)\,+\,J^k(t,x)\,\partial_kf(t,x)\big]
\,\chi(x)\,{\rm d}t\ =\ 0\,,
\label{rell}
\qqq
holding for each test function $\,f(t,x)$. \,Introducing a function
$\,h(t,y)\,$ defined by the relation $\,h(t,\Phi(t,x))=f(t,x)\,$ so that
\qq
&&(\partial_tf)(t,x)\,=\,(\partial_th)(t,\Phi(t,x))+(\partial_l h)
(t,\Phi(t,x))(\partial_t\Phi^l)(t,x)\,,\\
&&(\partial_kf)(t,x)\,=\,(\partial_l h)(t,\Phi(t,x))
(\partial_k\Phi^l)(t,x)\,,
\qqq
we may rewrite  (\ref{rell}) as the new identity
\qq
&&\int\big(\CJ^0(t,y)\,\partial_th(t,y)\,+\,\CJ^l(t,y)\,\partial_l
h(t,y)\big)\,\chi(y)\,{\rm d}t\ =\ 0\,,
\label{relll}\\
&&\CJ^0(t,\Phi(t,x))\ =\ J^0(t,x)\,,\label{CJCJ0}\\
&&\CJ^l(t,\Phi(t,x))\ =\
(\partial_t\Phi^l)\,J^0(t,x)+
(\partial_k\Phi^l)\,J^k(t,x)\,.
\label{CJCJl}
\qqq
Since Eq.\,(\ref{relll}) holds for any test function $\,h$, \,it
implies that the transformed currents are also conserved:
\qq
\partial_t\CJ^0\,+\,\nabla_l\CJ^l\,=\,0\,.
\label{trcons}
\qqq
Let us now discuss one by one the conservation laws related to infinitesimal symmetries
of the action functional (\ref{faction}).
\vskip 0.2cm

The time-translation invariance is assured by the absence of an explicit
time dependence of $\,L$. \,The conserved
current $\,(J^0(t,x),J^k(t,x))\,$ corresponding by the Noether theorem
to that symmetry has the form
\qq
&&J^0\ =\ \Big(-L\,+\,\frac{\partial L}{\partial(\partial_t\Phi^k)}\,
(\partial_t\Phi^k)\Big)\ =\ \frac{_1}{^2}\,g_{ij}(\Phi)\,(\partial_t\Phi^i)
(\partial_t\Phi^j)\,,\label{JJ0}\\
&&J^k\ =\ \ \frac{\partial L}{\partial(\partial_k\Phi^l)}\,
(\partial_t\Phi^l)\
=\ \lambda\,\Big[\Big(\frac{_{\partial\Phi}}{^{\partial x}}\Big)^{-1}\Big]^k_l
\,(\partial_t\Phi^l)\,,
\label{JJk}
\qqq
where we used the volume-preserving property of $\,\Phi$.
\,The corresponding modified currents $\,(\CJ^0(t,y),\CJ^l(t,y))\,$
of (\ref{CJCJ0}) and (\ref{CJCJl})
have the form:
\qq
&&\CJ^0\ =\ \frac{_1}{^2}\,g_{ij}\,v^i\,v^j\,,\\
&&\CJ^l\ =\ \left(\frac{_1}{^2}\,g_{ij}\,v^i\,v^j\,
+\,p\right)v^l\,.
\qqq
Their conservation implies the conservation in time of the energy
$\,E\,=\,\int\CJ^0(t,y)\,\chi(y)\,$
if the flux of the spatial current $\,(\CJ^l)\,$ vanishes at infinity.
\vskip 0.2cm

Let us pass to the symmetries induced by the left composition of $\,\Phi\,$
with diffeomorphisms $\,D\,$ preserving the Riemannian metric, \,i.e.
such that
\qq
g_{ij}(D(x))\,(\partial_kD^i)(x)\,(\partial_lD^j)(x)\
=\ g_{kl}(x)\,.
\qqq
A replacement $\,\Phi\mapsto D\circ\Phi\,$ does not change the Lagrangian
density and does not change the action functional if $\,D\,$ also preserves
the orientation. Infinitesimal diffeomorphisms preserving metric
correspond to Killing vector fields $\,X\,$ such that
$\,\nabla_iX_j+\nabla_jX_i=0$, \,see Eq.\,(\ref{LDv}).
\,By the Noether Theorem, such vector fields induce conserved currents
$\,(J^0_X(t,x),J^k_X(t,x))\,$ given by the formulae
\qq
J^0_X&=&-\frac{\partial L}{\partial(\partial_t\Phi^j)}X^j(\Phi)\
=\ -\,g_{ij}(\Phi)\,(\partial_t\Phi^i)\,X^j(\Phi)\,,\label{JX0}\\ \cr
J^k_X&=&-\,\frac{\partial L}{\partial(\partial_k\Phi^j)}\,X^j(\Phi)
\ =\ -\,\lambda\,\big[\big(\frac{_{\partial\Phi}}{^{\partial x}}\big)^{-1}]^k_j
\,X^j(\Phi)\,.\label{JXk}
\qqq
The transformation of Eqs.\,(\ref{CJCJ0}) and (\ref{CJCJl}) allows one to pass
to the modified conserved currents $\,(\CJ^0_X(t,y),\CJ^l_X(t,y))\,$ with
\qq
&&\CJ^0_X\ =\ -\,v_j\,X^j\,,\label{CJCJX0}\\
&&\CJ^l_X\ =\ -\,v_j\,X^j\,v^l\,-\,p\,X^l\,.
\label{CJCJXl}
\qqq
An example is  the constant vector field  inducing space
translations in the flat space, where the conservation of the corresponding
current expresses the fluid momentum conservation and is equivalent
to the Euler equation. Another flat-space example gives the conservation
of the fluid angular momentum obtained from the vector fields corresponding to
infinitesimal rotations with $\,X^i=x^k\delta^{li}-x^l\delta^{ik}\,$
for $\,k<l\,$ fixed.
\vskip 0.2cm

Finally,  consider the ``relabeling'' symmetries induced
by the right composition
of $\,\Phi\,$ with diffeomorphisms $\,D\,$ preserving the volume form
$\,\chi$, \,i.e. such that
\qq
\sqrt{g(D(x))}\,\m{\rm det}(\partial_iD^j(x)) =\ \sqrt{g(x)}\,.
\qqq
On the infinitesimal level, they correspond to
divergenceless vector fields $\,u\,$ and, by the Noether Theorem,
induce conserved currents $\,(J^0_u(t,x),J^k_u(t,x))\,$ with
\qq
&&J^0_u\ =\ \frac{\partial L}{\partial(\partial_t\Phi^j)}(\partial_l\Phi^j)u^l
\,=\,g_{ij}(\Phi)(\partial_t\Phi^i)(\partial_l\Phi^j)u^l\,,\cr\cr
&&J^k_u\ =\ -\,L\m u^k\,+\,
\frac{\partial L}{\partial(\partial_k\Phi^j)}(\partial_l\Phi^j)u^l\,=\,
\Big(\hspace{-0.08cm}-\frac{_1}{^2}\hspace{0.02cm}g_{ij}(\Phi)(\partial_t\Phi^i)
(\partial_t\Phi^j)+\lambda\Big)u^k\,.
\qqq
The last conservation laws are closely related to the laws of Euler 
hydrodynamics in the Lagrangian pictures. For example taking 
$\,u^l=\frac{1}{\sqrt{g}}\partial_k\eta^{kl}\,$ for $\eta^{kl}(x)
=-\eta^{lk}(x)$, \, one infers from the conservation of the integral
of $\,J_u^0\sqrt{g}\,$ that the quantity 
\qq
\big(\partial_iv_j-\partial_jv_i\big)(t,\Phi(t,x))\,(\partial_k\Phi^i)(t,x)
\,(\partial_l\Phi^j)(t,x)
\qqq
is time independent. Given the volume preserving property of the Lagrangian 
map $\,\Phi(t)$, \,the latter equation is equivalent
to Eq.\,(\ref{vorttr}). In three flat dimensions the above relations have 
a long history going back to Cauchy's work in the early XIXth century, see
\cite{FV}.
\vskip 0.1cm

For the two-dimensional case, there is an infinite-dimensional family of
conservation laws related to the currents
\qq
&&\CJ^0_f\ =\ f(\omega)\,,\cr
&&\CJ^l_f\ =\ \frac{_1}{^{\sqrt{g}}}\,f'(\omega)\,\psi\,
\epsilon^{lj}\partial_j\omega
\qqq
that lead to the conservation in time of integrals of arbitrary local
function $\,f\,$ of vorticity, also directly implied by Eq.\,(\ref{vorttr}).
The latter currents may be related to the $\Phi$-dependent relabeling 
symmetries. 
\vskip 0.1cm

In marked difference with the case of the Euler equation in a flat space
possessing also the Galilean symmetry, there does not seem to be a
replacement for it in the presence of curvature.

\nsection{Navier-Stokes equation in the geometric setup}
\label{sec:geomNS}

In the presence of viscosity and forcing, the Euler equation
should be replaced by the Navier-Stokes equation, which in the geometric setup
takes the form \cite{EM}
\qq
\da_tv^i+v^k\nabla_k v^i-\nu(Lv)^i\,=\, -g^{ij}\partial_jp\,+\,f^i\,,
\label{NSg}
\qqq
where $\nu\,$ is the kinematic viscosity and
$\,L=\Delta+2\hspace{0.02cm}{\rm Ric}\,$ with the Laplacian $\Delta$
acting on the vector fields  $v$ by the formula
\qq
(\Delta v)^i=g^{jk}\nabla_j\nabla_kv^i+g^{ij}
(\nabla_j\nabla_k-\nabla_k\nabla_j)v^k\,=\,g^{jk}\nabla_j\nabla_kv^i-
({\rm Ric}\,v)^i
\qqq
obtained from Eq.\,(\ref{laplvf}) and
$({\rm Ric}\,v)^i=({\rm Ric})^i_jv^j$,
so that using incompressibility of $\,v$, \,one obtains
\qq
(Lv)^i\,=\,g^{jk}\nabla_j\nabla_kv^i-g^{ij}
(\nabla_j\nabla_k-\nabla_k\nabla_j)v^k\,=\,g^{jk}\nabla_j\nabla_kv^i+g^{ij}
\nabla_k\nabla_jv^k,
\qqq
The action of $\,L\,$ preserves incompressibility
on Einstein manifolds where the Ricci tensor is proportional to the
metric one: $\,({\rm Ric})_{ij}=k\hspace{0.02cm}g_{ij}\,$ for some constant $\,k\,$
equal to the scalar curvature $\,S\,$ divided by the dimension $d$.
\,The force $\,f(t,x)\,$ is also supposed to be divergenceless.
Below, it will be assumed
to be random delta-correlated Gaussian process with mean zero and
covariance
\qq
\langle f^i(t_1,x_1)\,f^j(t_2,x_2)\rangle\,=\,\delta(t_1-t_2)\,C^{ij}(x_1,x_2).
\label{fcov}
\qqq
We shall use the It\^{o} convention in treating the differential equations
with the white-noise in time terms.
\,Multiplying the Navier-Stokes equation by $v$ and integrating
by parts, we obtain the energy balance:
\qq
\partial_tE\,=\,
\partial_t\int{_1\over^2}\Vert v\Vert^2\,\chi&=&-{_1\over^2}\nu\int
(\nabla_iv_j+\nabla_jv_i)(\nabla^iv^j+\nabla^jv^i)\,\chi\cr
&+&\int v_if^i\,\chi\,+\,{_1\over^2}\int g_{ij}c^{ij}\,\chi
\qqq
where $\,c^{ij}(x)\equiv C^{ij}(x,x)\,$ enters the It\^{o} term. \,The quantity
\qq
\varepsilon\ =\
\frac{_1}{^2}\nu(\nabla_iv_j+\nabla_jv_i)(\nabla^iv^j+\nabla^jv^i)
\label{dissip}
\qqq
is called the dissipation field and
\qq
\bar\iota\ =\ \frac{_1}{^2}\,g_{ij}\,c^{ij}
\label{inject}
\qqq
the mean energy injection rate (both space-dependent, in general).
\vskip 0.2cm

The vorticity
related to the incompressible velocity $v$ by Eqs.\,(\ref{vort}) and
(\ref{Delv}) evolves on Einstein manifolds according to the equation
\qq
\partial_t\omega_{j_1\dots j_{d-2}}\,+\,({\cal L}_v\omega)_{j_1\dots j_{d-2}}
-\,\nu((\Delta+\frac{_2}{^d}S)\omega)_{j_1\dots j_{d-2}}\,=\,*{\rm d}f^{\flat}\,.
\qqq

\nsection{Hyperbolic plane}
\label{sec:hyplane}

\noindent We  concentrate on a special kind of an
Einstein space: the hyperbolic plane $\,\NH_{\hspace{-0.01cm}_R}$.
\,Let us start from
the description of that space directly related to group $\,SL(2,\NR)\,$
and its Lie algebra $\,sl(2,\NR)$, whose basis is chosen to be
composed of the Pauli matrices,
\qq
\tau_1={_1\over^2}\sigma_1={_1\over^2}
\left(\begin{matrix}{0&1\cr1&0}\end{matrix}\right),\quad
\tau_2={_1\over^2}\sigma_3={_1\over^2}
\left(\begin{matrix}{1&0\cr0&{-1}}\end{matrix}\right),
\quad\tau_3={_1\over^2}i\sigma_2={_1\over^2}
\left(\begin{matrix}{0&1\cr-1&0}\end{matrix}\right)\,,\quad
\label{generat}
\qqq
with the commutation relations
\qq
[\tau_1,\tau_2]=-\tau_3\,,\quad\ [\tau_1,\tau_3]=-\tau_2\,,
\quad\ [\tau_2,\tau_3]=-\tau_1\,.
\label{comrel}
\qqq
Group $SL(2,\NR)$ acts on $sl(2,\NR)$ by the adjoint action
leaving invariant the quadratic form $\tr\,XY$, corresponding to the signature $(+,+,-)$:
\qq
\tr\,\tau_1^2\,=\,\tr\,\tau_2^2\,=\,-\tr\,\tau_3^2\,=\,{_1\over^2}\,,
\qquad\tr\,\tau_1\tau_2\,=\,\tr\,\tau_1\tau_3\,=\,\tr\,\tau_2\tau_3\,=\,0\,,
\qqq
For $\,R>0$, \,consider in $sl(2,\NR)$ the hypersurface
\qq
\NH_{\hspace{-0.01cm}_R}\,=\,\{\,\CX\in sl(2,\NR)\ |\ 2\,\tr\,\CX^2=-R^2,\ \,\tr\,\CX\tau_3<0\,\}
\label{NHR}
\qqq
In the coordinates $\,\CX=X^i\tau_i$, \,it is a hyperboloid given
by the equation
\qq
(X^1)^2+(X^2)^2+R^2=(X^3)^2\,,\qquad X^3>0\,.
\label{NHR1}
\qqq
The Lorentzian metric $\,({\rm d}X^1)^2+({\rm d}X^2)^2-({\rm d}X^3)^2\,$ 
induces on
$\,\NH_{\hspace{-0.01cm}_R}\,$ the Riemannian metric $\,g_R$. \,The Riemannian
distance $\,\delta_R(\CX_1,\CX_2)\,$ between two points
on $\,\NH_{\hspace{-0.01cm}_R}\,$
with coordinates $\,(X^i_1)\,$ and $\,(X_2^i)$, \,respectively, is given
by the formula
\qq
\cosh\hspace{-0.06cm}\frac{_{\delta_R(1,2)}}{^R}\ =\ \frac{_{X_1^3X_2^3-
X_1^1X_2^2-X_1^2X_2^2}}{^{R^2}}\,.
\label{chdist}
\qqq
One may parameterize $\,\NH_{\hspace{-0.01cm}_R}\,$ by the complex coordinate $\,z$, $\,|z|<1$,
\qq
X^1+iX^2\,=\,\frac{2Rz}{1-|z|^2}\,,\qquad X^3\,=\,
\frac{R(1+|z|^2)}{1-|z|^2}
\label{hyperb}
\qqq
In these coordinates, the Riemannian metric is
\qq
g_R\,=\,R^2\frac{4{\rm d}z{\rm d}\bar z}{(1-|z|^2)^2}
\label{metr2}
\qqq
and is conformally flat. The corresponding volume form is
\qq
\chi_R\,=\,R^2\frac{2i\,{\rm d}z\wedge{\rm d}\bar z}{(1-|z|^2)^2}\,.
\label{vol2}
\qqq
As we see, $\,R\,$ is the scale of the
metric with the dimension of length. In this parametrization
$\,\NH_{\hspace{-0.01cm}_R}\,$ is realized as the Poincare disc.
By the conformal transformation
$z = (w-i)/(w+i)$, $w=i(1+z)(1-z)$
one maps the unit disc in variable $z$ to the upper half-plane
$\,{\rm Im}\,w>0\,$ in the variable $\,w\,$ with the hyperbolic
metric
\qq
g_R\,=\,R^2\frac{{\rm d}w{\rm d}\bar w}{({\rm Im}\,w)^2}
\label{metr3}
\qqq
and the metric volume form
\qq
\chi_R\,=\,R^2\frac{i\,{\rm d}w\wedge{\rm d}\bar w}{2({\rm Im}\,w)^2}\,.
\label{vol3}
\qqq
In this parametrization of the hyperboloid,
\qq
X^1\,=\,R\,\frac{|w|^2-1}{2\,{\rm Im}\,w}\,,\qquad X^2\,=\,-R\,
\frac{{\rm Re}\,w}{{\rm Im}\,w}\,,
\qquad X^3\,=\,R\,\frac{|w|^2+1}{2\,{\rm Im}\,w}\,.
\qqq
The Riemannian distance between points $\,\CX_1\,$ and $\,\CX_2\,$
parameterized by $\,w_1\,$ and $\,w_2\,$ is
\qq
\delta_R(\CX_1,\CX_2)\ =\ R\,\ln\frac{|w_1-\bar w_2|+|w_1-w_2|}
{|w_1-\bar w_2|-|w_1-w_2|}\,.
\label{hypdist}
\qqq
Below, we shall drop  the subscript ``$R$'' from the
metric, the volume form and the distance function since it should be
clear from the context which is the underlying space $\,\NH_R$.
\,In the complex coordinates $\,(w,\bar w)$,
\qq
&&g_{ww}\,=\,0\,=\,g_{\bar w\bar w}\,,\quad g_{w\bar w}\,=\,\frac{_{-2R^2}}
{^{(w-\bar w)^2}}\,=\,g_{\bar w w}\,,\cr\cr
&&g^{ww}\,=\,0\,=\,g^{\bar w\bar w}\,,\quad g^{w\bar w}\,=\,
\frac{_{(w-\bar w)^2}}{^{-2R^2}}\,=\,g^{\bar w w}\,.
\qqq
and the only non-vanishing Christoffel symbols are
\qq
\Big\{{w\atop ww}\Big\}\,=\,-\frac{_{2}}{^{w-\bar w}}\,,\qquad
\Big\{{\bar w\atop \bar w\bar w}\Big\}\,=\,\frac{_{2}}{^{w-\bar w}}\,.
\label{Christ}
\qqq
The scalar curvature is $\,S=-2R^{-2}$, and the non-vanishing components of the Ricci tensor are
\qq
({\rm Ric})_{w\bar w}\,=\,2(w-\bar w)^{-2}\,=\,({\rm Ric})_{\bar w w}
\qqq

The Euler equation and  the incompressibility condition for the velocity components
$\,v_w=g_{w\bar w}v^{\bar w}\,$ and $\,v_{\bar w}=g_{\bar w w}v^w\,$
take the form
\qq
&&\partial_tv_w\,-\,\frac{_1}{^2}R^{-2}(w-\bar w)^2(v_w\partial_{\bar w}
+v_{\bar w}\partial_w)v_w-R^{-2}(w-\bar w)v_wv_{\bar w}\,=\,-\partial_wp\,,
\\ \cr
&&\partial_tv_{\bar w}\,-\,\frac{_1}{^2}R^{-2}(w-\bar w)^2(v_w\partial_{\bar w}
+v_{\bar w}\partial_w)v_{\bar w}+R^{-2}(w-\bar w)v_wv_{\bar w}\,
=\,-\partial_{\bar w}p\,,\\ \cr
&&\partial_wv_{\bar w}+\partial_{\bar w}v_w\,=\,0\,,\label{incomp2}
\qqq
The stream function
(\ref{psi2d}) and the vorticity (\ref{omega2d}) now satisfy  the
relations
\qq
v_w\,=\,-i\,\partial_w\psi\,,\qquad v_{\bar w}\,=\,i\,\partial_{\bar w}\psi\,.
\qquad\omega\,=\,\Delta\psi\,=\,-R^{-2}(w-\bar w)^2\partial_w
\partial_{\bar w}\psi\,,
\label{vpsiHR}
\qqq
In these terms, the evolution equation (\ref{evomega2d})  becomes
\qq
\partial_t\omega\,+\,\frac{_i}{^{2R^2}}(w-\bar w)^2\big[(\partial_{w}\psi)
(\partial_{\bar w}\omega)-(\partial_{\bar w}\psi)(\partial_w\omega)\big]\,.
\qqq
\vskip 0.2cm

\begin{figure}[h]
\begin{center}
\leavevmode
\includegraphics[width=7cm,height=4.8cm]{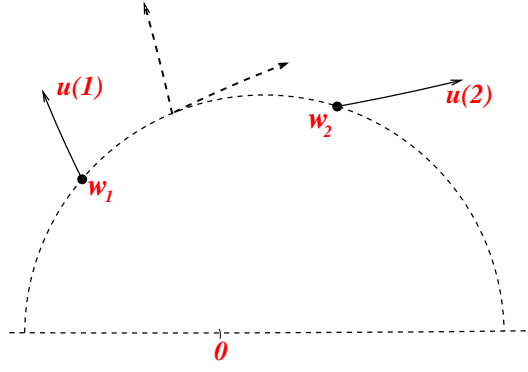}\\
\end{center}
\vskip -0.4cm
\caption{Intrinsic pairing of vectors}
\vskip 0.2cm
\label{fig:diamond}
\end{figure}

The adjoint action of the group $SL(2,\NR)$
\qq
X^i\tau_i\ \longmapsto\ \left(\begin{matrix}{a&b\cr c&d}\end{matrix}\right)
\,X^i\tau_i\,\left(\begin{matrix}{d&-b\cr-c&a}\end{matrix}\right)\ =\
{X'}^i\tau_i
\label{adjact}
\qqq
induces the action of the 3-dimensional Lorentz group $\,SO(2,1)$
\qq
X^i\ \longmapsto\ X'^i\,=\,\Lambda^i_{\ j}X^j\,.
\qqq
%Explicitly,
%\qq
%&&X'^1\,=\,\frac{_1}{^2}(a^2-b^2-c^2+d^2)X^1-(ab-cd)X^2+\frac{_1}{^2}
%(a^2+b^2-c^2-d^2)X^3\,,\\
%&&X'^2\,=\,-(ac-bd)X^1+(ad+bc)X^2-(ac+bd)X^3\,,\\
%&&X'^3\,=\,\frac{_1}{^2}(a^2-b^2+c^2-d^2)X^1-(ab+cd)X^2+\frac{_1}{^2}
%(a^2+b^2+c^2+d^2)X^3\,.\qquad
%\qqq
%The above formulae
that determines the 2-fold covering $\,SL(2,\NR)\rightarrow
SO(2,1)\,$ with the kernel composed of $\,\pm1$. \,The
Lorentz group $\,SO(2,1)\,$ acts effectively on the Lie algebra
$\,sl(2,1)\cong so(2,1)$. \,When restricted to the hyperboloid (\ref{hyperb}),
the action (\ref{adjact}) translates in variable $\,w\,$ to
the fractional action
$w\, \longmapsto\, (aw+b)/(cw+d)$ that preserves the metric (\ref{metr3}).
\,On the infinitesimal
level, the $\,sl(2,\NR)\,$ symmetry is induced by three vector fields
on the upper half-plane:
\qq
(X^w,X^{\bar w})\ =\ \begin{cases}{\,-(1,1)\,,\cr\,-(w,\bar w)\,,\cr
\,-(1+w^2,1+\bar w^2)}\,.
\end{cases}
\qqq
They correspond to three currents
of Eqs.\,(\ref{CJCJX0}) and (\ref{CJCJXl}):
\qq
\CJ^0(t,w,\bar w)&=&\begin{cases}{\,v_w+v_{\bar w}\,,\cr
\,wv_w+\bar wv_{\bar w}\,,\cr
\,(1+w^2)v_w+(1+\bar w^2)v_{\bar w}\,,}\end{cases}\cr\cr
\CJ^w_X(t,w,\bar w)&
=&\begin{cases}{\,(v_w+v_{\bar w})v^w+p\,,\cr\,
(wv_w+\bar wv_{\bar w})w+p\m w\,,\cr
\,((1+w^2)v_w+(1+\bar w^2)v_{\bar w})v^w+p(1+w^2)\,,}\end{cases}\cr\cr
\CJ^{\bar w}_X(t,w,\bar w)&
=&\begin{cases}{\,(v_w+v_{\bar w})v^{\bar w}\,,\cr
\,(wv_w+\bar wv_{\bar w})\bar w+p\m\bar w\,,\cr
\,((1+w^2)v_w+(1+\bar w^2)v_{\bar w})
v^{\bar w}+p(1+\bar w^2)}\end{cases}
\qqq
that satisfy the conservation relation
\qq
\partial_t\CJ^0_X+\big(\partial_w-\frac{_2}{^{w-\bar w}}\big)\CJ^w_X
+\big(\partial_{\bar w}+\frac{_2}{^{w-\bar w}}\big)\CJ^w_X\,=\,0\,.
\qqq
The quadratic form $\,(X^1)^2+(X^2)^2-(X^3)^2\,$ permits to pair
the vectors $\,u(1)\,$ and $\,u(2)\,$ tangent to the upper half-plane
at the points $\,w_1\,$ and $\,w_2$, \,respectively, \,in an
$\,SL(2,\NR)$-covariant way:
\qq
u(1)\centerdot u(2)\,=\,\frac{1}{2R^{2}}\Big((w_1-w_2)^2u_w(1)\,u_w(2)\,-\,
(w_1-\bar w_2)^2u_w(1)\,u_{\bar w}(2)\ +\ c.\,c.\,\Big).\ \quad
\label{copair}
\qqq
We shall call this paring ``extrinsic'' as it uses the embedding
of the hyperbolic plane into the Minkowski space. It turns out, however,
that this pairing is unphysical as it leads to correlation
functions of velocity that grow with the distance.
Therefore, we use instead an intrinsic way to pair
two vectors obtained by taking their scalar product after transporting
them by a parallel transport to a common point along the unique geodesic
(corresponding to an Euclidiean half-circle centered on the real line
in the upper half plane) joining the two attachment points,
see Fig.\,\ref{fig:diamond}.
\vskip 0.1cm
\vskip 0.2cm

\noindent This leads to the ``intrinsic'' pairing
\qq
u(1){_{^{\hspace{0.07cm}\diamond\,}}}
u(2)\ =\ \frac{{(w_1-\bar w_1)(w_2-\bar w_2)}}{{2R^2}}\,
\Big(\frac{{w_1-\bar w_2}}{{\bar w_1-w_2}}\,u_{w}(1)\,u_{\bar w}(2)\,+\ c.\,c.
\Big).
\label{copair2}
\qqq
At equal points, both pairings coincide with the Riemannian
scalar product.

\nsection{Harmonic analysis on $\,\NH_{\hspace{-0.01cm}_R}$}
\label{sec:harman}

The $\,SL(2,\NR)\,$ symmetry of functions or vector fields on
$\,\NH_{\hspace{-0.01cm}_R}\,$
may be studied with the help of harmonic analysis on $\,SL(2,\NR)$,
\,see \cite{Vilenkin}. \,In particular, in the space
$\,L^2(\NH_{\hspace{-0.01cm}_R})\,$ of functions
on $\,\NH_{\hspace{-0.01cm}_R}\,$ square-integrable with respect
to the volume measure (\ref{vol3}), the group
$\,SL(2,\NR)\,$ acts unitarily by the formula
\qq
(\gamma\psi)(w)\ =\ \psi(\gamma^{-1}w)\,.
\label{regact}
\qqq
The induced infinitesimal action of the Lie algebra $\,sl(2,\NR)\,$ is
given by the formula
$\tau_i\psi\ =\ \lim\limits_{\epsilon\to0}\ \ee^{\epsilon\tau_i}\psi$
and the Laplacian (\ref{Lapl}) by
\qq
\Delta\psi\ =\ (\tau_1^2+\tau_2^2-\tau_3^2)\m\psi\,.
\label{LaplH}
\qqq
For the reflection $\,P\,$ in the axis of imaginary $\,w\,$ acting
on $\,\NH_{\hspace{-0.01cm}_R}\,$ by
$Pw\,=\,-\bar w$ we define the induced action on functions by
$(P\psi)(w)\,=\,\pm\psi(-\bar w)$
with the sign $\,\pm\,$ for scalar or pseudo-scalar functions.
We shall also consider the space $\,L^2(\Vect)\,$ of
vector fields $\,v\,$ on $\,\NH_{\hspace{-0.01cm}_R}\,$ with the norm
squared $\,\Vert v\Vert^2$, \,see Eq.\,(\ref{normsq}).
\,Group $\,SL(2,\NR)\,$ acts an such vector fields by the formula
\qq
(\gamma v)^i(w)\ =\ (T\gamma) v(\gamma^{-1}v)\,,
\label{vact}
\qqq
where $\,T\gamma\,$ is the tangent map. In coordinates, if $\,\gamma w=(aw+b)/(cw+d)$ then
\qq
(\gamma v)^w(\gamma(w))\,=\,(cw+d)^{-2}v^w(w)\,,\qquad(\gamma v)^{\bar w}
(\gamma(w))\,=\,(c\bar w+d)^{-2}v^{\bar w}(w)\,.
\qqq
Action (\ref{vact}) preserves
$\,\Vert v\Vert^2$.  The reflection $\,P\,$ acts on the vector fields
$\,v\,$ by
\qq
(Pv)^w(w)\,=\,-v^{\bar w}(-\bar w)\,,\qquad (Pv)^{\bar w}(w)\,
=\,-v^{w}(-\bar w)
\qqq
and also conserves the norm. Both actions map
divergenceless vector fields to the vector fields with the same property.
\vskip 0.2cm

It is convenient to parameterize $\,\gamma\in SL(2,\NR)\,$ by
the variables $\,\phi\in[-\pi,\pi]$, $\,\xi\in[-2\pi,2\pi]$,
$\,{\rho}\in[0,+\infty)\,$ by writing
\qq
\gamma\ =\ \left(\begin{matrix}{\ \ \cos{\frac{\phi}{2}}&
\sin{\frac{\phi}{2}}\cr
-\sin{\frac{\phi}{2}}&\cos{\frac{\phi}{2}}}
\end{matrix}\right)\left(\begin{matrix}{\m\ee^{\frac{{\rho}}{2R}}&0\cr0&
\ee^{-\frac{{\rho}}{2R}}}
\end{matrix}\right)\left(\begin{matrix}{\ \ \cos{\frac{\xi}{2}}&
\sin{\frac{\xi}{2}}\cr-\sin{\frac{\xi}{2}}&\cos{\frac{\xi}{2}}}
\end{matrix}\right)\,\equiv\,\gamma_\phi \gamma_{\frac{_{\rho}}{^R}}
\gamma_\xi\,.
\label{gpara}
\qqq
By acting by $\,\gamma\,$ on $\,i\in\NH_{\hspace{-0.01cm}_R}$, \,we may parameterize $\,\NH_{\hspace{-0.01cm}_R}\,$ by
the variables $\,({\rho},\phi)\,$
(the fractional action of a pure rotation matrix preserves $\,i$).
\,In this parametrization, $\,{\rho}\,$ is the hyperbolic distance from
$\,i$. For fixed $\,{\rho}$, \,the points with different $\,\phi\,$
lie on the Euclidean circle in the upper half plane with the center
at $\,i\cosh({\rho}/{R})$ and with the Euclidean radius
$\,\sinh({\rho}/{R})$.
\,In this parametrization of $\,\NH_{\hspace{-0.01cm}_R}$, \,the hyperbolic metric and
the volume measure become
\qq
g\,=\,({\rm d}{\rho})^2+\,\m({\rm d}\phi)^2 R^2\sinh^2({\rho}/{R}),\qquad
\chi\,=\,R\m\sinh({\rho}/{R})\m\,{\rm d}{\rho}\wedge{\rm d}\phi\,.
\qqq
The reflection $\,P\,$ acts by sending $\,(\rho,\phi)\,$ to
$\,(\rho,-\phi)$.
\vskip 0.2cm

The unitary representation of $\,SL(2,\NR)\,$ in
$\,L^2(\NH_{\hspace{-0.01cm}_R},\chi)\,$
may be decomposed into the direct integral of irreducible
representations of the even principal series \cite{Vilenkin}:
\qq
L^2(\NH_{\hspace{-0.01cm}_R})\ =\
{\int\limits_0^\infty}\hspace{-0.05cm}{}^{\,^{^\oplus}}H_\sigma\,\,\sigma\,
\tanh(\pi\sigma)\,{\rm d}\sigma\,.
\label{directint}
\qqq
It is realized by the $\,\NH_{\hspace{-0.01cm}_R}\,$ Fourier transform
(in which we substitute $\,\sigma=Rk\,$
to make evident the relation to the flat space decomposition into the
polar harmonics that arises in the $\,R\to\infty\,$ limit):
\qq
\psi({\rho},\phi)\ =\
\sum\limits_{m=-\infty}^{\infty}\ee^{-im\phi}\int\limits_0^\infty
\CP^{-\frac{1}{2}+iR{k}}_{m0}(\cosh{\frac{_{\rho}}{^R}})\,\m
a_m({k})\,\m{k}\,
\tanh(\pi R{k})\,{\rm d}{k}\,,
\label{idecom}
\qqq
where
\qq
\CP^{\ell}_{mn}(\cosh{\frac{_{\rho}}{^R}})&=&\frac{_1}{^{2\pi}}
\int\limits_{-\pi}^{\pi}\left(\cosh{\frac{_{\rho}}{^{2R}}}+\ee^{i\vartheta}
\sinh{\frac{_{\rho}}{^{2R}}}\right)^{\ell+n}\cr
&&\quad\ \cdot\ \left(\cosh{\frac{_{\rho}}{^{2R}}}+\ee^{-i\vartheta}
\sinh{\frac{_{\rho}}{^{2R}}}\right)^{\ell-n}\ee^{i(m-n)\vartheta}\,{\rm d}\vartheta
\qqq
are the Legendre functions, and, in particular,
\qq
\CP^{\ell}_{m0}(\cosh{\frac{_{\rho}}{^R}})\
=\ \frac{_1}{^{2\pi}}\int\limits_{-\pi}^{\pi}
\left(\cosh{\frac{_{\rho}}{^R}}+\cos{\vartheta}\m\sinh{\frac{_{\rho}}{^R}}
\right)^\ell\ee^{im\vartheta}\,\m {\rm d}\vartheta\,.
\label{Pm0}
\qqq
Functions $\,\CP^\ell_{mn}\,$ satisfy the Legendre equation
\qq
(x^2-1)(\CP^{\ell}_{mn})''(x)\,+\,2\m x\,(\CP^{\ell}_{mn})'(x)\,-\,
\Big(\frac{_{m^2-2mn\m x+n^2}}{^{x^2-1}}+\ell(\ell+1)\Big)
\m\CP^{\ell}_{m0}(x)\ =\ 0\,.\quad
\label{LEq}
\qqq
and the relations
\qq
&&\CP^\ell_{(-m)\m(-n)}(x)\,=\,\CP^\ell_{mn}(x)\,,\qquad
\CP^\ell_{mn}(1)\,=\,\delta_{mn}\,,
\label{normal}\\
&&\CP_{mn}^{-\frac{1}{2}-i\sigma}(x)\ =\ (-1)^{m-n}\,
\frac{_{\Gamma(\frac{1}{2}+i\sigma+m)
\,\Gamma(\frac{1}{2}-i\sigma-m)}}{^{\Gamma(\frac{1}{2}-i\sigma+n)\,
\Gamma(\frac{1}{2}+i\sigma-n)}}\,\,\CP_{mn}^{-\frac{1}{2}+i\sigma}(x)\,.
\qqq
One has the inverse $\,\NH_{\hspace{-0.01cm}_R}\,$ Fourier transform formula:
\qq
a_m(k)\ =\ \frac{_1}{^{2\pi}}\int\limits_0^\infty R\sinh\hspace{-0.06cm}
\frac{_r}{^R}\,dr\int\limits_{-\pi}^\pi\overline{\ee^{-im\phi}\,\m
\CP^{-\frac{1}{2}+iRk}_{m0}(\cosh\hspace{-0.06cm}
\frac{_r}{^R})}\,\,
 \psi(r,\phi)\,\,{\rm d}\phi\,,
\label{amk}
\qqq
and the Plancherel formula:
\qq
\int\limits_0^\infty R\m\sinh{\hspace{-0.06cm}\frac{_{\rho}}{^R}}\m\,{\rm d}{\rho}
\int\limits_{-\pi}^{\pi}
|\psi({\rho},\phi)|^2\,\,{\rm d}\phi\ =\ {2\pi}
\sum\limits_{m=-\infty}^\infty\int\limits_0^\infty
|a_m({k})|^2\,{k}\tanh(\pi R{k})\,{\rm d}{k}\,.
\label{Planch}
\qqq
As in the flat space, the above Fourier transform and its inverse may be
generalized to broader classes of functions (and distributions) on
$\,\NH_{\hspace{-0.01cm}_R}$.
\vskip 0.2cm

The even principal series irreducible components
$\,H_{\hspace{-0.03cm}_{Rk}}\,$ of the $\,SL(2,\NR)\,$ action
(\ref{regact}) in $\,L^2(\NH_{\hspace{-0.01cm}_R})$, \,see Eq.\,(\ref{directint}),
\,correspond to fixed $\,\ell=-\frac{1}{2}+iR{k}$. \,They
are spanned by the functions
\qq
\psi^{k}_m({\rho},\phi)\,\equiv\,
\ee^{-im\phi}\,\m\CP^{-\frac{1}{2}+iR{k}}_{m0}(\cosh{\frac{_{\rho}}{^R}})\,
\label{pmr}
\qqq
that are eigenfunctions of the hyperbolic Laplacian (\ref{LaplH}) on scalars
\qq
\Delta&=&-R^{-2}
(w-\bar w)^2\partial_w
\partial_{\bar w}
\ =\ \sinh^{-1}{\hspace{-0.06cm}\frac{_{\rho}}{^R}}\,
\partial_{\rho}\sinh{\hspace{-0.06cm}\frac{_{\rho}}{^R}}\,
\partial_{\rho}+R^{-2}\sinh^{-2}\hspace{-0.06cm}{\frac{_r}{^R}}\,
\partial_\phi^2\cr
&=&R^{-2}\left(\partial_x(x^2-1)\partial_x\,+\,\frac{_1}{^{x^2-1}}\,
\partial_\phi^2\right)
\label{Delta}
\qqq
for $\,x=\cosh\hspace{-0.03cm}\frac{\rho}{R}\,$
with the eigenvalue $\,R^{-2}\ell(\ell+1)=-(\frac{1}{4}R^{-2}+{k}^2)$.
\,This follows from the Legendre equation (\ref{LEq}).
\,The left regular action of $\,SL(2,\NR)\,$ on functions
(\ref{pmr}) is given by the formulae:
\qq
(\gamma_{\phi_{_0}}\psi_m^{k})({\rho},\phi)&=&\ee^{im\phi_{_0}}\,\psi_m^{k}({\rho},
\phi)\,,\\
(\gamma_{\frac{{\rho}_{_0}}{R}}\psi_m^{k})({\rho},\phi)
&=&\sum\limits_{n=-\infty}^\infty
\overline{\CP^{-\frac{1}{2}
+iR{k}}_{nm}(\cosh\frac{_{{\rho}_{_0}}}{^R})}\,\,\psi_n^{k}({\rho},\phi)\,.
\qqq
As for the reflections,
\qq
(P\psi_m^k)(\rho,\phi)\,=\,\pm\m\psi_m^k(\rho,-\phi)\,
=\,\pm\m\psi_{-m}^k(\rho,\phi)\,.
\qqq
with the sign depending whether we use the scalar or the pseudo-scalar rule.
Hence function $\,P\psi\,$ corresponds in decomposition (\ref{idecom})
to the Fourier coefficients $\,(Pa)_m(k)\,=\,\pm\m a_{-m}(k)$.
\vskip 0.2cm

We would like to relate the actions of $\,SL(2,\NR)\,$ on functions and
on divergenceless vector fields. Every 1-form $\,v^\flat\,$ on
$\,\NH_{\hspace{-0.01cm}_R}\,$ satisfying the divergenceless condition
$\,{\rm d}^\dagger v^\flat=0\,$ may be written as $\,*{\rm d}\psi\,$ 
where $\,\psi\,$ is a
function on $\,\NH_{\hspace{-0.01cm}_R}$, \,see Eqs.\,(\ref{stream}),
(\ref{psi2d}) and (\ref{vpsiHR}). If $\,v^\flat\,$ corresponds
to an incompressible velocity field $\,v\,$ then $\,\psi\,$ is its stream
function. The relation between $\,\psi\,$ and $\,v\,$ commutes with the
$\,SL(2,\NR)$-action. \,It also commutes with the action of the reflection
$\,P\,$ if $\,\psi\,$ is a pseudo-scalar.
\,For the $\,L^2$-norm of the divergenceless vector field $\,v$, \,one
obtains:
\qq
\Vert v\Vert^2\ =\ ({\rm d}\psi,{\rm d}\psi)\,.
\label{strnorm}
\qqq
It follows that the Hilbert space $\,L^2(\Vect_\chi)\,$ of the
divergenceless vector fields $\,v\,$ with the $\,L^2$-norm may be
identified with the space of functions $\,\psi\,$ with the Sobolev norm
(\ref{strnorm}). There is a little catch here \cite{KM}. In fact,
$\,L^2(\Vect_\chi)\,$ splits into a direct orthogonal sum of two
subspaces
\qq
L^2(\Vect_\chi)\ =\ L^2_0(\Vect_\chi)\oplus \CH(\Vect_\chi)
\qqq
by the Hodge decomposition \cite{Car}. The square-integrable
incompressible vector fields $\,v\,$ in $\,L^2_0(\Vect_\chi)\,$
correspond to stream functions $\,\psi\,$ in
$\,L^2(\NH_{\hspace{-0.01cm}_R})$, \,whereas the ones in
$\,\CH(\Vect_\chi)\,$ are the harmonic ones such that
$\,{\rm d}v^\flat=0={\rm d}^\dagger v^\flat$. 
\,The latter may be obtained from harmonic
stream functions $\,\psi\,$ which do not vanish when $\,w\,$ tends
to the real line, as required for square-integrable functions.
By integration by parts,  the scalar product (\ref{strnorm})
may be rewritten for $\,v\in L^2_0(\Vect_\chi)\,$ as the matrix element
$(\psi,-\Delta\psi)\,\geq\,\frac{_{1}}{^4}R^{-2}
\Vert\psi\Vert^2$,
whereas for $\,v\in\CH(\Vect_\chi)\,$ as the boundary integral 
over the real axis:
$-i\int%\limits_{{\rm Im}\,w=0}
\psi (\partial_w-\partial_{\bar w})\psi
\,{\rm d}\hspace{0.01cm}{\rm Re}\,w$.
Harmonic vector fields \,are eliminated if we demand that
$\,v_w,\,v_{\bar w}\,$ tend to zero
when $\,{\rm Im}\,w\to0$, \,as we should do for the well-posedness
of the Euler and Navier-Stokes equations on $\,\NH_{\hspace{-0.02cm}_R}$,
as mentioned in Appendix \ref{appx:varprinc}, \,see also \cite{CC,KM}.
The decomposition (\ref{idecom}) induces a decomposition of
the non-harmonic subspace $\,L^2_0(\Vect_\chi)\,$
of square-integrable (complex) divergenceless vector fields into
the irreducible components of the even complementary series of the action
of $\,SL(2,\NR)\,$ (the harmonic space $\,\CH(\Vect_\chi)\,$ carries
the representation of the discrete series $\,\CD(1,\pm)$).
\vskip 0.2cm

Suppose that the stream functions $\,\psi\,$ form a random process
with a $\,SL(2,\NR)\,$-invariant distribution and equal-time
2-point function that, consequently, depends only on the 
hyperbolic distance $\,\delta_{12}$ between the points,
see Eq.\,(\ref{hypdist}),
\qq
\big\langle\,\psi(\CX_1)\,\psi(\CX_2)\,\big\rangle
\ \equiv\ \Psi(\cosh\frac{_{\delta_{12}}}{^R})\,.\label{fpp0}
\qqq
Under the Fourier transform, 
\qq
\Psi(\cosh\frac{_{\delta_{12}}}{^R})\ =\ 
\int\limits_0^\infty\CP^{iR{k}-\frac{1}{2} }_{00}
(\cosh\frac{_{\delta_{12}}}{^R})\m\,
f(k)\,{\rm d}{k}
\label{fpp}
\qqq
The positivity of the covariance means that $\,f(k)\geq 0$.
\,The equal-time correlation functions of the velocity
components are given by the formula
\qq
\big\langle\,v^{i_1}(\CX_1)\,v^{i_2}(\CX_2)\,\big\rangle\ =\
\epsilon^{i_1j_1}\epsilon^{i_2j_2}\frac{_1}
{^{\sqrt{g(\CX_1)}\m\sqrt{g(\CX_2)}}}\,\m\partial_{j_1}(1)\,\partial_{j_2}(2)\,
\big\langle\,\psi(\CX_1)\,\psi(\CX_2)\,\big\rangle\,.
\label{vel2pt}
\qqq
For the velocity 2-point function built using the extrinsic paring
(\ref{copair}), a straightforward although somewhat
tedious calculation gives
\qq
\big\langle\,v(\CX_1)\centerdot v(\CX_2)\,\big\rangle\ =\
-(\Delta\Psi)(\cosh\hspace{-0.06cm}\frac{_{\delta_{12}}}{^R})\,,
\label{<v.v>}
\qqq
where above and below,
\qq
(\Delta\Psi)(x)=R^{-2}\,\big[(x^2-1)\,
\partial_x\Psi''(x)+2x\Psi'(x)\big]\,,
\qqq
coinciding with the action of the Laplacian (\ref{Delta})
if we view $\,\Psi(x)\,$ as a function of the distance from 
$\,i\in\NH_{_R}$. \,Since the Legendre equation (\ref{LEq}) reduces 
for $\,\CP_{00}^{-\frac{1}{2}+iR{k}}\,$ to the relation
\qq
(x^2-1)\,(\CP_{00}^{-\frac{1}{2}+iR{k}})''(x)\,+\,2\m x\,
(\CP_{00}^{-\frac{1}{2}+iR{k}})'(x)\,
=\,-(\frac{_1}{^4}+R^2{k}^2)\,\CP_{00}^{-\frac{1}{2}+iR{k}}(x)\,,
\qqq
we obtain the identity
\qq
\big\langle\,v(\CX_1)\centerdot v(\CX_2)\,\big\rangle\ =\ 2\,\int\limits_0^\infty
\CP^{-\frac{1}{2}+iR{k}}_{00}(\cosh\hspace{-0.06cm}
\frac{_{\delta_{12}}}{^R})\m\,\CE({k})\,{\rm d}{k}\,,\qquad
\label{cov2pt}
\qqq
where
\qq
\CE(k)\,\equiv\,\frac{_1}{^2}\,({\frac{_1}{^4}R^{-2}+{k}^2})\,f(k)\,.
\label{spectrum}
\qqq
In particular, taking coinciding points, one gets for the mean energy
density with the help of (\ref{normal}) the spectral integral:
\qq
\frac{_1}{^2}\m\big\langle\,(g_{i_1i_2}v^{i_1}v^{i_2})(\CX)\,\big\rangle\ =\
\frac{_1}{^2}\m\big\langle\,v(\CX)\centerdot v(\CX)\,\big\rangle\ =\
\int\limits_0^\infty\CE({k})\,{\rm d}{k}\,.
\qqq
We see that the function $\,\CE({k})\geq0\,$ has the meaning of the
energy spectrum.
\,Similarly, for the velocity 2-point function based on the intrinsic
pairing (\ref{copair2}) using the parallel transport, one obtains
the expression
\qq
\big\langle\,v(\CX_1){_{^{\hspace{0.07cm}\diamond\,}}}
v(\CX_2)\,\big\rangle\ =\
-(\Delta^{\hspace{-0.095cm}^{_\diamond}}\Psi)(\cosh\hspace{-0.06cm}
\frac{_{\delta_{12}}}{^R})\,,
\label{<vstarv>}
\qqq
where
\qq
(\Delta^{\hspace{-0.095cm}^{_\diamond}}\Psi)(x)\,=\,R^{-2}\,[(x^2-1)\,
\Psi''(x)+(x+1)\,\Psi'(x)]\,.
\label{Lapdif}
\qqq
The 2-point function of the vorticity 
$\,\omega=\Delta\psi\,$ is given, in turn, by the formula
\qq
\big\langle\,\omega(\CX_1)\,\omega(\CX_2)\,\big\rangle&=&
\Delta(1)\,\Delta(2)\,\big\langle\,\psi(\CX_1)\,\psi(\CX_2)\,\big\rangle
\ =\ (\Delta^2\Psi)(\cosh\hspace{-0.06cm}
\frac{_{\delta_{12}}}{^R})\cr
&=&
\int\limits_0^\infty\CP^{-\frac{1}{2}+ik}_{00}(\cosh\hspace{-0.06cm}
\frac{_{\delta_{12}}}{^R})\m\,(\frac{_1}{^4}R^{-2}+k^2)\,2\m\CE({k})\,{\rm d}{k}\,,
\qqq
where the second equality results from the expression
(\ref{LaplH}) for the hyperbolic Laplacian and the $\,SL(2,\NR)\,$
invariance of the 2-point function of $\,\psi$.
\,In particular, the mean enstrophy density is
\qq
\frac{_1}{^2}\,\big\langle\,\omega^2(\CX)\,\big\rangle\ =\
\int\limits_0^\infty(\frac{_1}{^4}R^{-2}+k^2)\,\CE(k)\,{\rm d}k\,,
\qqq
i.e. $\,(\frac{1}{4}R^{-2}+k^2)\m\CE(k)\,$ is the enstrophy
spectrum.
\vskip 0.2cm

In fact, we shall be interested in random
$\,SL(2,\NR)$-covariant velocity processes taking values in incompressible
fields and obtained in various limiting regimes. Such processes may not
lift to $\,SL(2,\NR)$-covariant processes with values in stream functions.
If we assume, however, that the velocity equal-time 2-point functions
exist and are sufficiently regular ($C^1$ class is enough) then
there exist symmetric functions $\,\langle\psi(\CX_1)\,\psi(\CX_2)\rangle\,$
on $\,\NH_{\hspace{-0.01cm}_R}^2\,$ of positive type such that
Eq.\,(\ref{vel2pt}) holds. Besides,
\qq
\langle\psi(\CX_1)\,\psi(\CX_2)\rangle\,
=\,\Psi(\cosh\frac{_{\delta_{12}}}{^R})\ +\ \,\dots
\qqq
where the dots stand for terms that do not contribute to the velocity
2-point functions that determine function $\,\Psi\,$ uniquely
up to an overall constant. This is shown in Appendix \ref{appx:stream}.
The $\,SL(2,\NR)$-invariant part $\,\Psi\,$ might, however, be not
of the positive type. Formulae (\ref{<v.v>}) and (\ref{<vstarv>}) still
hold in this case.

\nsection{Navier-Stokes equation on the upper half-plane}
\label{sec:NSonH}

The Navier-Stokes equation on the upper half-plane takes the form:
\qq
&&\partial_tv_w\,-\,\frac{_1}{^2}R^{-2}(w-\bar w)^2(v_w\partial_{\bar w}
+v_{\bar w}\partial_w)v_w-R^{-2}(w-\bar w)v_wv_{\bar w}\cr\cr
&&+\,\nu R^{-2}\big[(w-\bar w)^2\partial_w\partial_{\bar w}v_w+2(w-\bar w)
\partial_{\bar w}v_w+2v_w\big]\,=\,-\partial_wp+f_w
\qqq
plus its complex conjugate and the incompressibility condition
(\ref{incomp2}). As before, we shall suppose that the force $\,f\,$
is incompressible and non-harmonic, random, white-noise in time,
see Eq.\,(\ref{fcov}), and that the spatial part of its covariance is
\qq
C^{i_1i_2}(\CX_1,\CX_2)\ =\ \,\epsilon^{i_1j_1}\epsilon^{i_2j_2}\frac{_1}
{^{\sqrt{g(\CX_1)}\m\sqrt{g(\CX_2)}}}\,\m\partial_{j_1}(1)\,\partial_{j_2}(2)
\,\,\CC\big(\cosh\hspace{-0.06cm}\frac{_{\delta_{12}}}{^R}\big)\,.
\label{8.2}
\qqq
The distribution of the force is then invariant
under the $\,SL(2,\NR)$-transformations and reflections of the vector
fields $\,f$. \,The mean injection rate
\qq
\bar\iota\ =\ \frac{_1}{^2}\,g^{j_1j_2}(\CX_1)\,\partial_{j_1}(1)\,
\partial_{j_2}(2)\,\CC\big(\cosh\hspace{-0.06cm}\frac{_{\delta_{12}}}{^R}\big)
\Big|_{_{\CX_2=\CX_1}}\label{inj1}
\qqq
is constant in space. A direct calculation shows that
\qq
\bar\iota\ =\ -R^{-2}\,\CC'(1)\,.
\label{bariota}
\qqq
Under such forcing and, say, zero initial condition for the velocities,
their distribution will be invariant under the $\,SL(2,\NR)$-transformations
and reflections of $\,v$.
\vskip 0.2cm

\nsection{Inverse cascade scenario}
\label{sec:invcasc}

We shall assume that the forcing is short-scale, $k_fR\gg1$, \,i.e. that
function $\,\CC\,$ in Eq.\,(\ref{8.2}) vanishes for the arguments
not very close to 1 and that
viscosity is so small that it does not effect the flow statistics
except at distances shorter than $1/k_f$. Let us recall that in such situation
in the infinite two-dimensional flat space one expects an inverse energy
cascade to develop \cite{Kr67,Batch}. In terms of the energy spectrum
\qq
\CE(t,k)\,=\,k^{-1}\hspace{-0.1cm}\int\ee^{-i\,{\rm k}\cdot x_{12}}\,
\big\langle\,v(x_1)\cdot v(x_2)\m\big\rangle\,{\rm d}x_{12}\,,
\qqq
where $k=|{\rm k}|$ and $x_{12}=x_1-x_2$, \,the energy injected around
the large modes $\,k\approx k_{f}\,$ starting at time $t=0$
\,flows to smaller $\,k\,$ building a stationary spectrum
$\,\CE_{stat}(k)\propto\bar\iota^{\hspace{0.02cm}2/3}k^{-5/3}\,$
for modes $\,k_f\gg k\gg k_t\sim(\bar\iota t^3)^{-{1/2}}$, \,with
the decay of $\,\CE(t,k)\,$
for $\,k\ll k_t\,$ and $\,k\gg k_f$.  \,This means that asymptotically
\qq
\CE(t,k)\ \approx\ t\,\bar\iota\,\delta(k)\,+\,\CE_{stat}(k)
\qqq
for large $\,t$, \,or, \,more formally, \,that
$\,\CE(t,k)-t\,\bar\iota\,\delta(k)\,\mathop{\longrightarrow}\limits_{t\to\infty}
\,\CE_{stat}(k)\,$
in the sense of distributions, where the distribution $\,k^{-5/3}\,$
is defined for small $\,k\,$ by analytic continuation of $\,k^\lambda\,$
from $\,{\rm Re}\,\lambda>-1$. \,\,On the other hand,
the enstrophy spectrum becomes stationary
\qq
k^2\,\CE(t,k)\ \mathop{\longrightarrow}\limits_{t\to\infty}\ k^2\,\CE_{stat}(k)
\qqq
and for very long times the enstrophy is dissipated only at short distances
(shorter than the injection scale).
\,In the position space,
\qq
\big\langle\,v(x_1)\cdot v(x_2)\m\big\rangle\ =\ t\,\m\bar\iota\,
\CF\Big(\frac{|x_{12}|^2}{\bar\iota\m t^3}\Big)
\label{scflat}
\qqq
for large times and distances. The inverse cascade means that the
scaling function is
$\CF(y)\ =\ 2-c\,y^{1/3}\,+\,o(y^{1/3})$ for small $\,y\,$ so that
\qq
\big\langle\,v(x_1)\cdot v(x_2)\m\big\rangle\ =\
2\,t\,\m\bar\iota-c\,\bar\iota^{2/3}|x_{12}|^{2/3}\label{vvflat}
\label{Estat}
\qqq
for $\,k_f^{-1}\ll|x_{12}|\ll(\bar\iota t^3)^{1/2}\,$ up to terms
that vanish when $\,t\to\infty$. \,At long times, the flat-space
covariance of the stream function is asymptotically
\qq
\big\langle\,\psi(x_1)\,\psi(x_2)\m\big\rangle\ =\
-\m\frac{_1}{^2}\,\bar\iota\,|x_1-x_2|^2\,t\,+\,c\,
{\bar\iota}^{\hspace{0.02cm}\frac{2}{3}}\,|x_1-x_2|^{\frac{8}{3}}\ +\ \dots
\label{flatinvc}
\qqq
with another constant $\,c$, \,where the dots represent terms
that vanish for long times and do not contribute the velocity correlation
function. \,When the viscosity is not infinitesimally
small, a similar inverse-cascade of energy seems to occur at sufficiently
small wavenumbers or long distances, but with the injection
rate $\,\bar\iota\,$ replaced by its fraction and the rest of the injected
energy dissipated at small or moderate distances.
\vskip 0.2cm

It is reasonable to expect that in the hyperbolic space
$\,\NH_{\hspace{-0.01cm}_R}$, \,the inverse cascade
of energy still takes place if $\,k_f\gg R^{-1}\,$ for times such that
$\,k_t\gg R^{-1}\,$ when one can ignore the effects of the curvature.
What happens next, is at the moment a matter of a guess. The inverse
cascade could stop after the time when it reaches $\,k\sim R^{-1}$. \,This
is what happens on a sphere of radius $\,R\,$ where there are no points
with distances $\,>\pi R$. \,After reaching the largest distance, the cascade
starts feeding energy into largest-scale flow that eventually
changes the spectrum for $\,k\gg R^{-1}$ as well \cite{XSF}.  In the hyperbolic plane,
however, there is more and more space available at distances $\,\gg R\,$
(the circumference of a circle of radius $\,\delta\,$ is equal to
$\,2\pi R\sinh(\delta/R)$ increasing
exponentially with the radius for $\delta\gg R$). \,Hence we see no
reason for the cascade to stop at $\,k\sim R^{-1}$.
\,We now explore the scenario of continued flow of energy
on $\,\NH_{_R}\,$ to long scales, \,assuming that at long times the
equal-time 2-point correlation function of the stream function takes
the form
\qq
\big\langle\psi(\CX_1)\,\psi(\CX_2)\big\rangle\
= \,t\,\Psi_0(\cosh\hspace{-0.06cm}\frac{_{\delta_{12}}}{^R})
+\,\Psi_{stat}(\cosh\hspace{-0.06cm}\frac{_{\delta_{12}}}{^R})\ +\ \dots
\label{invc}
\qqq
with
\qq
-\frac{_1}{_2}(\Delta\Psi_0)(1)\ =\ -R^{-2}\Psi_0'(1)\ =\ \bar\iota
\label{Psi0pr}
\qqq
fixed by the energy injection rate in the limit of vanishing viscosity.
\,From Eqs.\,(\ref{<v.v>}) and (\ref{<vstarv>}), we obtain then for "extrinsic" and "intrinsic" velocity 2-point functions invariant with respect to
$\,SL(2,\NR)$ (that are functions only of time and
the hyperbolic distance between the points):
\qq
\big\langle v(t,\CX_1)\centerdot v(t,\CX_2)\big\rangle&\,=\,&- \,t(\Delta\Psi_0)
(\cosh\hspace{-0.06cm}\frac{_{\delta_{12}}}{^R})\ -\
(\Delta\Psi_{stat})(\cosh\hspace{-0.06cm}\frac{_{\delta_{12}}}{^R})\,,
\label{invcv}\\
\big\langle v(t,\CX_1){_{^{\hspace{0.07cm}\diamond\,}}}v(t,\CX_2)\big\rangle&\,=\,&- \,t
(\Delta^{\hspace{-0.095cm}^{_\diamond}}\Psi_0)
(\cosh\hspace{-0.06cm}\frac{_{\delta_{12}}}{^R})\ -\
(\Delta^{\hspace{-0.095cm}^{_\diamond}}\Psi_{stat})(\cosh\hspace{-0.06cm}\frac{_{\delta_{12}}}{^R})\,.
\label{invcvv}
\qqq
Such relations should hold for asymptotically long times at distances
much longer than the forcing scale $\,\ell_f\equiv k_f^{-1}\ll R$. The
behavior of $\,\Psi_0\,$ and $\,\Psi_{stat}\,$ at distances much longer
than $\,\ell_f\,$ but
much smaller than $\,R\,$ should approximately be given by
Eq.\,(\ref{flatinvc}). \,We shall try to find in the sequel the asymptotic
behavior of those functions at distances much larger than $\,R$.
In particular, we shall see that $\,\Psi_0\,$ is very different than the
zero wave-number mode proportional to $\,P^{-\frac{1}{2}}_{00}$. \,The spectral
consequences of the assumption (\ref{invc}) will also be examined.
\vskip 0.2cm

Actually, similarly to (\ref{vvflat}), we may expect the overall
scaling
\qq
\big\langle v(t,\CX_1){_{^{\hspace{0.07cm}\diamond\,}}}
v(t,\CX_2)\big\rangle \,=\,\bar\iota\,t\,
\CF\Big(\frac{\delta_{12}^2}{\bar\iota\m t^3},\,\cosh \frac{\delta_{12}}{R}
\Big).
\label{vvcurved}
\qqq
for long times and long distances, see Sec.\,\ref{sec:Scal}.
In what follows, the most important
part of our assumption  is the finite limit of the scaling function
${\cal F}$ when the first argument is zero, i.e. the existence of $\,\Psi_0$.
\,The assumption on the stationarity of the first subleading term is
less relevant to the main subject of the paper, which is the nontrivial
structure of the growing 2-point function mode $\,\Psi_0$.

\nsection{Global flux relation}
\label{sec:GFR}

\noindent In the flat space, the inverse cascade pumps the lowest
$\,k=0\,$ Fourier mode which is constant in space. The stabilization
of the spectrum at non-zero $k$  translates
into the stabilization of the velocity 2$^{\rm nd}$ order structure
function. It also permits to extract an exact ``flux relation'', that holds at long times and
at distances much larger than the forcing scale $\,\ell_f$:
\qq
\partial_t\m\big\langle\,v(x_1)\cdot v(x_2)\,\big\rangle\ =\
-\,2\,\partial_{x_{12}^i}F^i(x_1,x_2)\,.
\qqq
Here
$F^i(x_1,x_2)\ =\ \big\langle\,v^i(x_1)\,(v(x_1)\cdot v(x_2))\,\big\rangle$
is a 2-point third-order velocity correlation function,
\,The flux relation, together
with the inverse-cascade scenario (\ref{Estat}), imposes on the position-space energy current across scales a linear growth with the distance:
\qq
\Theta(|x_{12}|)\ =\ -\,\frac{x_{12}}{|x_{12}|}\cdot F(1,2)\ =\
\frac{_1}{^2}\,\bar\iota\,|x_{12}|\,.
\qqq
\,In the present section,
we  examine analogous implications of the scenario of continuing
energy flow to large scales in the hyperbolic plane.
\vskip 0.2cm

We first consider the velocity 2-point function
$\,\big\langle\,v(t,\CX_1){_{^{\hspace{0.07cm}\diamond\,}}}
v(t,\CX_2)\,\big\rangle$ contracted in the
intrinsic way, \,see Eq.\,(\ref{<vstarv>}). \,For the time
derivative of that correlator, we have the expression
\qq
&&\partial_t\,\big\langle\,v(\CX_1){_{^{\hspace{0.07cm}\diamond }}}
v(\CX_2)\,\big\rangle%\cr\cr
\,=\,
\big\langle\,(\partial_tv)(\CX_1){_{^{\hspace{0.07cm}\diamond\,}}}
v(\CX_22)\,\big\rangle+
\big\langle\,v(\CX_1){_{^{\hspace{0.07cm}\diamond\,}}}
(\partial_tv)(\CX_2)\,\big\rangle
+\CG(\CX_1,\CX_2)\,,\qquad
\label{t23star}
\qqq
where
\qq
\CG(\CX_1,\CX_2)\,=\,
\frac{{(w_1-\bar w_1)(w_2-\bar w_2)}}{{2R^2}}\Big(
\,\frac{{w_1-\bar w_2}}{{\bar w_1-w_2}}\,C_{w\bar w}(\CX_1,\CX_2)\,
+\ c.\,c.\,\Big)
\qqq
is the It$\hat{\rm o}$ term written in terms of the spatial force covariance
(\ref{8.2}). In particular, $\,\CG(\CX,\CX)=2\m\bar\iota$, \,see
Eq.\,(\ref{inj1}).
Each term on the right hand side of Eq.\,(\ref{t23star}) is a scalar
function of two points invariant
under the $\,SL(2,\NR)$-action on $\,\NH_{\hspace{-0.01cm}_R}$. \,But for any
configuration of two points in $\,\NH_{\hspace{-0.01cm}_R}\,$ there exists
an element in $\,SL(2,\NR)\,$ that interchanges them. Indeed, any two points
$\,(w_1,w_2)\,$ may be mapped into any other 2-points with the same
value of the cross-ratio $\,
q=(w_1-\bar w_1)(w_2-\bar w_2)/((w_1-w_2)(\bar w_1-\bar w_2))\,<0$,
\,in particular into a pair $\,(\tau i,\tau^{-1}i)\,$ for $\,\tau>0\,$ such that
$\,q=-4(\tau-\tau^{-1})^{-2}$. \,Then the action of the rotation matrix
$\,\gamma_\pi\,$ exchanges the points. It follows that
the first two terms on the right hand side of Eq.\,(\ref{t23star}) are equal,
i.e. that
\qq
&&\partial_t\,\big\langle\,v(\CX_1){_{^{\hspace{0.07cm}\diamond\,}}}
v(\CX_2)\,\big\rangle\ =\
2\,\big\langle\,(\partial_tv)(\CX_1){_{^{\hspace{0.07cm}\diamond\,}}}
v(\CX_2)\,\big\rangle\,
+\,\CG(\CX_1,\CX_2)\cr\cr
&&=\ -\m2\,\big\langle\,V(\CX_1){_{^{\hspace{0.07cm}\diamond\,}}}
v(\CX_2)\,\big\rangle\,+\,
2\m\nu\,\big\langle\,(Lv)(\CX_1){_{^{\hspace{0.07cm}\diamond\,}}}
v(\CX_2)\,
\big\rangle\,+\,\CG(\CX_1,\CX_2)\,,
\label{t23}
\qqq
where we introduced a vector field, induced from $\,v$,
\qq
V^i\ =\ v^k\nabla_k v^i\ =\ \nabla_k(v^kv^i)\,,
\label{V1}
\qqq
which is the acceleration of a fluid element due to velocity spatial inhomogeneity.
\,We have dropped the contributions
from the pressure to the right hand side since the 2-point correlation
function
$\,\big\langle\,p(\CX_1)\,v^i(\CX_2)\m\rangle\,$ of a
scalar and a divergenceless
vector fields will be shown in a moment to vanish.
In complex coordinates,
\qq
V_w&=&\nabla_w(v^wv_w)+\nabla_{\bar w}(v^{\bar w}v_w)\cr\cr
&=&\partial_w(v^wv_w)+\Gamma_{ww}^wv^wv_w-\Gamma_{ww}^wv^wv_w
+\partial_{\bar w}(v^{\bar w}v_w)+\Gamma_{\bar w\bar w}^{\bar w}
v^{\bar w}v_w\\
&=&\partial_w(v^wv_w)
-\frac{_1}{^2}R^{-2}\partial_{\bar w}((w-\bar w)^2v_w^2)
-R^{-2}(w-\bar w)v_w^2\\
&=&\frac{_1}{^2}\partial_w(v^wv_w+v^{\bar w}v_{\bar w})
-\frac{_1}{^2}R^{-2}(w-\bar w)^2\partial_{\bar w}v_w^2\,,
\label{V2}
\qqq
where we have used the identity $\,v^wv_w=g_{w\bar w}v^wv^{\bar w}
=v^{\bar w}v_{\bar w}$.
\,The contribution of the first term on the right hand to
$\,\big\langle\,V(\CX_1){_{^{\hspace{0.07cm}\diamond\,}}}
v(\CX_2)\m\big\rangle\,$ involves again a correlation function
between a scalar and a divergenceless vector fields so it vanishes.
\,Hence
\qq
\big\langle\,V(\CX_1){_{^{\hspace{0.07cm}\diamond\,}}}
v(\CX_2)\,\big\rangle\, &=&\, \frac{1}{2R^{2}}\,
\Big(\,\frac{{(w_1-\bar w_1)^3(w_2-\bar w_2)}}{{|w_1-\bar w_2|^2}}\,
\partial_{\bar w_1}F^{\hspace{0.01cm}^{_\diamond}}_w(\CX_1.\CX_2)\,
+\ c.\,c.\,\Big)
\label{whn}
\qqq
%\vskip -0.1cm
\noindent for
%\vskip -0.1cm
\qq
F^{\hspace{0.01cm}^{_\diamond}}_w(\CX_1,\CX_2)&=&-\frac{_1}{^2}R^{-2}\,(w_1-\bar w_2)^2\,\big\langle
\,v_w^2(\CX_1)\,v_{\bar w}(\CX_2)\,\big\rangle\cr
&=&-R^2\,\big[\partial_{\bar w_1}\partial_{w_2}
\cosh( \delta_{12}/R )\big]\,
\big\langle\,v_w(\CX_1)\m v^{\bar w}(\CX_1)\ v^w(\CX_2)\,
\big\rangle\,,
\qqq
where the last equality follows from Eq.\,(\ref{hypdist})
by a straightforward calculation. \,We shall
denote by $\,F^{\hspace{0.01cm}^{_\diamond}}_{\bar w}(\CX_1,\CX_2)\,$ the
complex conjugate of $\,
F^{\hspace{0.01cm}^{_\diamond}}_w(\CX_1,\CX_2)$. \,These quantities involve
equal-time 3-velocity 2-point functions.
\vskip 0.2cm

The covariance of velocity correlation functions under the $\,SL(2,\NR)\,$
transformations and the reflection $\,P\,$ implies that
for $\,\gamma=\Big(\begin{matrix}{_a&\hspace{-0.2cm}_b\cr{}^c&
\hspace{-0.2cm}^d}\end{matrix}\Big)\in SL(2,\NR)$,
\qq
&&F^{\hspace{0.01cm}^{_\diamond}}_w(\gamma\CX_1,\gamma\CX_2)\
=\ (cw_1+d)^2\,F^{\hspace{0.01cm}^{_\diamond}}_w(\CX_1,\CX_2)
\,,\ \ \label{invFw} \\
&&F^{\hspace{0.01cm}^{_\diamond}}_{\bar w}(\gamma\CX_1,\gamma\CX_2)\ =\ (c\bar w_1+d)^2\,
F^{\hspace{0.01cm}^{_\diamond}}_{\bar w}(\CX_1,\CX_2)\,,
\label{invFbw}
\\&&
 F^{\hspace{0.01cm}^{_\diamond}}_w(P\CX_1,P\CX_2)\ =\ -\m
F^{\hspace{0.01cm}^{_\diamond}}_{\bar w}
(\CX_1,\CX_2)\,.
\label{invFP}
\qqq
This means that the quantity
$F^{\hspace{0.01cm}^{_\diamond}}\hspace{-0.05cm}(\CX_1,\CX_2)\,
=\,F^{\hspace{0.01cm}^{_\diamond}}_w(\CX_1,\CX_2)\,dw_1+F^{\hspace{0.01cm}^{_\diamond}}_{\bar w}
(\CX_1,\CX_2)\,d\bar w_1$
transforms covariantly under the $\,SL(2,\NR)\,$ transformations
as a 1-form in its dependence on point $\CX_1$ and a function in
its dependence on point $\CX_2$. \,Eq.\,(\ref{whn}) may be
rewritten in the more geometric way as
\qq
\big\langle\,V(\CX_1){_{^{\hspace{0.07cm}\diamond\,}}}
v(\CX_2)\m\big\rangle\ =\
\frac{2}{1
+\cosh(\delta_{12}/R)}\,\,{\rm d}^\dagger(1)\m
F^{\hspace{0.01cm}^{_\diamond}}\hspace{-0.05cm}(\CX_1,\CX_2)\,,
\label{whnstar}
\qqq
where $\,{\rm d}^\dagger(1)\,$ is the adjoint of the exterior derivative acting
on the first variable.
\vskip 0.2cm

In the inverse cascade scenario, one assumes that the forcing covariance,
and hence the function $\,\CG(\CX_1,\CX_2)\,$ vanishes for $\,\delta_{12}
\gg\ell\,$ and that the limit $\,\nu\to0\,$ may be taken. Under those
conditions, we infer from Eqs.\,(\ref{t23}) and (\ref{whnstar})
the flux relation
\qq
\partial_t\m\big\langle\,v(\CX_1){_{^{\hspace{0.07cm}\diamond\,}}}
v(\CX_2)\m\big\rangle\,=\,-\,
\frac{4}{1
+\cosh(\delta_{12}/R)}\,\,{\rm d}^\dagger(1)\m
F^{\hspace{0.01cm}^{_\diamond}}\hspace{-0.05cm}(\CX_1,\CX_2)\,.
\qqq
holding for $\,\delta_{12}\gg\ell$. \,In view of Eq.\,(\ref{invcvv}),
at long times this implies the identity
\qq
{\rm d}^\dagger(1)\m F^{\hspace{0.01cm}^{_\diamond}}\hspace{-0.05cm}(\CX_1,\CX_2)\
=\ \frac{_1}{^4}\big[1+\cosh(\delta_{12}/R)\big]\,
(\Delta^{\hspace{-0.095cm}^{_\diamond}}\Psi_0)
\big(\cosh(\delta_{12}/R))\,.
\label{GFRstar}
\qqq
It is a differential equation for the correlation functions
$\,F^{\hspace{0.01cm}^{_\diamond}}_w(1,2)\,$ and $\,F^{\hspace{0.01cm}^{_\diamond}}_{\bar w}(1,2)\,$
that we shall solve now.

Identities (\ref{invFw}) and (\ref{invFbw}) imply that
there must exist (in general, complex)
function $\,G^{\hspace{0.01cm}^{_\diamond}}(\cosh\hspace{-0.06cm}
\frac{_{\delta_{12}}}{^R})\,$
such that for $\,\CX_1\not=\CX_2\,$
\qq
F^{\hspace{0.01cm}^{_\diamond}}_w(\CX_1,\CX_2)\ =\ R^2\,G^{\hspace{0.01cm}^{_\diamond}}
(\cosh\hspace{-0.06cm}\frac{_{\delta_{12}}}{^R})
\,\,\partial_{w_1}\cosh\hspace{-0.06cm}\frac{_{\delta_{12}}}{^R}\,.
\label{srep}
\qqq
This follows from the fact that
\qq
\partial_{w_1}\m\cosh\hspace{-0.06cm}\frac{_{\delta_{12}}}{^R}&=&
\frac{2(\bar w_1-\bar w_2)(\bar w_1-w_2)}{(w_1-\bar w_1)^2(w_2-\bar w_2)}\,,
\label{pw1}
\qqq
does not vanish for
non-coinciding points and has the same
transformation properties under $\,SL(2,\NR)\,$ as
$\,F^{\hspace{0.01cm}^{_\diamond}}_w(\CX_1,\CX_2)\,$
hence
their ratio must be a function of the hyperbolic distance between
the points. Relation (\ref{invFP}) imposes that function
$\,G^{\hspace{0.01cm}^{_\diamond}}\,$
must be real. \,Substitution of relation (\ref{srep}) to
Eq.\,(\ref{GFRstar}) results in the differential equation
\qq
(x^2-1){G^{\hspace{0.01cm}^{_\diamond}}}'(x)+2xG^{\hspace{0.01cm}^{_\diamond}}(x)\,=\,\frac{_1}{^4}(x+1)
\Delta^{\hspace{-0.095cm}^{_\diamond}}\Psi_0\,.
\label{aneq}
\qqq
whose general solution is
\qq
G^{\hspace{0.01cm}^{_\diamond}}(x)\ =\ \frac{1}{4(x^2-1)}\Big(\int\limits_1^x
(x'+1)(\Delta^{\hspace{-0.095cm}^{_\diamond}}\Psi_0)(x')\,{\rm d}x'\ +\ const.\Big).
\qqq
The constant has to vanish if the solution for
$\,F^{\hspace{0.01cm}^{_\diamond}}_w(\CX_1,\CX_2)\,$
is to be regular at $\,x=1$, \,i.e. at the coinciding points, otherwise $\,F^{\hspace{0.01cm}^{_\diamond}}_w(\CX_1,\CX_2)\,$ would diverge as
$\,\delta_{12}^{-1}$.
\,We finally infer that %the 3-velocity correlation function
\qq
F^{\hspace{0.01cm}^{_\diamond}}_w(\CX_1,\CX_2)\ =\
\frac{R^2\m(\bar w_1-\bar w_2)(\bar w_1-w_2)}{2(w_1-\bar w_1)^2
(w_2-\bar w_2)}\,\,\frac{1}{\sinh^2(\delta_{12}/{ R})}\hspace{-0.3cm}
\int\limits_1^{\cosh(\delta_{12}/{ R})}\hspace{-0.3cm}(x'+1)
(\Delta^{\hspace{-0.095cm}^{_\diamond}}
\Psi_0)(x')\,{\rm d}x'\,.\quad
\label{solFst}
\qqq
\vskip 0.2cm

Denote by $\,\hat e_{1,2}\,$ the unit vector tangent at point $\CX_1$
to the geodesic joining that point to point $\CX_2$. \,Since the geodesics
on the upper half plane are half-circles centered at the real line,
it is easy to find that
\qq
\hat e^{\hspace{0.03cm}w}_{1,2}\,=\,\frac{{1}}{{2R}}\,
\frac{{(w_1-\bar w_1)(w_1-w_2)(w_1-\bar w_2)}}{{|w_1-w_2||w_1-\bar w_2|}}
\label{e1}
\qqq
and $\,\hat e^{\hspace{0.03cm}\bar w}_{1,2}\,$ is its complex conjugate.
Consider the scalar quantity, that we shall call ``intrinsic'' energy
current,
\qq
\Theta^{\hspace{0.01cm}^{_\diamond}}(\delta_{12})&\,
\equiv\,&\hat e^{\hspace{0.03cm}w}_{1,2}F^{\hspace{0.01cm}^{_\diamond}}
_w(\CX_1,\CX_2)+
\hat e^{\hspace{0.03cm}\bar w}_{1,2}F^{\hspace{0.01cm}^{_\diamond}}_{\bar w}
(\CX_1,\CX_2)\cr\cr
&\,=\,&
-R^2\,\hat e^{\hspace{0.03cm}w}_{1,2}\,\big(\partial_{\bar w_1}
\partial_{w_2}\cosh(\hspace{-0.06cm}
\frac{_{\delta_{12}}}{^R})\big)\,\big\langle\,v_w(\CX_1)\m
v^{\bar w}(\CX_1)\m v^w(\CX_2)\,
\big\rangle\,+\ c.\,c.\cr\cr
&\,\equiv\,&\hat e_{1,2}\cdot F^{\hspace{0.01cm}^{_\diamond}}
\hspace{-0.05cm}(\CX_1,\CX_2)\,,
\qqq
where the last expression contracts vector $\,\hat e_{1,2}\,$
with the 1-form at point $\,\CX_1$. \,Substituting expressions (\ref{e1})
and (\ref{solFst}), we obtain
the identity
\qq
\Theta^{\hspace{0.01cm}^{_\diamond}}\hspace{-0.03cm}(\delta_{12})\
=\ -\frac{R}{4}\,\frac{1}{\sinh(\delta_{12}/{ R})}\hspace{-0.2cm}\int\limits_1^{\cosh(\delta_{12}/{ R})}\hspace{-0.3cm}(x'+1)
(\Delta^{\hspace{-0.095cm}^{_\diamond}}
\Psi_0)(x')\,{\rm d}x'\,.
\label{Th12star}
\qqq
\vskip 0.2cm

To complete the above arguments, let us prove the vanishing of
2-point correlators between velocity and scalar quantities, used on the way.
The correlation functions
$
\hat F_w(\CX_1,\CX_2)$ (either $\big\langle\,v_w(\CX_1)\,p(\CX_2)\,\big\rangle
$
or
$\big\langle\,v_w(\CX_1)\,(v^wv_w+v^{\bar w}
v_{\bar w})(\CX_2)\,\big\rangle$)
and their complex conjugates $\,\hat F_{\bar w}(\CX_1,\CX_2)\,$ also possess
the symmetry properties (\ref{invFw}), (\ref{invFbw}) and (\ref{invFP}).
Hence $\,\hat F_w(\CX_1,\CX_2)\,$ may be represented as in (\ref{srep}) with
a real function $\,\hat G$. \,We also have the relation
\qq
\partial_{\bar w_1}\hat F_w(\CX_1,\CX_2)
+\partial_{w_1}\hat F_{\bar w}(\CX_1,\CX_2)\,=\,0\,,
\qqq
due to the incompressibility of $\,v$.  \,It implies that $\,\hat G\,$
satisfies the homogeneous version of Eq.\,(\ref{aneq})
whose general solution is proportional to $\,(x^2-1)^{-1}$. \,The
regularity at the coinciding points of $\,\hat F_w\,$ requires
that the proportionality constant be zero so that $\,\hat F_w(\CX_1,\CX_2)\,$
vanishes.
\vskip 0.2cm

For the extrinsically paired  2-point functions, similar
considerations lead to a simpler flux relation
\qq
\partial_t\m\big\langle\,v(\CX_1)\centerdot v(\CX_2)\m\big\rangle\,=\,
-\m2\,{\rm d}^\dagger(1)\m F^{\,^\centerdot}(\CX_1,\CX_2)
\qqq
with $\,F^{\,^\centerdot}(\CX_1,\CX_2)=F^{\,^\centerdot}_w(\CX_1,\CX_2)dw_1
+F^{\,^\centerdot}_{\bar w}(\CX_1,\CX_2)d\bar w_1\,$
for
\qq
F^{\,^\centerdot}_w(\CX_1,\CX_2)\ =\ \frac{_1}{^2}R^{-2}
\Big((w_1-w_2)^2\,\big\langle\,v_w^2(1)\,v_w(2)\,
\big\rangle\,-\,(w_1-\bar w_2)^2\,\big\langle\,v_w^2(1)\,v_{\bar w}(2)\,
\big\rangle\Big)\quad
\label{Fw}
\qqq
and complex conjugate $\,F^{\,^\centerdot}_{\bar w}$. \,The components
of $\,F^{\,^\centerdot}\,$ are related by Eq.\,(\ref{srep}) to a function
$\,G^{\,^\centerdot}(x)\,$ solving now
Eq.\,(\ref{aneq}) with the right hand side replaced by
$\,\frac{1}{2}\,(\Delta\Psi_0)(x)$.  \,Its solution gives
\qq
F^{\,^\centerdot}_w(\CX_1,\CX_2)\ =\
\frac{(\bar w_1-\bar w_2)(\bar w_1-w_2)}{(w_1-\bar w_1)^2
(w_2-\bar w_2)}\,\Psi'_0(\cosh\hspace{-0.06cm}
\frac{_{\delta_{12}}}{^R})\,.
\label{FluxG}
\qqq
The corresponding ``extrinsic'' energy current
\qq
\Theta^{\,^\centerdot}(\delta_{1.2})\ =\ \hat e_{1,2}\cdot F^{\,^\centerdot}
(\CX_1,\CX_2)\ =\ -\m\frac{1}{2R}\,\sinh\hspace{-0.06cm}
\frac{_{\delta_{12}}}{^R}\,
\Psi_0'\big(\cosh\hspace{-0.06cm}\frac{_{\delta_{12}}}{^R}\big)\,.
\label{Th12}
\qqq

\nsection{Scaling}
\label{sec:Scal}

Suppose that we rescale the distances by a constant $\,\lambda\,$
by changing the metric $\,g\,$ to $\,\tilde g=\lambda^2 g$. \,Introduce
\qq
\tilde v(t,y)&=&\lambda^\alpha v(\lambda^\alpha t,y)\,,\cr
\tilde f(t,y)&=&\lambda^{2\alpha}f(\lambda^\alpha t,y)\,,\cr
\tilde p(t,y)&=&\lambda^{2\alpha+2}p(\lambda^\alpha t,y)\,,\cr
\tilde\nu&=&\lambda^{\alpha+2}\nu\,.
\qqq
Then $\,\tilde v\,$ satisfies the Navier-Stokes equation (\ref{NSg})
for metric $\,\tilde g\,$ with viscosity $\,\tilde\nu\,$ pressure
$\,\tilde p\,$ and forcing $\,\tilde f\,$ if and only if $\,v\,$ satisfies
it for metric $\,g\,$ with
viscosity $\,\nu\,$ pressure $\,p\,$ and forcing $\,f$. \,If $\,f\,$
is a white noise in time with covariance (\ref{fcov}) then $\,\tilde f\,$
is also white in time with the covariance
\qq
\langle\tilde f^i(t_1,y_1)\,\tilde f^j(t_2,y_2)\rangle\,
=\,\delta(t_1-t_2)\,\tilde C^{ij}(y_1,y_2)\,,
\label{tfcov}
\qqq
where
$\tilde C^{ij}(y_1,y_2)\,=\,\lambda^{3\alpha}C^{ij}(y_1,y_2)$.
For the dissipation field (\ref{dissip}) and
the mean energy injection rate (\ref{inject}), we obtain
\qq
\tilde\epsilon\,=\,\lambda^{3\alpha+2}\epsilon\,,\qquad
\tilde{\bar\iota}\,=\,\lambda^{3\alpha+2}\bar\iota
\qqq
so that taking $\,\alpha=-\frac{2}{3}\,$ assures that they do not
change. \,Let us consider the equal-time correlation of velocities
\qq
F^{i_1\dots i_n}_{M,g,C,\nu}(t,x_1,\dots,x_n)\ =\ \Big\langle
\prod\limits_{j=1}^nv^{i_j}(t,x_j)\Big\rangle
\qqq
obtained by taking the initial condition $\,v(0,x)=0$. \,Two
basic (essentially tautological) properties of such correlation functions
is their diffeomorphism and scaling covariance. \,The first of those
states that for any diffeomorphism $\,D:M'\mapsto M\,$ of the manifolds
\qq
F^{i_1\dots i_n}_{M,g,C,\nu}(t,x_1,\dots,x_n)\ =\
\prod\limits_{m=1}^n(\partial_{i'_m}\hspace{-0.08cm}D^{i_m})(x'_m)\,
F^{i'_1\dots i'_n}_{M',D^*\hspace{-0.03cm}g,D^*C,\nu}(t,x'_1,\dots,x'_n)\,,\quad
\label{diffcov}
\qqq
for $\,x_i=D(x_i')\,$ and
\qq
&&(D^*\hspace{-0.06cm}g)_{i'_1i'_2}(x')
\,=\,(\partial_{i'_1}\hspace{-0.06cm}D^{i_1}\hspace{-0.05cm})(x')
\,(\partial_{i'_2}\hspace{-0.06cm}
D^{i_2}\hspace{-0.04cm})(x')\,\,g_{i_ii_2}(x)\,,\\ \cr
&&(\partial_{i'_1}\hspace{-0.06cm}D^{i_1}\hspace{-0.05cm})(x'_1)
\,(\partial_{i'_2}\hspace{-0.06cm}D^{i_2}\hspace{-0.04cm})(x'_2)\,\,
(D^*C)^{i'_1i'_2}(x'_1,x'_2)\,=\,C^{i_1i_2}(x_1,x_2)\,.\ \quad
\qqq
Identity (\ref{diffcov}) is assured by the geometric formulation
of the Navier-Stokes equation. \,On the other hand, the scaling properties
of solutions of the Navier-Stokes equation discussed above
imply the scaling relation:
\qq
F^{i_1\dots i_n}_{M,\lambda^2g,C,\nu}
(t,x_1,\dots,x_n)\ =\
\lambda^{\alpha n}\,F^{i_1\dots i_n}_{M,g,\lambda^{-3\alpha}C,\lambda^{-\alpha-2}\nu}
(\lambda^\alpha t,x_1,\dots,x_n)\,.\label{scaling}
\qqq
\vskip 0.2cm

Let us first recall how these two tautological properties are used in
the flat space $\,\NR^d\,$ with $\,g=g_0=({\rm d}x^i)^2\,,$ and with
the spatial forcing covariance $\,C^{ij}_\ell(x,y)=C^{ij}(\frac{_{x-y}}{^\ell})$.
One considers the diffeomorphism $\,D(x)=\lambda\hspace{0.02cm}x\,$
for which relation (\ref{diffcov}) reads
\qq
F^{i_1\dots i_n}_{\NR^d,g_0,C_{\ell},\nu}(t,\lambda x_1,\dots,
\lambda x_n)\ =\ \lambda^n\,F^{i_1\dots i_n}_{\NR^d,
\lambda^2g_0,\lambda^{-2}C_{\ell/\lambda},\nu}(t,x_1,\dots,x_n)\,.
\qqq
Using subsequently the scaling property (\ref{scaling})
\qq
F^{i_1\dots i_n}_{\NR^d,\lambda^2g_0,\lambda^{-2}C_{\ell/\lambda},\nu}
(t,x_1,\dots,x_n)\ 
%&&\hspace{3cm}
=\ \lambda^{\alpha n}\,F^{i_1\dots i_n}_{\NR^d,g_0,
\lambda^{-3\alpha-2}C_{\ell/\lambda},
\lambda^{-\alpha-2}\nu}
(\lambda^\alpha t,x_1,\dots,x_n)\,,\qquad
\qqq
one infers that
\qq
F^{i_1\dots i_n}_{\NR^d,g_0,C_{\ell},\nu}(t,\lambda x_1,\dots,
\lambda x_n)\ =\ \lambda^{(1+\alpha)n}\,F^{i_1\dots i_n}_{\NR^d,g_0,
\lambda^{-3\alpha-2}C_{\ell/\lambda},\lambda^{-\alpha-2}\nu}(\lambda^\alpha t;
x_1,\dots,x_n)\,.\qquad
\qqq
In particular, \,taking $\,\alpha=-\frac{2}{3}\,$ gives:
\qq
F^{i_1\dots i_n}_{\NR^d,g_0,C_{\ell},\nu}(\lambda^{\frac{2}{3}}t,\lambda x_1,\dots,
\lambda x_n)\ =\
\lambda^{\frac{n}{3}}\,F^{i_1\dots i_n}_{\NR^d,g_0,C_{\ell/\lambda},\lambda^{-\frac{4}{3}}\nu}
(t;x_1,\dots,x_n)\,.
\label{scaleflat}
\qqq
In a flat space, perturbative
renormalization-group calculations support the assumption that the limit $\,\nu\to0\,$ of the
correlation functions exists. The behavior at the forcing scale
$\,\ell\,$ depends on the dimension. For  $\,d\geq3$, experimental and numerical data show that the
moments of velocity difference 
(the velocity structure functions) of all orders except 3 scale with anomalous
powers of $\,\ell\,$ when $\,\ell\to\infty$ and $\,\nu\to0\,$.
For $\,d=2$, the data suggest that the limits $\,\nu\to0\,$ and $\,\ell\to 0\,$ may be taken for the moments of velocity
differences. If this is true, then relation
(\ref{scaleflat}) implies an asymptotic scaling of the correlation
functions in the inverse cascade domain. Note that Eq.\,(\ref{scflat})
is a consequence of such a scaling.
\vskip 0.2cm

Should there be an extension of such conjectures to
the curved spaces, in particular, in two-dimensions?
\,The limit of the forcing scale sent to zero still makes sense
(in that limit, $\,C^{ij}(x,x')\,$ concentrates on the diagonal
with the injection rate $\,\bar\iota(x)\,$ kept constant).
The main problem is what classes of correlation functions to consider. In a flat space, Galilean invariance suggested considering moments of velocity differences.  What is the natural class
in the absence of the Galilean invariance of the Navier-Stokes equations?
\,One possible way could be to consider the correlation functions of
covariant derivatives of $\,v$, \,but is it a good solution?
\vskip 0.2cm

Let us study more closely the case of the hyperbolic plane.
The main problem concerns the behavior of the theory at long
distances that we may formulate as the question
about the asymptotics at $\,\lambda\to\infty\,$ of
the correlation functions
\qq
F^{i_1\dots i_n}_{\NH_{\hspace{-0.02cm}_R},g,C,\nu}(t;\lambda\CX_1,\dots,\lambda\CX_n)\,,
\label{dilat}
\qqq
where $\,\CX_1,\dots,\CX_n$, \,with $\,\CX_m=X^i_m\tau_i$, \,are
$\,n\,$ points in Lie algebra $\,sl(2,\NR)$. \,The problem, however, is
that the dilation by $\,\lambda\not=1\,$ does not preserve the hyperboloids
$\,\NH_{\hspace{-0.02cm}_{R}}$. \,If we want that $\,\lambda\CX_m\,$
belong to $\,\NH_{\hspace{-0.02cm}_{R}}$, \, we have to choose points
$\,\CX_m\,$ in a $\,\lambda$-dependent way so that
$\,\CX_m\in\NH_{\hspace{-0.02cm}_{R/\lambda}}$, \,but we shall do it so that
they tend to limits in the cone $\,\NH_{\hspace{-0.01cm}_0}\,$ when
$\,\lambda\to\infty$. \,In this case, the distance between points
$\,\lambda\CX_m\in\NH_{\hspace{-0.02cm}_{R}}\,$ in general positions
will grow with $\,\lambda\,$ and correlation functions (\ref{dilat})
will. indeed, test the long-distance behavior of the theory.
\vskip 0.2cm

\,Recall that the metric $\,g\,$ on $\,\NH_{\hspace{-0.01cm}_R}\,$ is induced
from the metric $\,({\rm d}X^1)^2+({\rm d}X^2)-({\rm d}X^3)\,$ 
on $\,sl(2,\NR)\,$ and
the forcing covariance $\,C^{ij}\,$ is given by Eq.\,(\ref{8.2}).
\,The application of identity (\ref{diffcov}) to the diffeomorphism
\qq
\NH_{\hspace{-0.02cm}_{R/\lambda}}\ni\CX\ \mathop{\longmapsto}
\limits^{D_\lambda}\ \lambda\CX\in  \NH_{\hspace{-0.01cm}_{R}}
\qqq
results in the relation
\qq
F^{i_1\dots i_n}_{\NH_{\hspace{-0.02cm}_R},g,C,\nu}(t;\lambda\CX_1,\dots,\lambda\CX_n)
\ =\ \prod\limits_{m=1}^n(\partial_{i'_m}\hspace{-0.06cm}D_\lambda^{i_m}
\hspace{-0.04cm})(\CX_m)
\,\,F^{i'_1\dots i'_n}_{\NH_{\hspace{-0.02cm}_{R/\lambda}},
D_\lambda^*g,D_\lambda^*C,\nu}(t;\CX_1,\dots,\CX_n)\,.\qquad
\label{lambdasc}
\qqq
Note that $\,D_\lambda^*g_R=\lambda^2g_{R/\lambda}$.  \,If the forcing covariance
$\,C^{i_1i_2}\,$ is determined by a scalar function
$\,\CC(\cosh\hspace{-0.06cm}\frac{_{\delta_{12}}}{^R})\,$ from Eq.\,(\ref{8.2})
then for points $\,\CX'_1,\CX'_2\in\NH_{\hspace{-0.01cm}_{R/\lambda}}$,
\qq
\lambda^2(D_\lambda^*C)^{i'_1i'_2}(\CX_1',\CX_2')\ =\ \lambda^{-2}\,
\epsilon^{i'_1j'_1}\epsilon^{i'_2j'_2}\frac{_1}
{^{\sqrt{g(\CX_1')}\m\sqrt{g(\CX_2')}}}\,
\m\partial_{j'_1}(1')\,\partial_{j'_2}(2')
\,\,\CC(\cosh\hspace{-0.06cm}\frac{_{\delta_{1'2'}}}{^{R/\lambda}})\,,\quad
\label{rescC}
\qqq
where we employed the identities
$\,\delta(D_\lambda(\CX_1'),D_\lambda(\CX_2'))=
\lambda\,\delta_{1'2'}\,$ and $\,\sqrt{(D_\lambda^*g)(\CX_1')}=
\lambda^2\sqrt{g(\CX_1')}$. \,Thus the forcing covariance
$\,\lambda^2D_\lambda^*C\,$ corresponds to the scalar function
$\,\lambda^{-2}\m\CC(\cosh\hspace{-0.06cm}\frac{_{\delta_{1',2'}}}
{^{R/\lambda}})\,$ and to the same energy injection rate
$\,\bar\iota\,$ as $\,C^{i_1i_2}$, \,see Eq.\,(\ref{bariota}).
\,Upon relabeling the dependence on the forcing covariance by the function
$\,\CC$, \,we obtain from the identities (\ref{lambdasc}) and
(\ref{scaling}) with $\,\alpha=-\frac{2}{3}\,$ the scaling relation
\qq
&&F^{i_1\dots i_n}_{\NH_{\hspace{-0.02cm}_R},g,\CC,\nu}
(\lambda^{\frac{2}{3}}t;\lambda\CX_1,\dots,\lambda\CX_n)
\cr
&&\qquad\quad=\ \lambda^{-\frac{2}{3}n}\prod\limits_{m=1}^n
(\partial_{i'_m}\hspace{-0.08cm}D_\lambda^{i_m}\hspace{-0.04cm})(\CX_m)\,\,
F^{i'_1\dots i'_n}_{\NH_{\hspace{-0.02cm}_{R/\lambda}},g,
\lambda^{-2}\CC,\lambda^{-\frac{4}{3}}\nu}
(t;\CX_1,\dots,\CX_n)\,.\qquad
\label{scalR}
\qqq
It seems then that the asymptotic behavior of the right hand side in the
limit $\,\lambda\to\infty\,$ should be
determined by the theory on the limiting space $\,\NH_0\,$ with the
vanishing viscosity and the forcing covariance obtained in the limit from
$\,\lambda^2D_\lambda^*C$. \,The next section is devoted to
a detailed discussion of the ideal hydrodynamics on $\,\NH_{_0}$.

\nsection{Euler equation on $\,\NH_{_0}$}
\label{sec:R=0}

\subsection{Limiting geometry}

When $\,R\to\infty$, \,the theory on $\,\NH_{_R}\,$ (written e.g.
in coordinates $\,(\rho,\phi)\,$ of Sec.\,\ref{sec:harman})
\,tends to the one on $\,\NR^2\,$ (written
in the radial coordinates), \,at least for finite times. \,Let us see what
can we tell about the theory in the opposite limit $\,R\to0$.
\,Due to the scaling properties discussed at the end of the last section,
such a limit should provide, at least to some extent, the asymptotic
behavior of the theory on $\,\NH_{\hspace{0.01cm}_R}\,$
at long distances $\,\gg R$.
\,When $\,R\to0$, the hyperboloids $\,(X^1)^2+(X^2)^2+R^2=(X^3)^2\,$ with
$\,X_3>0\,$ approach the (upper light-)cone, see Fig.\,\ref{fig:lightcone}.
The theory on $\,\NH_{\hspace{-0.01cm}_R}\,$ should then tend to the one
on the cone $\,\NH_{_0}$. \,In order to parameterize the family
of spaces $\,\NH_{\hspace{-0.01cm}_R}$ with $\,R\geq0$,  it will be convenient
to use the coordinates $\,(r,\varphi)$,
$\,0\leq r<\infty$, $\,-\pi\leq\varphi\leq\pi\,$ with
\qq
X^1=r\cos\varphi\,,\quad\ X^2=r\sin\varphi\,,\quad\ X^3=\sqrt{R^2
+r^2}
\qqq
For $\,R=0$, these
coordinates (with $\,r>0$) \,parameterize the (upper light-) cone
$\,\NH_0\,$ defined by the equation
$\,(X^1)^2+(X^2)^2=(X^3)^2\,$ and the inequality $\,X^3>0$.
\,In the variables $\,(r,\varphi)$,
\,the Riemannian metric on $\,\NH_{\hspace{-0.01cm}_R}\,$ is
\qq
g\ =\ \frac{R^2({\rm d}r)^2+(R^2+r^2)r^2({\rm d}\varphi)^2}{R^2+r^2}
\qqq
and the metric volume element is
\qq
\chi\ =\ \frac{R\m r}{\sqrt{R^2+r^2}}\,{\rm d}r\wedge {\rm d}\varphi\,.
\qqq
When $\,R\to0$, \,the metric becomes degenerate with the null
radial direction,
\qq
g|_{R=0}\ =\ r^2({\rm d}\varphi)^2\,.
\label{metric0}
\qqq
The metric volume also degenerates, but divided
by $\,R$, \,has a non-singular limit
\qq
R^{-1}\chi|_{R=0}\ =\ {\rm d}r\wedge {\rm d}\varphi\,\equiv\,\chi_0.
\label{volume0}
\qqq

\subsection{Variational principle at $\,R=0$}

The action functional $\,S(\Phi,\lambda)\,$ of
Sec.\,\ref{sec:geodflow} divided by $\,R\,$ has a limit when $\,R\to0\,$
\qq
&&R^{-1}\m S(\Phi,\lambda)|_{R=0}%\cr\cr&&
=\int L%(\Phi^r(t,r,\varphi),
%\lambda(t,r,\varphi),
%\partial_t\Phi^\varphi(t,r,\varphi),\partial_r\Phi(t,r,\varphi),\partial_\varphi
%\Phi(t,r,\varphi))\ 
\,\,{\rm d}r\wedge{\rm d}\varphi\wedge{\rm d}t\,,\qquad\quad
\label{S0}
%\qqq\qq
\\&& L\left(\Phi^r(r,\varphi),
\partial_t\Phi^\varphi(r,\varphi),\partial_r\Phi(t,r,\varphi),\partial_\varphi
\Phi(t,r,\varphi)\right)\cr\cr
&&\quad=\ \frac{_1}{^2}(\Phi^r)^2
(\partial_t\Phi^\varphi)^2\,+\,\lambda\left[(\partial_r\Phi^r)
(\partial_\varphi\Phi^\varphi)-(\partial_\varphi\Phi^r)
(\partial_r\Phi^\varphi)-1\right].\qquad
\label{Ldens0}
\qqq
By the variational principle, the action (\ref{S0}) should give
rise to the Euler
equation on $\,\NH_{\hspace{-0.01cm}_0}$. \,Here, we shall study directly
the limit when $\,R\to0\,$ of the Euler equations on
$\,\NH_{\hspace{-0.01cm}_R}\,$ in coordinates $\,(r,\varphi)$. \,The result
will be the same (in a slightly different presentation).
\vskip 0.2cm

In the coordinates $\,(r,\varphi)$, \,the matrix of inverse metric on
$\,\NH_{\hspace{-0.01cm}_R}\,$ is
\qq
(g^{ij})\ =\ \left(\begin{matrix}{\frac{R^2+r^2}{R^2}&0\cr
0&\frac{1}{r^2}}\end{matrix}\right)
\qqq
and the non-zero Christoffel symbols are:
\qq
\Big\{{_r\atop^{r\m r}}\Big\}\,=\,-\m\frac{_r}{^{R^2+r^2}}\,,\quad
\Big\{{_r\atop^{\varphi\m\varphi}}\Big\}\,=\,-\m\frac{_{r(R^2+r^2)}}{^{R^2}}\,,
\quad\Big\{{_\varphi\atop^{r\m\varphi}}\Big\}\,=\,\frac{_1}{^r}\,=\,
\Big\{{_\varphi\atop^{\varphi\m r}}\Big\}.
\qqq
Hence, in coordinates
$\,(r,\varphi)\,$ the Euler equations (\ref{Euler}) take the form:
\qq
&&\partial_t v^r+(v^r\partial_r+v^\varphi\partial_\varphi)v^r
\,-\,\frac{_r}{^{R^2+r^2}}(v^r)^2\,-\,\frac{_{r(R^2+r^2)}}{^{R^2}}
(v^\varphi)^2\,=\,-\,\frac{_{R^2+r^2}}{^{R^2}}\m\partial_rp\,,\quad
\label{tvr}\\ \cr
&&\partial_t v^\varphi+(v^r\partial_r+v^\varphi\partial_\varphi)v^\varphi
\,+\,\frac{_2}{^r}v^rv^\varphi\,=\,-\,
\frac{_1}{^{r^2}}\m\partial_\varphi p\,,
\label{tvp}
\qqq
and the incompressibility condition is
\qq
\partial_r\Big(\frac{_r}{^{\sqrt{R^2+r^2}}}v^r\Big)+\partial_\varphi\Big(
\frac{_r}{^{\sqrt{R^2+r^2}}}v^\varphi\Big)\,=\,0\,.
\label{inR}
\qqq
When $\,R\to0$, \,the terms of order $\,R^{-2}\,$ from Eq.\,(\ref{tvr})
give the condition
\qq
r(v^{\varphi})^2\,=\,\partial_rp\,,
\label{cons1}
\qqq
whereas the terms of order $\,1\,$ give, with the use of the last equation,
\qq
\partial_tv^r+(v^\varphi\partial_\varphi+v^r\partial_r)v^r-r^{-1}(v^r)^2\
=\ -r^2\partial_rq\,,
\label{1214}
\qqq
where $\,q=\partial_{R^2}p|_{R=0}$. \,Eq.\,(\ref{tvp}) keeps its form in
the limit $\,R\to0$, \,whereas the incompressibility condition becomes
\qq
\partial_\varphi v^\varphi+\partial_rv^r\ =\ 0\,.
\label{inc0}
\qqq
One can always find a function
$\,q\,$ such that Eq.\,(\ref{1214}) holds, so we may drop it (it does
not even appear if we work directly with the $\,R=0\,$ variational principle).
Hence at $\,R=0$, \,the Euler equation reduces to relations
(\ref{tvp}), (\ref{cons1}), and (\ref{inc0}).
\,In terms of the stream function $\,\psi\,$ such that
\qq
v^r\,=\,-\partial_\varphi\psi\,,\qquad v^\varphi\,=\,\partial_r\psi\,,
\label{vpsiR0}
\qqq
and the vorticity $\,\omega=\partial_rr^2v^\varphi=\partial_rr^2\partial_r\psi$,
\,the $\,R=0\,$ Euler equation takes the form
\qq
\partial_t\omega\,+\,(\partial_r\psi)(\partial_\varphi\omega)\,-\,
(\partial_\varphi\psi)(\partial_r\omega)\,=\,0\,.
\label{evomegaHR0}
\qqq

\subsection{Symmetries of the Euler equation on $\,\NH_{_0}$}

What are the symmetries of the $\,R=0\,$ theory? \,Since the Lagrangian
density (\ref{Ldens0}) is independent of time, the time translation invariance
gives rise to the conserved currents
\qq
&&\CJ^0\ =\ \frac{_1}{^2}\m r^2(v^\varphi)^2\,,\label{CJ00}\\
&&\CJ^r\ =\ (\frac{_1}{^2}\m r^2(v^\varphi)^2+p)v^r\,,\label{CJr0}\\
&&\CJ^\varphi\ =\ (\frac{_1}{^2}\m r^2(v^\varphi)^2+p)v^\varphi\,.
\label{CJp0}
\qqq
$\,\CJ^0(r,\varphi)\,$ is the energy density. Note that the component
$\,v^r\,$ of velocity does not contribute to it.
\vskip 0.2cm

The degenerate metric (\ref{metric0}) has an infinite-dimensional
group of isometries $\,\Diff\m S^1$,
\,preserving also the
volume form (\ref{volume0}), given by the transformations
\qq
(r,\varphi)\ \longrightarrow\ (\frac{_r}{^{D'(\varphi)}},D(\varphi))
\label{Diff}
\qqq
where $\,\varphi\to D(\varphi)\,$ is a diffeomorphism of a circle.
\,In particular, the adjoint action (\ref{adjact}) of $\,SL(2,\NR)\,$
restricted to the cone $\,\NH_{_0}\,$ gives a 3-parameter
family of such transformations. \,The composition of the maps $\,\Phi\,$
with the transformations (\ref{Diff})
induces their action on the velocity fields that
send $\,v\,$ to $\,v'\,$ with
\qq
&&v'^r(t,\frac{_r}{^{D'(\varphi)}},D(\varphi))\ =\ \frac{_1}{^{D'(\varphi)}}\m
v^r(t,r,\varphi)\,-\,\frac{_{D''(\varphi)\, r}}{^{D'(\varphi)^2}}
\m v^\varphi(t,r,\varphi)
\,,\label{diff1}\\ \cr
&&v'^\varphi(t,\frac{_r}{^{D'(\varphi)}},D(\varphi))\ =\ D'(\varphi)\,
v^\varphi(t,r,\varphi)\,.\label{diff2}
\qqq
On the other hand, the stream function and the vorticity on $\,\NH_{_0}\,$
transform as scalar functions
\qq
\psi'(t,\frac{_r}{^{D'(\varphi)}},D(\varphi))\ =\ \psi(t,r,\varphi)\,,\qquad
\omega'(t,\frac{_r}{^{D'(\varphi)}},D(\varphi))\ =\ \omega(t,r,\varphi)\,,
\qqq
and it is easy to see that the evolution (\ref{evomegaHR0}) commutes
with such transformations. \,The infinitesimal transformations
(\ref{Diff}) are given by the Killing vector fields
\qq
X(r,\varphi)\ =\ (X^r(r,\varphi),\m X^\varphi(r,\varphi))\
=\ (-r\m \zeta'(\varphi),\m\zeta(\varphi))\,.
\label{Killi}
\qqq
In particular, for the infinitesimal transformations induced
by matrix $\,\Big(\begin{matrix}{_\alpha&\hspace{-0.3cm}_\beta\cr
^\gamma&\hspace{-0.3cm}^{-\alpha}}\end{matrix}\Big)\in sl(2,\NR)$,
\qq
\zeta(\varphi)\ =\ -2\alpha\sin\varphi-(\beta+\gamma)\cos\varphi+\beta-\gamma\,.
\qqq
The Killing vectors (\ref{Killi}) induce symmetries of the variational
problem with the action functional (\ref{S0}). They correspond to
the conserved currents
\qq
&&\CJ^0_X\ =\ -r^2 v^\varphi\zeta\,,\cr
&&\CJ^r_X\ =\ -r^2v^rv^\varphi\zeta\,+\,rp\m\zeta'\,,\cr
&&\CJ^\varphi_X\ =\ -(r^2(v^\varphi)^2+p)\m\zeta\,,
\qqq
see Eq.\,(\ref{CJCJX0}) and (\ref{CJCJXl}).
\,This can be alternatively presented with purely radial flux:
\,since $\,r^2(v^\varphi)^2+p=r\partial_rp+p=\partial_r(rp)$,
\,we may also take
\qq
&&\CJ^r_X\ =\ -r^2v^rv^\varphi\zeta\,+\,r(\partial_\varphi p)\zeta\,,\cr
&&\CJ^\varphi_X\ =0\,,
\qqq
which shows that at every point there is no exchange of the invariant
$\,r^2 v^\varphi\,$ along the angle, only along the radius.
\vskip 0.2cm

In Appendix \ref{appx:harman0}, we discuss harmonic analysis on
$\,\NH_{_0}$, how it arises from that on $\,\NH_{_R}$ and how the
enhanced $\,\Diff\m S^1\,$ symmetry
appears in that context.

\nsection{Scaling limit of the $\,\NH_{_R}\,$ theory}
\label{sec:scalim}

\noindent In the variables $\,(r,\varphi)$, \,scaling relation
(\ref{scalR}) may be put into the form
\qq
\lambda^{\frac{2}{3}n-m}
\,F^{i_1\dots i_n}_{\NH_{_R},\CC}(\lambda^{\frac{2}{3}}
t;\lambda r_1,\varphi_1;\dots;\lambda
r_n,\varphi_n)\ =\
F^{i_1\dots i_n}_{\NH_{_{R/\lambda}},\lambda^{-2}\CC}(t;
r_1,\varphi_1;\dots;r_n,\varphi_n)
\label{scalRl}
\qqq
if $\,m\,$ of the indices $\,i_k\,$ are equal to $\,r\,$ and
$\,(n-m)\,$ to $\,\varphi$. \,The hyperboloids $\,\NH_{_R}\,$ and
$\,\NH_{_{R/\lambda}}\,$ are taken with their Riemannian metrics $\,g\,$
that we dropped from the notation. \,The limit $\,\nu\to0\,$ was presumed.
In the limit $\,\lambda\to\infty$, \,the correlation function
\qq
F^{i_1\dots i_n}_{\NH_{_{R/\lambda}},\lambda^{-2}\CC}
(t;r_1,\varphi;\dots;r_n,\varphi_n)
\label{Roverl}
\qqq
should correspond to the forced correlation functions of the theory
on $\,\NH_{_0}\,$ but, as we shall see, the limiting procedure
(\ref{Roverl}) is somewhat delicate, at least for some components
of the correlation functions. We also should not expect that the
limiting correlation functions with forcing respect the $\,\Diff\m S^1\,$ 
symmetry present in the unforced theory
at $\,R=0$. \,The aim of the present section is to study these issues
in more detail.

\subsection{Forcing in the limit $\,\lambda\to\infty$}

We shall first consider the behavior of the spatial covariance (\ref{rescC})
on $\,\NH_{_{R/\lambda}}\,$ that corresponds to the scalar function
$\,\lambda^{-2}
\CC(\cosh\hspace{-0.06cm}\frac{_{\delta_{1'2'}}}{^{R/\lambda}})$.
\,Let us examine how the distance $\,\delta_{1'2'}\,$ between the points
$\,\CX'_1,\CX'_2\in\NH_{\hspace{-0.01cm}_{R/\lambda}}\,$ with fixed coordinates
$\,(r_1,\varphi_1)\,$ and $\,(r_2,\varphi_2)\,$ behaves when
$\,\lambda\to\infty$. \,According to Eq.\,(\ref{chdist}),
\qq
\cosh\hspace{-0.06cm}\frac{_{\delta_{1'2'}}}{^{R/\lambda}}&=&
\frac{_{r_1r_2\left(\sqrt{1+(R/\lambda)^2r_1^{-2}}
\sqrt{1+(R/\lambda)^2r_2^{-2}}\,-\,\cos(\varphi_1-\varphi_2)\right)}}
{^{(R/\lambda)^2}}\cr
&=&
\frac{_{r_1r_2\left(\sqrt{1+(R/\lambda)^2r_1^{-2}}
\sqrt{1+(R/\lambda)^2r_2^{-2}}\,-\,1\right)}}{^{(R/\lambda)^2}}\,+\,
\frac{_{r_1r_2(1-\cos(\varphi_1-\varphi_2))}}{^{(R/\lambda)^2}}\,.
\label{coshd}
\qqq
It follows that for $\,\varphi_1=\varphi_2$,
\qq
\cosh\hspace{-0.06cm}\frac{_{\delta_{1'2'}}}{^{R/\lambda}}\ \
\mathop{\longrightarrow}\limits_{\lambda\to\infty}\ \ \,
\frac{_1}{^2}\big(\frac{_{r_1}}{^{r_2}}+
\frac{_{r_2}}{^{r_1}}\big)\,.
\qqq
On the other hand, if $\,\varphi_1\not=\varphi_2\,$ then
$\,\cosh\hspace{-0.06cm}\frac{_{\delta_{1'2'}}}{^{R/\lambda}}\,$ diverges as $\,\CO(\lambda^2)$
when $\,\lambda\to\infty\,$. \,That divergence
leads at $\,R/\lambda\to0\,$ to the decorrelation
of the forcing covariance for different angles $\,\varphi$ if the function
$\,\CC\,$ in (\ref{8.2}) is of compact support.
\,For force correlation on $\,\NH_{\hspace{-0.01cm}_{R/\lambda}}\,$
given by Eq.\,(\ref{8.2}), \,we obtain the following limiting behavior
when $\,\lambda\to\infty\,$:
\qq
\lambda^{-2}C^{rr}(\CX_1',\CX_2')&=&
\begin{cases}{\,-\,\frac{_{\lambda^2\m r_1r_2}}
{^{R^4}}\,\CC'\left(\frac{_1}{^2}(\frac{_{r_1}}{^{r_2}}
+\frac{_{r_2}}{^{r_1}})\right)\cr
\,-\frac{_1}{^{2R^2}}(\frac{_{r_1}}{^{r_2}}+\frac{_{r_2}}{^{r_1}})\,
\CC'\left(\frac{_1}{^2}(\frac{_{r_1}}{^{r_2}}+\frac{_{r_2}}{^{r_1}})\right)\cr
\,+\frac{_1}{^{8R^2}}(\frac{_{r_1}}{^{r_2}}-\frac{_{r_2}}{^{r_1}})^2\,
\CC''\left(\frac{_1}{^2}(\frac{_{r_1}}{^{r_2}}+\frac{_{r_2}}
{^{r_1}})\right)\,+\,
o(1)\quad\ \,{\rm if}\quad\ \varphi_1=\varphi_2\,,\qquad\cr\cr
\,\,o(1)\hspace{6.54cm}\,{\rm if}\quad\ \varphi_1\not=\varphi_2\,.}\end{cases}
\label{Crrinf}\\ \cr
\lambda^{-2}C^{r\varphi}(\CX_1',\CX_2')&=&o(1)\,,\label{Crpinf}\\ \cr
\lambda^{-2}C^{\varphi\varphi}(\CX_1',\CX_2')&=&\begin{cases}{-\,\frac{_1}
{^{2R^2\m r_1r_2}}(\frac{_{r_1}}{^{r_2}}+\frac{_{r_2}}{^{r_1}})\,
\CC'(\frac{_1}{^2}(\frac{_{r_1}}{^{r_2}}+\frac{_{r_2}}{^{r_1}}))\cr
-\,\frac{_1}{^{4R^2\m r_1r_2}}\,(\frac{_{r_1}}{^{r_2}}-\frac{_{r_2}}{^{r_1}})^2
\CC''(\frac{_1}{^2}(\frac{_{r_1}}{^{r_2}}+\frac{_{r_2}}{^{r_1}}))\,+\,
o(1)\ \ {\rm if}\quad\ \varphi_1=\varphi_2\,,\ \cr\cr
\,\,o(1)\hspace{6.73cm}{\rm if}\quad\ \varphi_1\not=\varphi_2\,,}\end{cases}
\qquad\label{Cppinf}
\qqq
see Appendix \ref{appx:rescforce}. Hence the rescaled forcing covariance
tends to zero at non-collinear points, but its behavior is very non-uniform
when one approaches collinearity. The $\,\CO(\lambda^2)\,$ divergence
of the covariance of the radial component of force is consistent with
the appearance of the $\,\CO((R/\lambda)^{-2})\,$ term in the radial
component of the Euler equation when $\,\lambda\to\infty$.
\,On the other hand, the Euler equation for the angular component of
velocity had a finite limit when $\,(R/\lambda)\to0$ and so does the covariance
of the angular component of force.
\vskip 0.2cm

It is instructive to inspect how the limit $\,\lambda\to\infty\,$
works for the mean energy balance given by the identity
\qq
\partial_t\,\frac{_1}{^2}\,\big\langle\,\Vert v\Vert^2(\CX')\,
\big\rangle\ =\ \frac{_1}{^2}
\Big(\frac{_{(R/\lambda)^2}}{^{(R/\lambda)^2+r^2}}\m \lambda^{-2}\m
C^{rr}(\CX',\CX')\,+\,r^2\m
\lambda^{-2}\m C^{\varphi\varphi}(\CX',\CX')\Big)\ =\ \bar\iota\,,
\label{hfhf}
\qqq
where we ignored for the moment the viscous term.
The middle term is equal to $\,-R^{-2}\CC'(1)$, with equal contributions
$\,\frac{1}{2}\bar\iota\,$ from the
$\,C^{rr}\,$ and $\,C^{\varphi\varphi}\,$ terms.
\,How is this reproduced in the $\,R=0\,$ theory? Suppose that
the latter theory were given
by the stochastic equation
\qq
\partial_t v^\varphi+(v^r\partial_r+v^\varphi\partial_\varphi)v^\varphi
\,+\,\frac{_2}{^r}v^rv^\varphi\,=\,-\,
\frac{_1}{^{r^2}}\m\partial_\varphi p\,+\,f^\varphi\,,
\label{tvpf}
\qqq
with $\,f^\varphi(t,r,\varphi)\,$ white in time with the covariance
\qq
\!\!\!\!&&\big\langle\,f^\varphi(t_1,r_1,\varphi_1)\,
f^\varphi(t_2,r_2,\varphi_2)\,\big\rangle%\cr\cr&&\quad
= \delta(t_1-t_2) 
\begin{cases}{ -\frac{_1}{^2}(\frac{_1}{^{r_1^2}}+\frac{_1}{^{r_2^2}})\,
\CC'(\frac{_1}{^2}(\frac{_{r_1}}{^{r_2}}+\frac{_{r_2}}{^{r_1}}))\nonumber\cr
- \frac{_{r_1r_2}}{^4} (\frac{_1}{^{r_1^2}}-\frac{_1}{^{r_2^2}})^2
\CC''(\frac{_1}{^2}(\frac{_{r_1}}{^{r_2}}+\frac{_{r_2}}{^{r_1}}))
\ \ {\rm if}\ \ \varphi_1=\varphi_2\,, \nonumber\cr 
\,0\hspace{5.25cm}{\rm if}\ \ \varphi_1\not=\varphi_2\,,}\end{cases}
\qqq
i.e. the one obtained in the limit $\,R\to0$, \,see Eq.\,(\ref{Cppinf}).
If Eq.\,(\ref{tvpf}) is augmented by the constraint
(\ref{cons1}) and the incompressibility condition (\ref{inc0}) then
we obtain the mean energy balance as follows:
\qq
\partial_t\,\frac{_1}{^2}\,\big\langle\,r^2(v^\varphi)^2\,\big\rangle\ =\
\big\langle\,r^2v^\varphi(\partial_tv^\varphi)\,\big\rangle\,-\,\frac{_1}{^2}
\m \CC'(1)\,
=\,-\frac{_1}{^2}\m\CC'(1)\,=\,\frac{_1}{^2}\,\bar\iota\,.
\label{lhs}
\qqq
The contribution $\,-\frac{_1}{^2}\m\CC'(1)\,$ is the It\^{o} term and the
second equality results from the identity
\qq
r^2v^\varphi(\partial_tv^\varphi)&=&
\partial_r\CJ^r+\,\partial_\varphi\CJ^\varphi\,+\,
v^r(\partial_rp-r(v^\varphi)^2)\cr\cr
&&+\,(\frac{_1}{^2}r^2(v^\varphi)^2+p)(\partial_rv^r+\partial_\varphi v^\varphi)
\,+\,v^\varphi f^\varphi
\qqq
with the current given by Eqs.\,(\ref{CJr0}) and (\ref{CJp0}). Under the assumptions, the expectation of the right hand side
vanishes (the one of $\,v^\varphi f^\varphi\,$ due to the use of the
It\^{o} convention). We infer that only half of the energy injection rate
is pumped into the $\,R=0\,$ mean energy density $\,\frac{1}{2}\big\langle
r^2(v^\varphi)^2\big\rangle$. \,The other half has still to be carried by the
$\,v^r\,$ components forced with the covariance that diverges along
the collinear directions.

\subsection{Scaling at $\,R=0$}

For $\,R=0$, \,define
\qq
&&\tilde v^r(t,\lambda r,\varphi)\ =\ \lambda^{\alpha+1}
\m v^r(\lambda^\alpha t,r,\varphi)\,,\qquad
\tilde v^\varphi(t,\lambda r,\varphi)\ =\ \lambda^{\alpha}
\m v^r(\lambda^\alpha t,r,\varphi)\,,\\
&&\tilde p(t,\lambda r,\varphi)\ \ \m=\ \lambda^{2\alpha+2}
\m p(\lambda^\alpha t,r,\varphi)\,,\qquad
\tilde f^\varphi(t,\lambda r,\varphi)\ =\ \lambda^{2\alpha}
\m f^\varphi(\lambda^\alpha t,r,\varphi)\,,\qquad
\qqq
for $\,\alpha=- {2}/{3}$. \,Then the fields with ``$tilde$'' correspond
to the $\,R=0\,$ theory with the same forcing as those without ``$tilde$''.
It follows that at $\,R=0\,$ the velocity correlation functions satisfy
\qq
\!\!\!\!&&\lambda^{\frac{2}{3}n-m}\,F_{\NH_{_0},\CC}^{i_1\dots i_n}
(\lambda^{\frac{2}{3}}t;\lambda r_1,\varphi_1;
\dots;\lambda r_n,\varphi_n)
=F_{\NH_{_0},\CC}^{i_1\dots i_n}(t;r_1,\varphi_1;\dots;r_n,\varphi_n)\,,
\label{scal0}
\qqq
if $\,m\,$ of the indices $\,i_k\,$ are equal to $\,r\,$ and $\,(n-m)\,$
to $\,\varphi$. \,This relation would follow from the identity
(\ref{scalRl}) but only under the condition that the latter has a limit when
$\,\lambda\to\infty$, \,which is not always the case.
\vskip 0.2cm

Let us consider the 2-point function of the velocity. The only
$\,SL(2,\NR)$-invariant of two non-colinear points on the cone
$\,\NH_{_0}\,$ is
\qq
-\frac{_{(\CX^1_1\CX^1_2+\CX^2_1\CX^2_2-\CX^3_1\CX^3_2)}}{^{L^2}}
\ =\ \frac{_{r_1r_2
\,(1-\cos(\varphi_1-\varphi_2))}}{^{L^2}}\ \equiv x\,,\qquad
\label{x}
\qqq
where $\,L\,$ is an arbitrary length scale introduced to render $\,x\,$
dimensionless.
\,When two points are colinear, \,i.e. $\varphi_1=\varphi_2$, \,then their only
invariant is $\,r_1/r_2$.
$\,SL(2,\NR)$-covariant 2-point functions of the velocity field satisfying
the incompressibility condition (\ref{inc0})
have the form
\qq
\big\langle\,v^{r_1}\,v^{r_2}\,\big\rangle\,&=&
\partial_{\varphi_1}\partial_{\varphi_2}F(t,x)
\ =\ (x\,F'(t,x)+x^2\m F''(t,x)\cr
&&\hspace{4.67cm}-\,\frac{_{r_1r_2}}{^{L^2}}\,(F'(t,x)+2\m x\,F''(t,x))\,,
\label{vrvr0}\\
\big\langle\,v^{r_1}\,v^{\varphi_2}\,\big\rangle\m&=&-\m
\partial_{\varphi_1}\partial_{r_2}F(t,x)%\cr
\ =\ -\m\frac{_{r_1\,\sin(\varphi_1-\varphi_2)}}
{^{L^2}}\,(F'(t,x)\,+\,x\, F''(t,x))\,,
\qquad\label{vrvp0}\\
\big\langle\,v^{\varphi_1}\,v^{\varphi_2}\,\big\rangle&=&
\partial_{r_1}\partial_{r_2}F(t,x)%\cr
\ =\ \frac{_1}{^{r_1r_2}}\,(x\,F'(t,x)\,+\,x^2\m F''(t,x))\,.
\label{vpvp0}
\qqq
There are two explicit solutions satisfying
scaling (\ref{scal0}). \,The first one is proportional to $\,t$,
\,modulo a logarithmic correction:
\qq
F(t,x)\ =\ \left(\frac{_1}{^2}A\m(\ln{x})^2+B\ln x+C\right)t\,-\,3A(\ln{x})\,.
t\ln{t}
\qqq
That leads to
\qq
\big\langle\,v^{r_1}\,v^{r_2}\,\big\rangle\,&=&\left(A\,-\,
\frac{_{2A-A\m\ln{x}-B}}
{^{1-\cos(\varphi_1-\varphi_2)}}\right)t\,-\,
\frac{_{3\m A}}{^{1-\cos(\varphi_1-\varphi_2)}}\,t\ln{t}\,,\\
\big\langle\,v^{r_1}\,v^{\varphi_2}\,\big\rangle\m&=&-\,\frac{_{A\,
\sin(\varphi_1-\varphi_2)}}{^{r_2\,(1-\cos(\varphi_1-\varphi_2))}}\,t\,,\\
\big\langle\,v^{\varphi_1}\,v^{\varphi_2}\,\big\rangle&=&\frac{_A}{^{r_1r_2}}
\,t\,.
\qqq
The second sollution is time-independent:
\qq
F(t,x)\ =\ 9\m D\,x^{\frac{1}{3}}\,+\,E
\qqq
leading to
\qq
\big\langle\,v^{r_1}\,v^{r_2}\,\big\rangle\,&=&\frac{_{D\,(2-\cos(\varphi_1-\varphi_2)}}
{^{1-\cos(\varphi_1-\varphi_2)}}\,x^{\frac{1}{3}}\,,\\
\big\langle\,v^{r_1}\,v^{\varphi_2}\,\big\rangle\m&=&-\,\frac{_{D\,
\sin(\varphi_1-\varphi_2)}}{^{r_2\,(1-\cos(\varphi_1-\varphi_2))}}
\,x^{\frac{1}{3}}\,,\\
\big\langle\,v^{\varphi_1}\,v^{\varphi_2}\,\big\rangle&=&\frac{_D}{^{r_1r_2}}
\,x^{\frac{1}{3}}\,.
\qqq
Both solutions preserve the $\,SL(2,\NR)$-symmetry but violate the
$\,\Diff\m S^1\,$ one.

\subsection{Velocity 2-point functions}

Let us return to the theory on $\,\NH_{_R}\,$
with the equal-time covariance of the stream function
\qq
\big\langle\psi(t,\CX_1)\,\psi(t,\CX_2)\big\rangle\
=\ \Psi(t,x)\ +\ \dots\,,
\label{strfctcov}
\qqq
where now
\qq
\!\!\!x\ \equiv\ \cosh\big(\delta_{12}/R\big)&=&
\frac{r_1r_2}{ {R^2}}\left(\sqrt{1+R^2r_1^{-2}}
\sqrt{1+R^2r_2^{-2}}\,-\,\cos(\varphi_1-\varphi_2)\right)\,,
\label{xR}
\qqq
see Eq.\,(\ref{chdist}), and the dots do not contribute
to the velocity correlators. The explicit expressions for the rescaled
version of the latter in terms of $\,\Psi\,$ are given
in Appendix \ref{appx:rescvel}.
Recall that in Sec.\,\ref{sec:invcasc} we have postulated the inverse
cascade scenario with the long-time behavior of the equal-time covariance
of stream function behaving according to Eq.\,(\ref{invc}). \,Now we may
check the consistency of that scenario with the postulated
long-time-long-distance asymptotics imposed by the existence of the
scaling limit at $\m\lambda\to\infty\m$ of the rescaled correlation
functions (\ref{scalRl}). Assuming that for large $x$
\qq
\Psi_0(x)\quad&\,\approx\,&\frac{1}{2}A(\ln{x})^2+B\ln{x}+C\,,\label{asympsi0}\\
\Psi_{stat}(x)&\,\approx\,&9Dx^{1/3}+E\,,\label{asympsistat}
\qqq
together with the first and second derivatives, \,we infer from the formulae
(\ref{CrrR})-(\ref{CppR}) of Appendix \ref{appx:rescvel} that
\qq
&&\lambda^{\frac{4}{3}-2}\,F^{rr}_{\NH_{_R},\CC}(\lambda^{\frac{2}{3}}t;
\lambda r_1,\varphi_1;\lambda r_2,\varphi_2)\ =\ \frac{_{2\m A}}{^{R^2(1-
\cos(\varphi_1-\varphi_2))}}\,\ln{\lambda}%\cr&&\hspace{6.5cm}
 +\
\frac{_1}{^{R^2}}\left(A\,-\,
\frac{_{2A-A\m\ln{x}-B}}
{^{1-\cos(\varphi_1-\varphi_2)}}\right)t\cr\cr&& +\,o(\lambda)\m t %\hspace{6.5cm}
+ \frac{_{D\,(2-\cos(\varphi_1-\varphi_2)}}
{^{R^2(1-\cos(\varphi_1-\varphi_2))}}\,x^{\frac{1}{3}}\ +\ o(\lambda)\,,
\label{vrvrinf}\\ \cr
&&\lambda^{\frac{4}{3}-1}\m F^{r\varphi}_{\NH_{_R},\CC}
(\lambda^{\frac{2}{3}}t;\lambda r_1,
\varphi_1;\lambda r_2,\varphi_2)\,\ =\
\frac{_1}{^{R^2}}\frac{_{A\,
\sin(\varphi_1-\varphi_2)}}{^{r_2\,(1-\cos(\varphi_1-\varphi_2))}}\,t\
+\ o(\lambda)\m t\cr\cr
&&\hspace{6.5cm}-\ \frac{_{D\,
\sin(\varphi_1-\varphi_2)}}{^{R^2\,r_2\,(1-\cos(\varphi_1-\varphi_2))}}
\,x^{\frac{1}{3}}\ +\ o(\lambda)\,,\label{vrvpinf}\\ \cr
&&\lambda^{\frac{4}{3}}\m F^{\varphi\varphi}_{\NH_{_R},
\CC}(\lambda^{\frac{2}{3}}t;\lambda r_1,\varphi_1;\lambda r_2,\varphi_2)
\,\quad\ =\
\frac{_A}{^{R^2r_1r_2}}\,t\ +\ o(\lambda)\m t%\cr&&\hspace{6.5cm}
+\ \frac{_D}{^{R^2r_1r_2}}\,x^{\frac{1}{3}}\ +\ o(\lambda)\,\,
\label{vpvpinf}
\qqq
for large $\,\lambda$. \,Here $\,x\,$ is given by Eq.\,(\ref{x})
with $ L=R$. \,In other words, the scaling limit exists, except
for the 2-point function of the radial velocity diverging
logarithmically. That divergence could be canceled by a
$\,\sim t\ln{t}\,$ term as in the $\,R=0\,$ theory, but appearance
of such a term in $\,R>0\,$ theory is inconsistent since
\qq
\frac{_1}{^2}\frac{_{R^2}}{^{R^2+r^2}}\,\big\langle\,(v^r)^2(t)\,\big\rangle\
=\ \frac{_1}{^2}\,\bar\iota\,t,
\qqq
see Eq.\,(\ref{hfhf}) and the discussion that follows it.
\,Since the 2-point function of $\,v^r\,$ at coinciding points bounds
the one at non-coinciding points by the Schwartz inequality, then the 
latter cannot grow with time faster than linearly (this argument does 
not work at $\,R=0\,$ where $\,v^r\,$ does 
not contribute to the energy balance).
\,The logarithmic divergence with $\,\lambda\,$ on the right hand side of
Eq.\,(\ref{vrvrinf}) may rather be related to the collinear
divergence of the rescaled covariance of the radial force, see
Eq.\,(\ref{Crrinf}). Further checks of that explanation should be performed.
\vskip 0.2cm

Let us also look at the scaling limit of the velocity 2-point functions
at the colinear points with $\,\varphi_1=\varphi_2$, \,where
\qq
\lim\limits_{\lambda\to\infty}\
\cosh\hspace{-0.06cm}\frac{_{\delta_{1'2'}}}{^{R/\lambda}}\ =\ \frac{_1}{^2}
\Big(\frac{_{r_1}}{^{r_2}}+\frac{_{r_2}}{^{r_1}}\Big)\,.
\qqq
We have:
\qq
&&\lim\limits_{\lambda\to\infty}\ \,\lambda^{\frac{4}{3}}\,
F^{\varphi\varphi}_{\NH_{_R},C}(\lambda^{\frac{2}{3}}t;
\lambda r_1,\varphi;\lambda r_2,\varphi)\cr\cr
&&=\
\,-\m\frac{_{1}}{^{4\m r_1r_2\m R^2}}\Big[2\m
(\frac{_{r_1}}{^{r_2}}+\frac{_{r_2}}{^{r_1}})\,\Psi_{0}'(\frac{_1}{^2}
(\frac{_{r_1}}{^{r_2}}+
\frac{_{r_2}}{^{r_1}}))\,
+\,(\frac{_{r_1}}{^{r_2}}-
\frac{_{r_2}}{^{r_1}})^2\,\Psi_{0}''(
\frac{_1}{^2}(\frac{_{r_1}}{^{r_2}}+
\frac{_{r_2}}{^{r_1}}))\Big]t\,.\qquad
\qqq
The term involving $\,\Psi_{stat}\,$ behaves as
$\,\CO(\lambda^{-\frac{2}{3}})\,$ giving no contribution here.
The mixed radial-angular velocity correlator vanishes at the collinear
configuration, whereas the rescaled radial one is
\qq
&&\lambda^{\frac{4}{3}-2}\,F^{rr}_{\NH_{_R},\CC}(\lambda^{\frac{2}{3}}t;
\lambda r_1,\varphi,\lambda r_2,\varphi)\ =\ -\,
\frac{_{r_1r_2\,\sqrt{1+(R/\lambda)^2r_1^{-2}}\,
\sqrt{1+(R/\lambda)^2r_2^{-2}}}}{^{R^4}}\cr\cr
&&\hspace{5cm}\times\,\Big[\lambda^{2}\m\Psi_0'(\cosh\hspace{-0.06cm}
\frac{_{\delta_{12}}}{^{R/\lambda}})\,t\,+\,\lambda^{\frac{4}{3}}
\Psi_{stat}'(\cosh\hspace{-0.06cm}
\frac{_{\delta_{12}}}{^{R/\lambda}})\Big].\qquad
\qqq
The right hand side diverges when $\,\lambda\to\infty$, \,confirming
the singular character of the radial velocity correlators in this limit.
\vskip 0.2cm

\begin{figure}[h]
\begin{center}
\leavevmode
\hspace{0.05cm}
{
      \begin{minipage}{0.44\textwidth}
        \includegraphics[width=7cm,height=6cm]{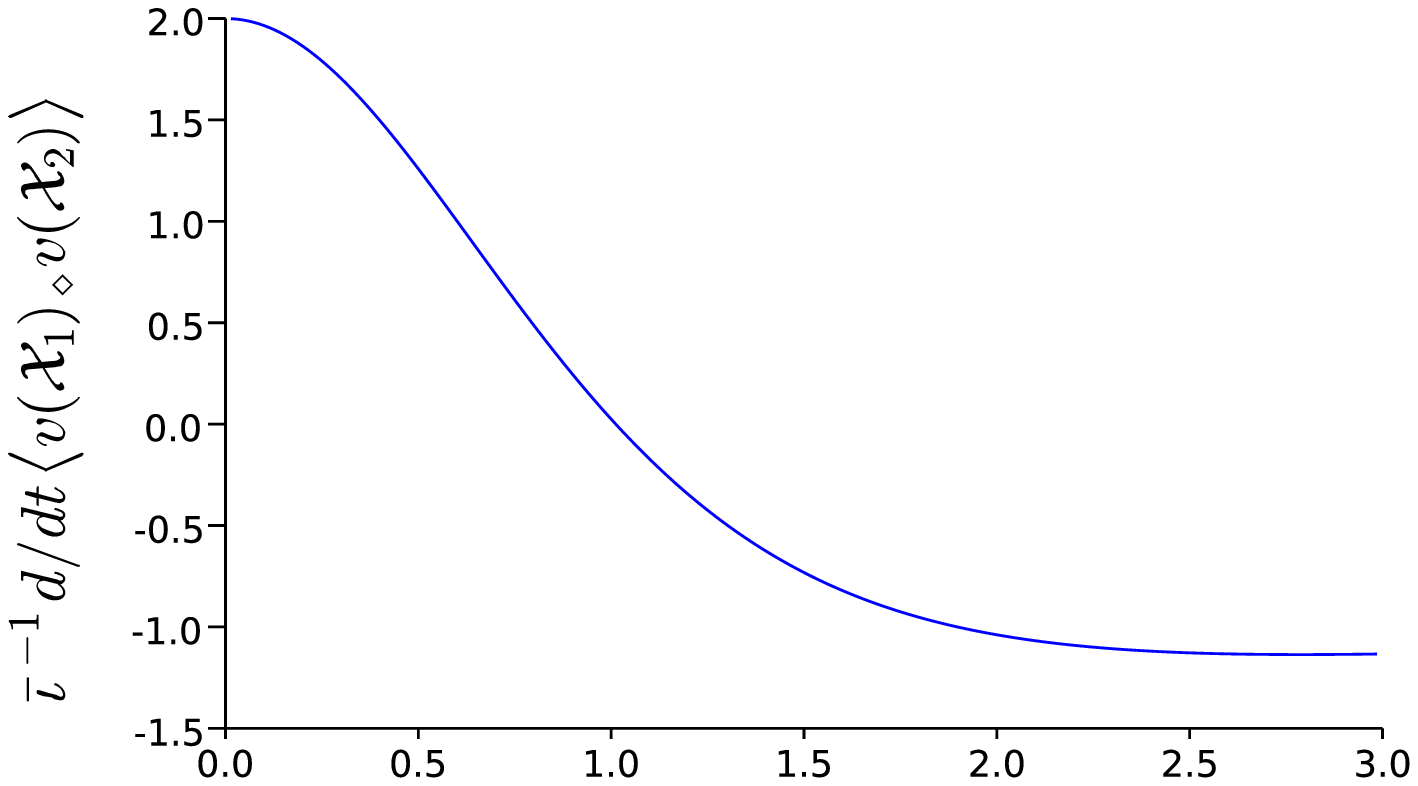}\\
        \vspace{-0.8cm} \strut
        \caption{Intrinsic growing 2-pt mode}
        \label{fig:intr.mode}
        \end{minipage}}
    \hspace*{0.8cm}
{
      \begin{minipage}{0.44\textwidth}
        \includegraphics[width=7cm,height=6cm]{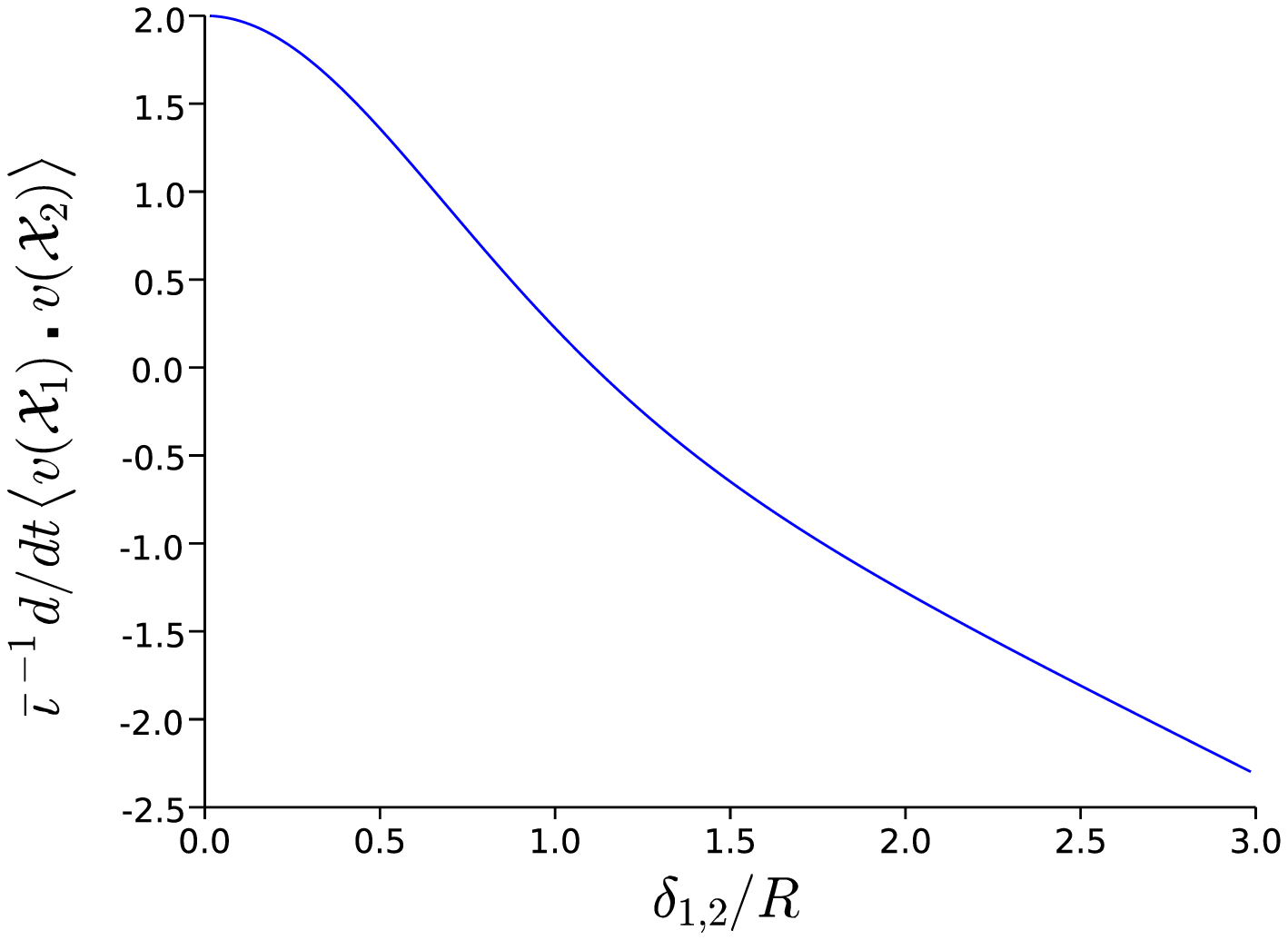}\\
        \vspace{-0.8cm} \strut
        \caption{Extrinsic growing 2-pt mode}
        \label{fig:extr.mode}
        \end{minipage}}
    \hspace*{15pt}
    \hspace{14pt}
\end{center}
\end{figure}

In all, \,the above analysis shows the consistency of the 
scenario (\ref{invc}) with the asymptotic behavior 
of the modes $\,\Psi_0\,$ and $\,\Psi_{stat}$ for large $\,x\,$ given 
by relations (\ref{asympsi0}) and (\ref{asympsistat}) and their first 
two derivatives, in particular, of the asymptotic equalities
\qq
\Psi_0'(x)\ \approx\ A\,\frac{\ln(x)}{x}+B\,\frac{1}{x}\,,\qquad
\Psi_0''(x)\ \approx\ A\,\frac{1-\ln(x)}{x^2}-B\,\frac{1}{x^2}
\label{approx}
\qqq
If we extrapolate this asymptotic behavior
to $\,O(1)\,$ values of $\,x\,$ then relation (\ref{Psi0pr}) fixes
the value of $\,B\,$ to \,$-R^2\bar\iota\,$ whereas relation
(\ref{lhs}) for the scaling limit (\ref{vpvpinf}) of the
$\,\langle v^\varphi\,v^\varphi\rangle\,$ correlation function
fixes $\,A\,$ to the value $\,R^2\bar\iota$. \,With such an
extrapolation, the growing-in-time contributions to the
"intrinsic" and "extrinsic" invariant velocity 2-point
functions are given by the relations
\qq
\partial_t\big\langle\,v(\CX_1){_{^{\hspace{0.07cm}\diamond\,}}}
v(\CX_2)\,\big\rangle&\,
=\,&\bar\iota\,\Big(-1+\frac{1-\ln{x}}{x}+\frac{2-\ln{x}}{x^2}\Big)\,,
\\
\partial_t\big\langle\,v(\CX_1)\centerdot v(\CX_2)\,\big\rangle\,&\,
=\,&\bar\iota\,\Big(-\ln{x}+\frac{2-\ln{x}}{x^2}\Big)\,,
\qqq
for $\,x=\cosh(\delta_{12}/R)$. \, Figures\,\ref{fig:intr.mode} 
and \ref{fig:extr.mode}
show the dependence of the respective expressions
on $ {\delta_{12}}/{R}$.
\vskip 0.2cm

\noindent The time derivatives of both
invariant 2-point functions are equal to $\,2\bar\iota\,$ at the coinciding
points, in accordance with the energy balance,
and stay close to that value, to which they would be locked
in the flat space, at $\,\delta_{12}\ll R$.
\,They change sign around
$\,\delta_{12}\approx R\,$ showing that velocities become anti-correlated 
at the distances exceeding the curvature radius. Continuing growth with 
the distance of the absolute value of the extrinsic function shows that 
this is not the proper object to describe velocity correlations at different 
points. We therefore focus on physically more meaningful 
intrinsic 2-point function, which tends to the constant $\,-\bar\iota\,$ 
at large distances. That means that, when compared using the parallel 
transport along geodesics, the velocities stay anti-correlated at long 
times and long distances by half of the value at coinciding points. Most 
probably,
that behavior of the correlation function may be interpreted as follows. Our
energy condensing mode is not a coherent motion, but a correlation function
contribution, i.e. it is built of separate pieces of flow. Those look like
a uniform flow only at distances shorter than the radius of curvature. 
Indeed, inverse cascade in a flat space produces ever-increasing pieces 
of almost uniform velocity, which
locally look like jets. However, when the transversal extent of a jet
overgrows the curvature radius $\,R$, \,the jet splits into two.
Anti-correlation at larger distances means that they are circular motions.
The fact that anti-correlation persists to arbitrary large (ever-increasing
with time) scale means that there are vortex rings with arbitrary large
radius, but the fact that the positive correlation is only until $\,R\,$
means that those vortices are actually narrow rings with the width of
order $\,R$.  \,Incidentally, the value of the correlation at large
distances being {\it half} of the value at zero follows from the picture
of a jet (at distances smaller than $R$) splitting into {\it two} vortices.
The fact that we do not grow ever-increasing pieces of a uniform flow
is likely related to the absence of Galilean invariance. Why curved space
favors rings is unclear. It would be interesting to look at concrete
examples of such flows on the hyperbolic plane.
\vskip 0.2cm

\begin{figure}[h]
\begin{center}
\leavevmode
{
      \begin{minipage}{0.43\textwidth}
        \includegraphics[width=7cm,height=6cm]{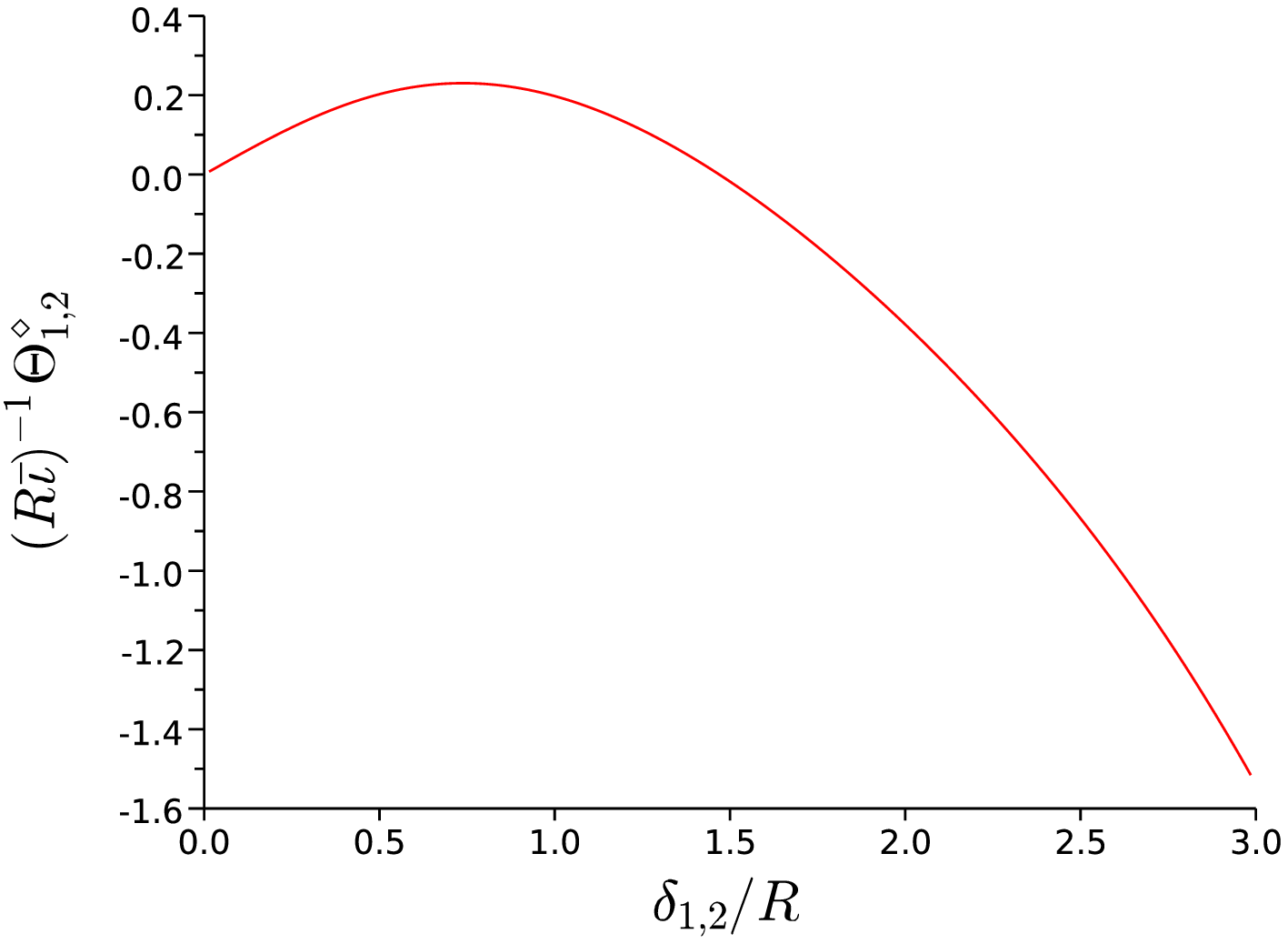}\\
        \vspace{-0.8cm} \strut
        \caption{Intrinsic energy current}
        \label{fig:intr.ecurr}
        \end{minipage}}
    \hspace*{0.8cm}
{
      \begin{minipage}{0.43\textwidth}
        \includegraphics[width=7cm,height=6cm]{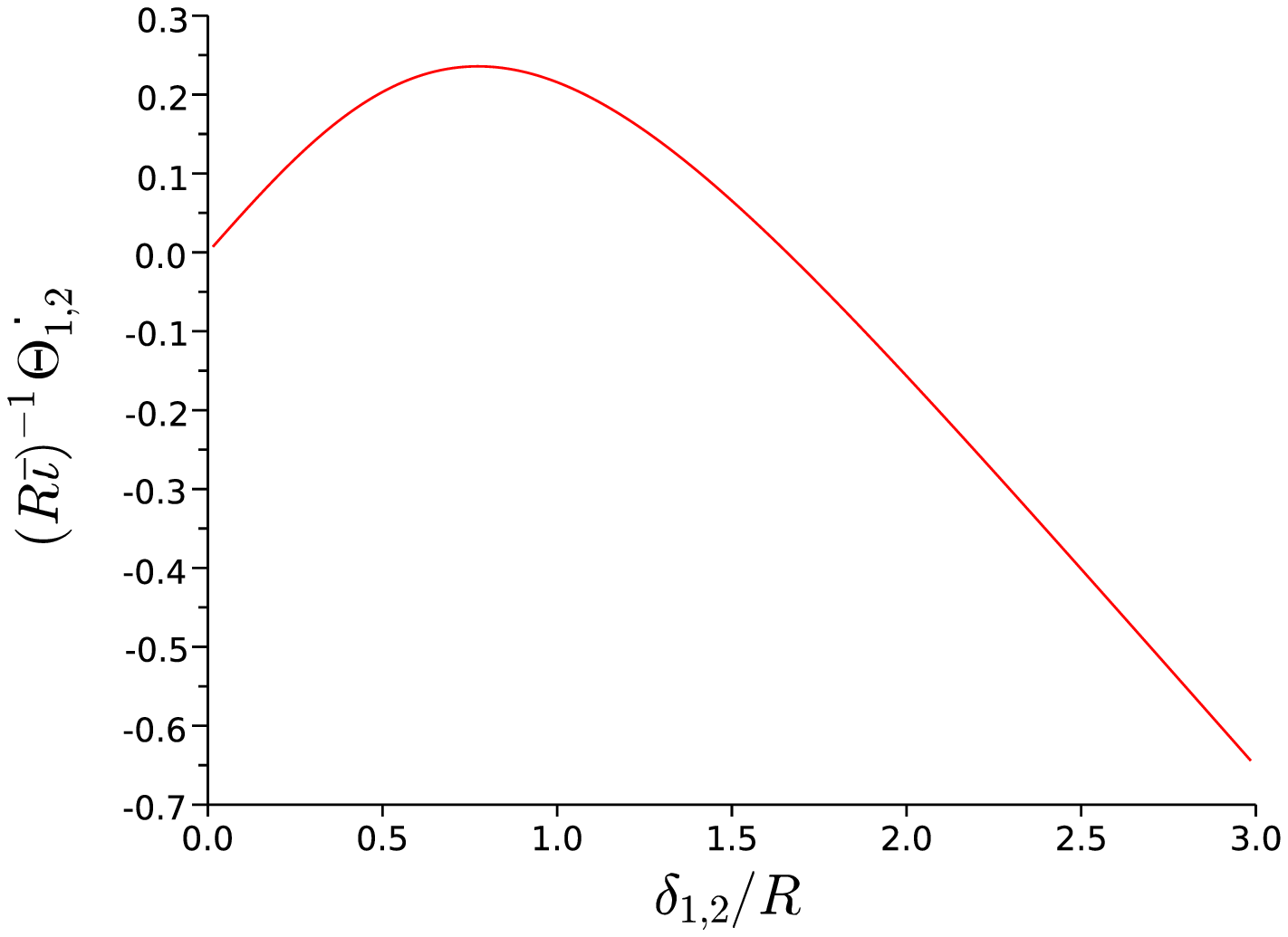}\\
      \vspace{-0.8cm} \strut
        \caption{Extrinsic energy current}
        \label{fig:extr.ecurr}
        \end{minipage}}
    \hspace*{15pt}
    \hspace{14pt}
\end{center}
\end{figure}

Similarly, Figures\,\ref{fig:intr.ecurr} and \ref{fig:extr.ecurr} 
show the dependence on $ {\delta_{12}}/{R}\,$ 
of the "intrinsic" and "extrinsic" energy current of Eqs.\,(\ref{Th12star})
and (\ref{Th12}) given, under the extrapolation assumption, by
the expressions
\qq
\Theta^{\hspace{0.01cm}^{_\diamond}}(\delta_{12})&\,=\,&\frac{R\,\bar\iota}{4}\,
(x^2-1)^{^{-\frac{1}{2}}}
\Big[\frac{1-x^2}{2}+(3-x)\ln{x}+x-(\ln{x})^2+\frac{\ln{x}}{x}
-\frac{1}{x}\Big],\qquad\\
\Theta^{\,^\centerdot}(\delta_{12})&\,=\,&\frac{R\,\bar\iota}{2}\,(x^2-1)^{^{\frac{1}{2}}}\,
\frac{1-\ln{x}}{x}\,.
\qqq
\vskip 0.2cm

\vskip 0.2cm

\noindent Both expressions behave as $\,\bar\iota\,\delta_{12}/2\,$
for distances $\,\ll R$, \,which is the flat space behavior. For larger
distances both reach a maximum and eventually turn negative. The
absolute value of the intrinsic current increases then exponentially in
$\,\delta_{12}\,$ at long distances whereas that of the extrinsic current
behaves asymptotically as $\,\,\bar\iota\,\delta_{12}/2$.
\,The behavior of the intrinsic (i.e. physically meaningful) energy current
suggests that at large distances it is proportional (with a logarithmic
accuracy) to the circumference $\,R\exp(\delta_{12}/R)\,$ rather than
to the radius $\,\delta_{12}$. \,To interpret it, recall that the flux
through the scale $\delta$ is the energy density $v_\delta^2$ at that scale
divided by the turn-over time $t_\delta\simeq\ell/v_\delta$, and the energy
flux constancy predicts $v_\delta^3\propto \ell$.  \,Identifying
$\,v_\delta^3\,$ with $\,\Theta^{\hspace{0.01cm}^{_\diamond}}(\delta)\,$ we see that
$\ell$ must be taken as the circumference rather than the radius in the 
relation for the turnover time. That probably
makes sense since what mostly contributes to the velocity difference between
two points (and the energy at this scale) is an eddy, i.e. a flow along
circumference having these two points at opposite ends.
\vskip 0.2cm

What is the meaning of this saturation of current at intermediate distances?
In a flat space, divergence of the current is the energy transfer rate which
is a scale-independent constant for a cascade. In our case, we see that the
transfer rate decreases with increase of the distance and changes sign
(when current saturates) at a distance comparable to the radius of curvature.
This seems consistent with  the non-cascade behavior at longer distances,
namely the growth of the energy condensing autocorrelation mode carried
by large-scale motions.
\vskip 0.2cm

The contribution of the stationary mode $\,\Psi_{stat}\,$ to both
invariant 2-point functions is expected to have the flat space
behavior  $\,\propto-{\bar\iota}^{2/3}\,\delta_{12}^{2/3}\,$  for
$\,\delta_{12}\ll R\,$ whereas for large distances its behavior should
be $\,\propto-R^{2/3}{\bar\iota}^{2/3}\,x^{1/3}\,$ leading to an
exponential decrease when $\,\delta_{12}\to\infty$.
\vskip 0.2cm

What can we infer about the energy spectrum $\,\CE(t,k)$ from  the behavior of $\,\Psi_{0}\,$ and $\,\Psi_{stat}\,$ at large $\,x\,$? \,Suppose that
$\,\Psi_0\,$ and $\,\Psi_{stat}\,$ behave asymptotically as in
Eqs.\,(\ref{asympsi0}) and (\ref{asympsistat}) up to a sum of negative,
possibly fractional, powers of $\,x$. \,In particular, both modes are
very different than the zero wavenumber mode $\,P^{-\frac{1}{2}}_{00}(x)\,$
that decreases as $\,x^{-1/2}$, see (\ref{GrR}). \,We show in Appendix
\ref{appx:spectr} that under these assumptions, the contributions
of $\,\Psi_0\,$ and $\,\Psi_{stat}\,$ to the energy spectrum are analytic
around the positive real axis including $\,k=0$, \,in contrast with the
flat space behavior where $\,\CE_0(t,k)\propto t\hspace{0.02cm}\delta(k)\,$
and $\,\CE_{stat}(k)\propto k^{-5/3}$. \,In particular, the stationary
autocorrelation mode then contains finite energy (per unit area),
unlike in the flat space case.

\nsection{Conclusions}

\noindent We considered an inverse cascade scenario for the
two-dimensional turbulence in the hyperbolic plane described by the
Navier-Stokes equation with a random Gaussian white-in time forcing
operating on scales much shorter than the curvature radius $\,R$.
The main focus was on the possible structure of the mode of the equal-time
2-point  velocity function growing linearly in time. On distances
much smaller than the curvature radius, where one can ignore the
curvature effects, such a mode should behave similarly to the energy
condensing constant mode in a flat space. Long-distance long-time scaling 
limit of the theory on hyperbolic plane lives on the conical surface. By 
analyzing such theory,  we have found the long-distance behavior of the 
energy condensing mode and of the energy current across scales.
Their behavior indicates that the energy transfer on scales longer
than $\,R\,$ loses the cascade character where the energy
flux is constant across scales and the spectrum stabilizes up to smaller
and smaller wave numbers. Instead, the spectral density in all scales
eventually grows linearly in time and the energy current across
scales reaches a maximum around the curvature radius distance.
The analytic results seem to indicate that the energy is cumulated
in ring-like vortices of arbitrary diameter but of width of order $\,R$.
While these results have not brought any direct progress in understanding the
conformal-invariant sector of the inverse cascade visible in the
statistics of the zero-vorticity lines, yet we exhibited on the way
the $\,\Diff\m S^1$-symmetry of the
unforced Euler equation in the asymptotic geometry at infinity of
the hyperbolic plane. This infinite-dimensional 
conformal symmetry in one dimension appears, however, to be broken 
by the forcing.
\,The analysis that we performed here was limited to few consistency
checks on and few implications of the inverse-cascade scenario assumed 
for the hyperbolic-plane turbulence. It calls for numerical simulations 
that could confirm or refute the picture put forward here. Such simulations 
would have to adapt the numerical
schemes for flat-space two-dimensional turbulence to the context
of the hyperbolic geometry, which is not a trivial task. The problem
of turbulence in the hyperbolic plane might seem somewhat academic
as such geometry is not realized by surfaces in three-dimensional
Euclidean space \cite{H}. Nevertheless, the basic results about
the influence of the negative curvature on the behavior of the cascade
at distances longer than the curvature radius may have some bearing
on the fate of the inverse cascade in soap bubbles and films that form surfaces
with negative curvature. Measuring such curvature effects in the soap
film turbulence \cite{KeGo} would be a challenge for experimentalists.

\nsection{Appendices}
\begin{appendix}

\setcounter{section}{0}
\setcounter{equation}{0}
\def\thesection{{\small A}}
\def\theequation{A.\arabic{equation}}
\section{\small\bf Some notions of Riemannian geometry}
\label{appx:Riemgeom}

\noindent Let $M$ be a $d$-dimensional oriented manifold equipped with a
Riemannian metric $\,g_{ij}{\rm d}x^i{\rm d}x^j$. \,Such a metric may 
be used to lower
the indices of the contravariant tensor fields, \,e.g. $v_i=g_{ij}v^j$.
\,One has $\,v^j=g^{ij}v_i$, \,where $\,(g^{ij})\,$ is the matrix inverse
to $\,(g_{ij})$. \,The metric together with the orientation induces
the volume form $\,\chi=\sqrt{g}\,{\rm d}x^1\wedge\cdots\wedge{\rm d}x^d
\equiv\sqrt{g}\,{\rm d}x$, 
\,where
$\,\sqrt{g}\,$ is the shorthand notation for $\,\sqrt{{\rm det}(g_{ij})}$.
\,A Riemannian metric induces the Levi-Civita connection acting on tensor
fields by
\qq
&&\nabla_it_{k_1\dots k_r}^{j_1\dots j_p}\cr 
&&=\,\partial_it_{k_1\dots k_r}^{j_1\dots j_p}
\,+\,\Gamma_{il}^{j_1}t_{k_1\dots k_r}^{lj_2\dots j_p}
+ \cdots +\Gamma_{il}^{j_p}
t_{k_1\dots k_r}^{j_1\dots j_{p-1}l}
\,-\,\Gamma_{ik_1}^lt_{lk_2\dots k_r}^{j_1\dots j_p}- 
\cdots -\Gamma_{ik_r}^lt_{k_1\dots k_{r-1}l}^{j_1\dots j_p}\,,\ \qquad
\qqq
where $\,\Gamma^k_{ij}\,$ are given by the Christoffel symbols
\qq
\Gamma^k_{ij}\equiv\{{_k\atop^{i\s j}}\}\s=\s
{_1\over^2}\m g^{kn}(\da_i g_{jn}+\da_j g_{in}-\da_n g_{ij})\s=\,
\{{_k\atop^{j\s i}}\}.
\qqq
For the Levi-Civita connection, the covariant derivatives of the tensors
$g_{ij}$ and $g^{ij}$ vanish so that $\nabla_i$ commutes with the raising and
lowering of indices. The Riemann curvature tensor is defined by the formula
\qq
(\nabla_k\nabla_l-\nabla_l\nabla_k)t^i\,=\,R^i_{\,jkl}t^j\qqq 
and Ricci tensor by
\qq({\rm Ric})_{jl}\,=\,R^i_{\,jil}\,=\,({\rm Ric})_{lj}\,.\qqq
The scalar curvature is
$ S\,=\,g^{jl}({\rm Ric})_{jl}$.
The totally antisymmetric tensors may be identified with
the differential forms by the correspondence
\qq
(\eta_{k_1\dots k_r})\quad\leftrightarrow\quad
\eta={_1\over^{r!}}\,\eta_{k_1\dots k_r}{\rm d}x^{k_1}
\wedge\cdots\wedge{\rm d}x^{k_r}.
\qqq
The exterior derivative in terms of this identification is given by
the formula
\qq
({\rm d}\eta)_{k_0k_1\dots k_r}&=&\da_{k_0}\eta_{k_1\dots k_r}
-\partial_{k_1}\eta_{k_0k_2\dots k_r}\ \cdots\ (-1)^r\da_{k_r}
\eta_{k_0\dots k_{r-1}}\cr
&=&\nabla_{k_0}\eta_{k_1\dots k_r}-\nabla_{k_1}\eta_{k_0k_2\dots k_r}\ \cdots\
(-1)^r\nabla_{k_r}\eta_{k_0\dots k_{r-1}}.
\label{extder}
\qqq
The Hodge star on $r$-forms is defined by
\qq
(*\eta)_{i_1\dots i_{d-r}}\,=\,{_1\over^{r!}}\,
\eta^{j_1\dots j_r}\,\epsilon_{j_1\dots
j_ri_1\dots i_{d-r}}\,\sqrt{g}\,.
\label{H*}
\qqq
It satisfies $\,*^2=(-1)^{(d-r)r}$. \,The $L^2$ scalar product
of $r$-forms is given by
\qq
(\eta,\xi)\ =\,\int\eta\wedge*\xi\ =\ {_1\over^{r!}}\int
\eta^{i_1\dots i_r}\,\xi_{i_1\dots i_r}\,\chi\ =\ (\xi,\eta)\
=\ (*\eta,*\xi).
\qqq
The (formal) adjoint of the exterior derivative $d$ mapping $r$-forms
to $(r+1)$-forms is the operator
\qq
{\rm d}^\dagger\,=\,(-1)^{r+1}*^{-1}\hspace{-0.1cm}{\rm d}\,*
\label{ddag}
\qqq
mapping $(r+1)$-forms to $r$-forms. \,In terms of the tensor components,
\qq
({\rm d}^\dagger\eta)_{i_1\dots i_r}\,=\,-\nabla_{i}\eta^i_{\ i_1\dots i_r}\,.
\label{ddag1}
\qqq
The (negative) Laplace-Beltrami operator is
\qq
\Delta\,=\,-({\rm d}^\dagger{\rm d}+{\rm d}\hspace{0.02cm}{\rm d}^\dagger)\,.
\label{LaplB}
\qqq
In the action on functions, it gives
\qq
\Delta\psi\ =\ \frac{_1}{^{\sqrt g}}\,\partial_i\,g^{ij}\sqrt{g}\,\partial_j\psi\,.
\label{Lapl}
\qqq
The Laplace-Beltrami operator act also on vector
fields $\,v\,$ by the formula
\qq
(\Delta v)^{\flat}\,=\,\Delta v^{\flat}\,,
\label{laplvf}
\qqq
where $\,v^{\flat}\,$ is the 1-form represented by $(v_i)$ associated 
to the vector field represented by $\,(v^i)$.
\,We also need the Lie derivative of tensors w.r.t. vector
fields $\,u\,$ given by the formulae
\qq
&&(\CL_uv)^i\,=\,[u,v]^i\,=\,u^j\partial_jv^i-v^j\partial_ju^i\,=\,
u^j\nabla_jv^i-v^j\nabla_ju^i\,,\label{LDv}\\ \cr
&&(\CL_u\eta)_i\,=\,(\partial_iu^j)\eta_j+u^j(\partial_j\eta_i)
\,=\,(\nabla_iu^j)\eta_j+u^j(\nabla_j\eta_i)\,,
\label{LDf}
\qqq
in the action on the vector fields and 1-forms, respectively, and by the
Leibniz rule on higher tensors. In particular, the vector fields $\,u\,$ are called Killing vectors if $\,\CL_u\,$ annihilate the metric tensor $\,g$:
\qq
(\CL_ug)_{ij}=(\nabla_iu^k)g_{kj}+(\nabla_ju^k)g_{ik}\,=\,\nabla_iu_j+
\nabla_ju_i\,=\,0\,,
\qqq
We also consider divergenceless
vector fields $\,v\,$ defined by the condition $\,\CL_v\chi=0$.
In coordinates, the condition is equivalent to the
conditions $\,\partial_k(v^k\sqrt{g})=0\,$ or $\,\nabla_kv^k=\sqrt{g}^{-1}
\partial_k(v^k\sqrt{g})=0$. \,Yet another form of this condition
is the requirement that $\,{\rm d}^\dagger v^\flat=0$. \,Locally, every 1-form
satisfying such condition can be written via $(d-2)$-form $\,\psi\,$:
\qq
v^\flat\,=\,*{\rm d}\psi
\label{stream}
\qqq

\setcounter{section}{0}
\setcounter{equation}{0}
\def\thesection{{\small B}}
\def\theequation{B.\arabic{equation}}

\section{\small Variation principle for Euler equation}
\label{appx:varprinc}

\noindent It will be convenient
to perform the extremization over all diffeomorphisms imposing
the volume-preservation constraint by adding
to the action a term
\qq
S_{Lm}(\Phi,\lambda)\ =\ \int\lambda\,(\Phi^*_t\chi-\chi)\,{\rm d}t
\qqq
with a Lagrange multiplier $\,\lambda(t,x)$.
\,For the variation of $\,S_g\,$ at volume-preserving $\,\Phi$,
\,we obtain\footnote{field theorists will notice a similarity
of the following calculation to those in nonlinear sigma models}:
\qq
&&\delta S_g(\Phi)\s
=\s{_1\over^2}\s\delta\int g_{ij}(y)
\s v^i(t,y)\s v^j(t,y)\s\s\chi(y)\s\s{\rm d}t \cr&&
=\s{_1\over^2}\s\delta\int g_{ij}(\Phi(t,x))\s\s\da_t\Phi^i(t,x)\s\s
\da_t\Phi^j(t,x)\s\s\chi(x)\s\s{\rm d}t\cr
&&=\s{_1\over^2}\int\da_k g_{ij}(\Phi(t,x))\s\s\delta\Phi^k(t,x)\s\s
\da_t\Phi^i(t,x)\s\s\da_t\Phi^j(t,x)\s\s\chi(x)\s\s{\rm d}t\cr
&&\hspace{4cm}+\s\int g_{ij}(\Phi(t,x))\s\s
\da_t\Phi^i(t,x)\s\s\da_t\delta
\Phi^j(t,x)\s\s\chi(x)\s\s{\rm d}t\cr
&&=\int\delta\Phi^j(t,x)\bigg[{_1\over^2}\s\da_j g_{ik}(y)\s\m
v^i(t,y)\s\m v^k(t,y)\s-\s\da_k g_{ij}(y)\s\m v^i(t,y)\s\m
v^k(t,y)\cr
&&\hspace{3.4cm}-\s g_{ij}(y)\left(\m\da_t v^i(t,y)\m+\m v^k(t,y)\m
\da_k v^i(t,y)\right)\bigg]\chi(y)\s\s{\rm d}t\cr
&&=\s-\int g_{ij}\s\m u^j\bigg[\da_t v^i\s+\s v^k\da_k v^i\s
+\s{_1\over^2}\m g^{in}(\da_k g_{ln}+\da_l g_{kn}-\da_n g_{lk})\s v^k
\s v^l\bigg]\chi\s\m{\rm d}t\cr
&&=\s -\int g_{ij}\,u^j\,(\da_tv^i+v^k\nabla_kv^i)\,\chi\,{\rm d}t\,,
\label{121}
\qqq
where $\,u^j(t,\Phi(t,x))=\delta\Phi^j(t,x)\,$ and $\,\nabla_k\,$
is the covariant derivative with respect
to the Levi-Civita connection.
\,On the other hand, the variation of $\,S_{Lm}\,$
is given by
\qq
&&\delta S_{Lm}(\Phi,\lambda)\ =
\ \delta\int\lambda(t,x)\left(\sqrt{g(\Phi(t,x))}
\,\frac{{\partial(\Phi(t,x))}}
{{\partial(x)}}\,-\,\sqrt{g(x)}\right){\rm d}x\,{\rm d}t\cr\cr
&&=\int(\delta\lambda)(t,x)\left(\sqrt{g(\Phi(t,x))}\,
\frac{{\partial(\Phi(x))}}{{\partial(x)}}\,-\,\sqrt{g(x)}\right){\rm d}x
\,{\rm d}t\cr\cr
&&+\int\lambda(t,x)\,\Big(\frac{_1}{^2}g^{ij}(\Phi(t,x))
(\partial_kg_{ji})(\Phi(t,x))\,(\delta\Phi^k)(t,x)\,\cr\cr
&&\qquad\qquad\quad\ +\,\big[\big(\frac{_{\partial\Phi(t,x)}}
{^{\partial x}}\big)^{-1}
\big]^j_k\,(\partial_j\delta\Phi^k)(t,x)\Big)\,
\sqrt{g(\Phi(t,x))}\,
\frac{{\partial(\Phi(t,x))}}{{\partial(x)}}\,{\rm d}x\,{\rm d}t\cr\cr
&&=\int(\delta\lambda)(t,x)\left(\sqrt{g(\Phi(t,x))}\,
\frac{{\partial(\Phi(x))}}{{\partial(x)}}\,-\,\sqrt{g(x)}\right){\rm d}x
\,{\rm d}t\cr\cr
&&+\int p(t,y)\left(\frac{_1}{^2}g^{ij}(y)(\partial_kg_{ij}(y)\,u^k(t,y)\,
+\,(\partial_ku^k)(t,y)\right)\sqrt{g(y)}\,{\rm d}y\,{\rm d}t\,,
\qqq
where we have introduced the pressure by the formula
\qq
p(t,\Phi(t,x))=\lambda(t,x)\,.
\label{pressure}
\qqq
Integrating by parts in the last term,
we finally obtain
\qq
&&\delta S_{Lm}(\Phi,\lambda)\
=\int(\delta\lambda)(t,x)\left(\sqrt{g(\Phi(t,x))}\,
\frac{{\partial(\Phi(x))}}{{\partial(x)}}\,-\,\sqrt{g(x)}\right){\rm d}x
\,{\rm d}t\cr\cr
&&-\int\,u^k(t,y)\,(\partial_kp)(t,y)\,\sqrt{g(y)}\,{\rm d}y\,{\rm d}t\,.
\label{212}
\qqq
Equating to zero the variation $\,\delta S(\Phi,\lambda)\,$ for
$\,S(\Phi,\lambda)=S_g(\Phi)+S_{Lm}(\Phi,\lambda)$, \,we obtain
the (generalized) Euler equation (\ref{Euler})
together with the volume-preserving condition on $\,\Phi(t)\,$ which,
in terms of the velocity is the incompressibility condition (\ref{incomp}).
\,On noncompact manifolds, these should be accompanied
by the decay conditions at infinity that assure that the integrations
by parts above may be performed and eliminate the solutions
$\,v^i\,=\,g^{ij}\partial_jf\,$ with $\,\Delta f=g^{ij}\nabla_i\partial_jf=0\,$
and arbitrary $t$-dependence (solving also the unforced Navier-Stokes
equations on Einstein manifolds, see \cite{CC,KM}).
\vskip 0.2cm

\setcounter{section}{0}
\setcounter{equation}{0}
\def\thesection{{\small C}}
\def\theequation{C.\arabic{equation}}
\section{\small Velocity covariance in  terms of stream functions}
\label{appx:stream}

$\NH_{\hspace{-0.02cm}_R}\,$ is a contractible space. On the other hand,
due to incompressibility
of velocity, the equal-time 2-point function $\,\langle(*v^\flat)(\CX_1)
\otimes(*v^\flat(\CX_2)\rangle$ is a closed 1-form in its dependence
on each of the two points. We may then obtain
a stream-function correlator $\,\langle\psi(\CX_1)\,\psi(\CX_2)\rangle\,$
satisfying Eq.\,(\ref{vel2pt}) by integrating $\,\langle(*v^\flat)(\CX_1)
\otimes(*v^\flat(\CX_2)\rangle\,$ in each variable from a fixed point
of $\,\NH_{\hspace{-0.02cm}_R}\,$ to $\,\CX_1\,$ and $\,\CX_2$,
\,respectively. Function $\,\langle\psi(\CX_1)\,\psi(\CX_2)\rangle\,$
obtained this way is symmetric and of positive type, but it depends
on the choice of the initial point of integration. As a result,
the $\,SL(2,\NR)$-covariance of the velocity 2-point function does not
imply the $\,SL(2,\NR)\,$ invariance of $\,\langle\psi(\CX_1)\,
\psi(\CX_2)\rangle$. \,However, for $\,\gamma\in SL(2,\NR)$, \,the function
\qq
\langle\psi(\gamma\CX_1)\psi(\gamma\CX_2)\,\rangle\,-\,
\langle\psi(\CX_1)\,\psi(\CX_2)\rangle\ \equiv\ f_\gamma(\CX_1,\CX_2)
\label{fgam}
\qqq
is annihilated by the product of exterior derivatives $\,d(1)d(2)\,$
in $\,\CX_1\,$ and $\,\CX_2$. \,It has then to be of the form
$\,f^1_\gamma(\CX_1)+f^2_\gamma(\CX_2)\,$ where, by symmetry, we may take
$\,f^1_\gamma(\CX)=f^2_\gamma(\CX)=\frac{1}{2}\,f_\gamma(\CX,\CX)$.
\,Definition (\ref{fgam}) implies the cocycle condition
\qq
f^1_{\gamma_1\gamma_2}(\CX)\,=\,f^1_{\gamma_1}(\gamma_2\CX)
+f^1_{\gamma_2}(\CX)\,.
\label{cocyc}
\qqq
Taking, in particular, $\,\CX=\CX_0\,$ corresponding to $\,w=i\,$ and
$\,\gamma_2\in SO(2,\NR)$, \,which is the stabilizer subgroup of
$\,\CX_0$, \,we obtain
\qq
f^1_{\gamma_1\gamma_2}(\CX_0)\,=\,f^1_{\gamma_1}(\CX_0)
+f^1_{\gamma_2}(\CX_0)\,.
\label{CX0}
\qqq
Taking also $\,\gamma_1\in SO(2,\NR)$, \,we infer that
$\,f^1_\gamma(\CX_0)\,$ is an additive function of $\,\gamma\in SO(2,\NR)$,
\,so it must be equal to zero when restricted to such $\gamma$. \,Consequently,
for general $\,\gamma_1$,
\,one has $\,f^1_{\gamma_1\gamma_2}(\CX_0)\,=\,f^1_{\gamma_1}(\CX_0)$, 
\,so that $\,f^1_\gamma(\CX_0)\,$ defines a function $\,g(\CX)\,$
on $\,\NH_{\hspace{-0.02cm}_R}\cong SL(2,\NR)/SO(2,\NR)\,$
(vanishing at $\,\CX=\CX_0$) \,and it determines
$\,f^1_\gamma(\CX)\,$ by the formula
\qq
f^1_{\gamma_1}(\gamma_2\CX_0)\,=\,g(\gamma_1\gamma_2\CX_0)-
g(\gamma_2\CX_0)
\qqq
following from the cocycle condition (\ref{cocyc}). This, in fact,
provides a general solution of this condition if we consider arbitrary
functions $\,g(\CX)\,$ on $\,\NH_{\hspace{-0.02cm}_R}$.
\,Note that the symmetric function
$\,\Psi(\CX_1,\CX_2)=
\big\langle\,\psi(\CX_1)\,\psi(\CX_2)\m\big\rangle-g(\CX_1)-g(\CX_2)\,$
satisfies the relation
$\,\Psi(\gamma\CX_1,\gamma\CX_2)=\Psi(\CX_1,\CX_2)$,
\,hence it depends only on the hyperbolic distance $\,\delta_{12}$.
\,It defines the same velocity correlators if used in formula
(\ref{vel2pt}) instead of $\,\big\langle\psi(\CX_1)\,
\psi(\CX_2)\big\rangle$.
If $\,\Psi'\,$ is another such functions, then
\qq
\Psi'(\CX_1,\CX_2)\,=\,\Psi(\CX_1,\CX_2)\,+\,g^1(\CX_1)+g^1(\CX_2)\,.
\qqq
for some function $\,g^1(\CX)$. \,The $SL(2,\NR)$-covariance of $\,\Psi\,$
and $\,\Psi'\,$ implies then that
\qq
g^1(\gamma\CX_1)-g^1(\CX_1)\,=\,-\,g^1(\gamma\CX_2)+g^1(\CX_2)\,=\,c_\gamma\,,
\qqq
but for each $\,\gamma\,$ there is $\,\CX_1\,$ such that
$\,\gamma\CX_1=\CX_1\,$ and we infer that $\,c_\gamma=0\,$
so that function $\,g^1\,$ is constant.

\setcounter{section}{0}
\setcounter{equation}{0}
\def\thesection{{\small D}}
\def\theequation{D.\arabic{equation}}
\section{\small Forcing covariance on $\,\NH_{\hspace{-0.01cm}_{R/\lambda}}$}
\label{appx:rescforce}

Eq.\,(\ref{8.2}) implies the formulae:
\qq
&&\lambda^{-2}C^{rr}(\CX_1',\CX_2')\ =\ \frac{_{\sqrt{1+(R/\lambda)^2r_1^{-2}}
\,\sqrt{1+(R/\lambda)^2r_2^{-2}}}}{^{R^2}}\,\,\m
\partial_{\varphi_1}\partial_{\varphi_2}\CC(\cosh\hspace{-0.06cm}
\frac{_{\delta_{1'2'}}}{^{R/\lambda}})\cr\cr
&&=\ -\,\frac{_{\sqrt{1+(R/\lambda)^2r_1^{-2}}\,\sqrt{1+(R/\lambda)^2
r_2^{-2}}}}{^{R^2}}\,\Big(\frac{_{r_1r_2
\cos(\varphi_1-\varphi_2)}}{^{(R/\lambda)^2}}
\,\CC'(\cosh\hspace{-0.06cm}\frac{_{\delta_{1'2'}}}{^{R/\lambda}})\cr
&&\hspace{6cm}+\,\frac{_{r_1^2r_2^2\m\sin^2(\varphi_1-\varphi_2)}}
{^{(R/\lambda)^4}}\,
\CC''(\cosh\hspace{-0.06cm}\frac{_{\delta_{1'2'}}}{^{R/\lambda}})\Big).
\qquad\qquad\label{CrrR}\\ \cr
&&\lambda^{-2}C^{r\varphi}(\CX_1',\CX_2')\
=\ -\,\frac{_{\sqrt{1+(R/\lambda)^2r_1^{-2}}\,\sqrt{1+(R/\lambda)^2r_2^{-2}}
}}{^{R^2}}\,\m\partial_{\varphi_1}\partial_{r_2}\CC(\cosh\hspace{-0.06cm}
\frac{_{\delta_{1'2'}}}{^{R/\lambda}})\cr
&&=\ -\,\frac{_{\lambda^2\m r_1\sin(\varphi_1-\varphi_2)}}{^{R^4}}\,
\CC'(\cosh\hspace{-0.06cm}\frac{_{\delta_{1'2'}}}{^{R/\lambda}})\cr
&&\hspace{0.6cm}-\,\frac{_{\lambda^4\m r_1^2r_2\m
\sin(\varphi_1-\varphi_2)\Big(
\sqrt{1+(R/\lambda)^2r_1^{-2}}-\sqrt{1+(R/\lambda)^2r_2^{-2}}
\,\cos(\varphi_1-\varphi_2)\Big)}}{^{R^4\,
\sqrt{1+(R/\lambda)^2r_2^{-2}}}}\m\,
\CC''(\cosh\hspace{-0.06cm}\frac{_{\delta_{1'2'}}}{^{R/\lambda}})
\,.\qquad\qquad\label{CrpR}\\ \cr
&&\lambda^{-2}C^{\varphi\varphi}(\CX_1',\CX_2')\
=\ \,\frac{_{
\sqrt{1+(R/\lambda)^2r_1^{-2}}\,\sqrt{1+(R/\lambda)^2r_2^{-2}}}}{^{R^2}}\m\,
\partial_{r_1}\partial_{r_2}\CC(\cosh\hspace{-0.06cm}
\frac{_{\delta_{1'2'}}}{^{R/\lambda}})\cr
&&=\ \,\frac{_{\lambda^2\Big(1-\sqrt{1+(R/\lambda)^2r_1^{-2}}\,
\sqrt{1+(R/\lambda)^2r_2^{-2}}\,
\cos(\varphi_1-\varphi_2)\Big)}}{^{R^4}}\m\,
\CC'(\cosh\hspace{-0.06cm}\frac{_{\delta_{1'2'}}}{^{R/\lambda}})\cr
&&+\,
\frac{_{\lambda^4\m r_1r_2\Big(\sqrt{1+(R/\lambda)^2r_2^{-2}}-
\sqrt{1+(R/\lambda)^2r_1^{-2}}\,\cos(\varphi_1-\varphi_2)\Big)
\Big(\sqrt{1+(R/\lambda)^2r_1^{-2}}
-\sqrt{1+(R/\lambda)^2r_2^{-2}}\,\cos(\varphi_1-\varphi_2)\Big)}}
{^{R^6}}\cr
&&\hspace{9.5cm}\cdot\,\CC''(\cosh\hspace{-0.06cm}\frac{_{\delta_{1'2'}}}
{^{R/\lambda}})\,.\qquad
\label{CppR}
\qqq
The asymptotic behavior (\ref{Crrinf}), (\ref{Crpinf}) and (\ref{Cppinf})
follow from these expressions in a straightforward way.

\setcounter{section}{0}
\setcounter{equation}{0}
\def\thesection{{\small E}}
\def\theequation{E.\arabic{equation}}
\section{\small Rescaled velocity 2-point functions
on $\,\NH_{\hspace{-0.01cm}_{R}}$}
\label{appx:rescvel}

\noindent Assuming the form (\ref{strfctcov}) of the 2-point correlator
of the stream function on $\NH_{\hspace{-0.01cm}_{R}}$, \,the
rescaled version of the velocity 2-point functions given
by Eq.\,(\ref{vel2pt}) may be written in the form:
\qq
&&\lambda^{\frac{4}{3}-2}\m F^{rr}_{\NH_{_R},\CC}(\lambda^{\frac{2}{3}}t;
\lambda r_1,\varphi_1;
\lambda r_2,\varphi_2)\ =\ \lambda^{-\frac{2}{3}}
\frac{_{\sqrt{1+(R/\lambda)^2r_1^{-2}}\,\sqrt{1+(R/\lambda)^2r_2^{-2}}}}
{^{R^2}}\Bigg[x\,\Psi'(\lambda^{\frac{2}{3}}t,x)\cr\cr
&&\hspace{0.5cm}
+\ x^2\m\Psi''(\lambda^{\frac{2}{3}}t,x)\,
-\,\frac{_{r_1r_2\,\sqrt{1+(R/\lambda)^2r_1^{-2}}\,
\sqrt{1+(R/\lambda)^2r_2^{-2}}}}{^{(R/\lambda)^2}}
\Big(\Psi'(\lambda^{\frac{2}{3}}t,x)\,
+\,2\m x\,\Psi''(\lambda^{\frac{2}{3}}t,x)\Big)\cr\cr
&&\hspace{7.8cm}+\,\left(1+\frac{_{r_1^2+r_2^2}}{^{(R/\lambda)^2}}
\right)\Psi''(\lambda^{\frac{2}{3}}t,x)\Bigg],\label{vrvrl}\\ \cr
&&\lambda^{\frac{4}{3}-1}\m F^{r\varphi}_{\NH_{_R},\CC}
(\lambda^{\frac{2}{3}}t;\lambda r_1,
\varphi_1;\lambda r_2,\varphi_2)\ =\
-\m\lambda^{-\frac{2}{3}}\,\frac{_{\sqrt{1+(R/\lambda)^2r_1^{-2}}
\,\sqrt{1+(R/\lambda)^2r_2^{-2}}\,r_1\sin(\varphi_1-\varphi_2)}}
{^{R^2(R/\lambda)^2}}\cr\cr
&&\hspace{2.25cm}\times\,\Bigg[\Psi'(\lambda^{\frac{2}{3}}t,x)\,
+\,x\,\Psi''(\lambda^{\frac{2}{3}}t,x)\,
-\,\frac{_{r_1\,\sqrt{1+(R/\lambda)^2r_1^{-2}}}}{^{r_2\,
\sqrt{1+(R/\lambda)^2r_2^{-2}}}}\,
\Psi''(\lambda^{\frac{2}{3}}t,x)\Bigg],\label{vrvpl}\\ \cr
&&\lambda^{\frac{4}{3}}\m F^{\varphi\varphi}_{\NH_{_R},
\CC}(\lambda^{\frac{2}{3}}t;\lambda r_1,\varphi_1;\lambda r_2,\varphi_2)\ =\
\lambda^{-\frac{2}{3}}\,\frac{_{\sqrt{1+(R/\lambda)^2r_1^{-2}}\,
\sqrt{1+(R/\lambda)^2r_2^{-2}}}}{^{r_1r_2\,R^2}}\cr\cr
&&\hspace{1cm}\times\,\Bigg[
\Big(x\,+\,\frac{_{r_1r_2\,(1-(1+(R/\lambda)^2r_1^{-2})(1+(R/\lambda)^2
r_2^{-2}))}}{^{(R/\lambda)^2\,\sqrt{1+(R/\lambda)^2r_1^{-2}}\,
\sqrt{1+(R/\lambda)^2r_2^{-2}}}}\Big)\,\Psi'(\lambda^{\frac{2}{3}}t,x)\cr\cr
&&\hspace{2.6cm}+\,\Big(x-\frac{_{r_1\,\sqrt{1+(R/\lambda)^2r_1^{-2}}}}
{^{r_2\,\sqrt{1+(R/\lambda)^2r_2^{-2}}}}\Big)
\Big(x-\frac{_{r_2\,\sqrt{1+(R/\lambda)^2r_2^{-2}}}}{^{r_1\,
\sqrt{1+(R/\lambda)^2r_1^{-2}}}}\Big)\,
\Psi''(\lambda^{\frac{2}{3}}t,x)\Bigg],\label{vpvpl}
\qqq
where $\,x\,$ is given by the right hand side of Eq.\,(\ref{xR}) with $\,R\,$
replaced by $\,R/\lambda$. \,The above expressions should be compared
to Eqs.\,(\ref{vrvr0}), (\ref{vrvp0}) and (\ref{vpvp0}) for the velocity
correlators on $\,\NH_{\hspace{-0.01cm}_0}$.

\setcounter{section}{0}
\setcounter{equation}{0}
\def\thesection{{\small F}}
\def\theequation{F.\arabic{equation}}
\section{\small Harmonic analysis on $\,\NH_{\hspace{-0.01cm}_R}\,$
in the $\,R\to0\,$ limit}
\label{appx:harman0}

\noindent To study the harmonic analysis on $\,\NH_{\hspace{-0.01cm}_R}\,$ in
the limit $\,R\to0$, \,we rewrite it using the coordinates
$\,(r,\varphi)$.
\,The harmonic analysis in space $\,L^2(\NH_{\hspace{-0.01cm}_R})\,$
is realized by the formulae that combine the Gel'fand-Graev
transformation $\,\psi\mapsto\psi^{GG}\,$ from functions on the
hyperboloid $\,\NH_{\hspace{-0.01cm}_R}\,$ to functions on the
cone $\,\NH_{_0}\,$
\qq
\psi^{GG}(Y)\ =\ \int\limits_{\NH_{_R}}\,\delta(Y^1X^1+Y^2X^2-Y^3X^3+R)\,
\,\psi(X)\,\,\chi(X)
\qqq
with the Fourier transform in the angular direction and Mellin transform
in the radial one  \cite{Vilenkin}. \,In the coordinates
$\,(r,\varphi)\,$ it is given by the relations:
\qq
\psi(r,\varphi)&=&\frac{_1}{^{2\pi R}}\sum\limits_{m=-\infty}^\infty
\int\limits_0^\infty\Big(\int\limits_{-\pi}^\pi
{\rm d}\vartheta\int\limits_0^\infty s^{-\frac{1}{2}-i\sigma}\,\ee^{-im\vartheta}
\,\delta\big(s\m r\cos(\vartheta-\varphi)-s\sqrt{R^2+r^2}+R\big)\,\,ds\Big)\cr
&&\hspace{6.5cm}\times\ a_m(\sigma)\,\,\sigma\,\tanh(\pi\sigma)\,{\rm d}
\sigma\cr
&=&\frac{_1}{^{2\pi R^2}}\!\sum\limits_{m=-\infty}^\infty\!\!\ee^{-im\varphi}
\int\limits_0^\infty\!
\Big(\int\limits_{-\pi}^\pi\ee^{-im\vartheta}\,\big[\frac{_{\sqrt{R^2+r^2}}}
{^R}-\frac{_r}{^R}\cos\vartheta
\big]^{i\sigma-\frac{1}{2}}\,{\rm d}\vartheta\Big)%\cr&&\hspace{8cm}\cdot\
a_m(\sigma)\,\sigma\,\tanh(\pi\sigma)\,{\rm d}\sigma\cr
&=&\frac{_1}{^{R^2}}\sum\limits_{m=-\infty}^\infty\ee^{-im\varphi}
\int\limits_0^\infty\CP^{-\frac{1}{2}+i\sigma}_{m0}(\frac{_{\sqrt{R^2+r^2}}}
{^R})\,\,a_m(\sigma)\,\sigma\,\tanh(\pi\sigma)\,{\rm d}\sigma\,,
\label{GGpsi}
\\ \cr
a_m(\sigma)&=&
\frac{_R}{^{4\pi^2}}\int\limits_{-\pi}^\pi\ee^{im\vartheta}\,{\rm d}\vartheta\int
\limits_0^\infty s^{-\frac{1}{2}+i\sigma}\m ds \cr&&\quad\times\,
\int\delta\big(s\m r\cos(\vartheta-\varphi)-
s\sqrt{R^2+r^2}+R\big)\,\psi(r,\varphi)
\,\m\frac{_{R\m r}}{^{\sqrt{R^2+r^2}}}\,{\rm d}r\wedge{\rm d}\varphi\cr
&=&\frac{_1}{^{4\pi^2}}
\int\ee^{im\varphi}\Big(\int\limits_{-\pi}^\pi\ee^{im\theta}
\big[\frac{_{\sqrt{R^2+r^2}}}{^R}-\frac{_r}{^R}
\cos\vartheta\big]^{-\frac{1}{2}-i\sigma}\,d\theta\Big)\,
\psi(r,\varphi)\,\m\frac{_{R\m r}}
{^{\sqrt{R^2+r^2}}}\,{\rm d}r\wedge{\rm d}\varphi\cr
&=&\frac{_1}{^{2\pi}}\int\ee^{im\varphi}\,\,\CP^{-\frac{1}{2}-i\sigma}_{m0}
(\frac{_{\sqrt{R^2+r^2}}}{^R})\,\,
\psi(r,\varphi)\,\m\frac{_{R\m r}}
{^{\sqrt{R^2+r^2}}}\,{\rm d}r\wedge{\rm d}\varphi\,,\label{GGam}\\ \cr\cr
&&\hspace{-1.7cm}\int|\psi(r,\varphi)|^2\,\,\frac{_{R\m r}}{^{\sqrt{R^2+r^2}}}
\,{\rm d}r\wedge{\rm d}\varphi\ 
=\ \frac{_{2\pi}}{^{R^2}}\sum\limits_{m=-\infty}^\infty
\int\limits_0^\infty
|a_m(\sigma)|^2\,\,\sigma\,\tanh\pi\sigma\,{\rm d}\sigma\,,\label{GGPlanch}
\qqq
which reproduce Eqs.\,(\ref{idecom}), (\ref{amk}) and (\ref{Planch})
upon setting $\,\sigma=Rk$.
\vskip 0.2cm

To study the  $\,R\to0\,$ limit, we use the asymptotic expansion
from \cite{GrRy} holding for large positive $\,x$:
\qq
\!\!&&\CP^{-\frac{1}{2}+i\sigma}_{m0}(x)=\frac{_{\Gamma(\frac{1}{2}+i\sigma)x^{-\frac{1}{2}}}}
{^{\Gamma(\frac{1}{2}+i\sigma+m)}} \Big(\frac{_{2^{-\frac{1}{2}+i\sigma} 
\Gamma(i\sigma)}}{^{\pi^{\frac{1}{2}} \Gamma(\frac{1}{2}+i\sigma-m)}}
 x^{i\sigma} + \frac{_{2^{-\frac{1}{2}-i\sigma} 
\Gamma(-i\sigma)}}{^{\pi^{\frac{1}{2}} 
\Gamma(\frac{1}{2}-i\sigma-m)}}
 x^{-i\sigma}\Big)%\qquad\cr\cr&&\hspace{9.9cm}\cdot\,
\left(1+\CO(x^{-2})\right)\,.\quad\label{GrR}
\qqq
Substituting this to Eq.\,(\ref{GGpsi}) and using the relation
$\,\sigma \tanh(\pi\sigma) = {|\Gamma(\frac{1}{2}+i\sigma)|^2} 
 /{|\Gamma(i\sigma)|^2} $, \,we obtain
\qq
\psi(r,\varphi)&=&
\frac{_1}{^{R^2}}\sum\limits_{m=-\infty}^\infty\ee^{-im\varphi}
\int\limits_0^\infty\Big(
\frac{_{(-1)^m\,2^{-\frac{1}{2}+i\sigma}\,
\Gamma(\frac{1}{2}-i\sigma)\,(\frac{1}{2}-i\sigma)\m\cdots\m
(\frac{1}{2}-i\sigma+m-1)}}
{^{\pi^{\frac{1}{2}}\,\Gamma(-i\sigma)\,(\frac{1}{2}+i\sigma)
\m\cdots\m(\frac{1}{2}+i\sigma+m-1)}}\,
(\frac{_{\sqrt{R^2+r^2}}}{^R})^{-\frac{1}{2}+i\sigma}\cr
&&+\,\frac{_{(-1)^m\,2^{-\frac{1}{2}-i\sigma}\,
\Gamma(\frac{1}{2}+i\sigma)}}{^{\pi^{\frac{1}{2}}\,
\Gamma(i\sigma)}}
\,(\frac{_{\sqrt{R^2+r^2}}}{^R})^{-\frac{1}{2}-i\sigma}\Big)
\,a_m(\sigma)\,\,{\rm d}\sigma\ \left(1+\CO(R^{2})\right)\,,\cr\cr
&=&\sum\limits_{m=-\infty}^\infty
\ee^{-im\varphi}\int\limits_{-\infty}^\infty r^{^{-\frac{1}{2}+i\sigma}}\,
\,\tilde a_m(\sigma)\,\,{\rm d}\sigma\ \left(1+\CO(R^{2})\right)\,,
\label{GGpsi2}
\qqq
where we have set
\qq
\tilde a_m(\sigma)\ =\ \frac{_{(-1)^m\,2^{-\frac{1}{2}+i\sigma}\,
\Gamma(\frac{1}{2}-i\sigma)}}
{^{\pi^{\frac{1}{2}}\,\Gamma(-i\sigma)}}\,R^{-\frac{3}{2}-i\sigma}\,
\cdot\begin{cases}{
\frac{_{(\frac{1}{2}-i\sigma)\,\cdots\,
(\frac{1}{2}-i\sigma+m-1)}}
{^{(\frac{1}{2}+i\sigma)\,\cdots\,(\frac{1}{2}+i\sigma+m-1)}}
\,a_m(\sigma)\quad\ {\rm if}
\quad\sigma>0\,,\cr\cr
a_m(-\sigma)\hspace{3.25cm}{\rm if}
\quad\sigma<0\,.\quad}\end{cases}
\qqq
Hence, the limiting decomposition at $\,R=0\,$ of functions
on the cone $\,\NH_{_0}\,$ is given by the Mellin transform in $\,r\,$
combined with the Fourier transform in the angular variable $\,\varphi$:
\qq
&&\psi(r,\varphi)\ =\ \sum\limits_{m=-\infty}^\infty\ee^{-im\varphi}
\int\limits_{-\infty}^\infty r^{-\frac{1}{2}+i\sigma}\,\tilde a_m(\sigma)\,
{\rm d}\sigma
\,\label{GGpsi0}\\ \cr
&&\tilde a_m(\sigma)\ =\ \frac{_1}{^{4\pi^2}}\int
\ee^{im\varphi}\,r^{-\frac{1}{2}-i\sigma}\,\psi(r,\varphi)\,{\rm d}r
\wedge{\rm d}\varphi\,,
\label{GGam0}\\ \cr
&&\int|\psi(r,\varphi)|^2\,{\rm d}r\wedge{\rm d}\varphi\ =\ 4\pi^2\int
|\tilde a(\sigma)|^2\,{\rm d}\sigma\,.\label{GGPlanch0}
\qqq
Action (\ref{Diff}) of $\,\Diff\m S^1\,$
on $\,\NH_{_0}\,$ induces the unitary action of that infinite-dimensional
group in the space $\,L^2(\NH_{_0})\,$ of functions on $\,\NH_{_0}\,$
square-integrable with respect to the volume measure $\,\chi_0$,
\,defined by the formula
\qq
\psi\,\mapsto\,D\psi\,,\qquad
(D\psi)(r,\varphi)\,=\,\psi(\frac{_r}{^{(D^{-1})'(\varphi)}},D^{-1}(\varphi))\,,
\qqq
Note that this action commutes with the Laplacian
$\,\Delta=\partial_r r^2\partial_r$. \,The functions
$\,\ee^{im\varphi}\,r^{-\frac{1}{2}+i\sigma}\,$ are eigenvectors of $\,\Delta\,$
corresponding to the eigenvalue $-(\frac{1}{4}+\sigma^2)$. \,The decomposition
(\ref{GGpsi0}) realizes the decomposition of $\,L^2(\NH_{_0})\,$ into
the spectral eigenspaces of $\,\Delta\,$ and, at the same time, the
decomposition of the unitary representation
of $\,\Diff\m S^1\,$ in
$\,L^2(\NH_{_0})\,$ into the irreducible components:
\qq
L^2(\NH_{_0})\ =\ {\int\limits_{-\infty}^\infty}\hspace{-0.07cm}{}^{\,^{^\oplus}}
H^0_\sigma\,\,{\rm d}\sigma\,.
\label{directint0}
\qqq
Unlike for $\,R>0$, \,where only $\,\sigma>0\,$ appeared in the decomposition
(\ref{directint}), \,now $\,\sigma>0\,$ and $\,\sigma<0\,$ give rise to
inequivalent representations.

\setcounter{section}{0}
\setcounter{equation}{0}
\def\thesection{{\small G}}
\def\theequation{G.\arabic{equation}}
\section{\small Real analyticity of the spectrum}
\label{appx:spectr}

Spectrum $\,\CE(k)\,$ is related to the stream function correlator
$\,\Psi\,$ by the Fourier transform on $\,\NH_{\hspace{-0.01cm}_R}\,$
(\ref{fpp}) and Eq.\,(\ref{spectrum}) which imply the inverse transform 
relation
\qq
\frac{{8\m\CE_{stat}(k)}}{{(1+(2Rk)^2)\,k\,\tanh(\pi Rk)}}\ =\,
\int_1^\infty\overline{\CP^{-\frac{1}{2}+iRk}_{00}(x)}
\m\,\Psi(x)\,\m{\rm d}x\,,
\label{invtr}
\qqq
where the integral has to be interpreted in the appropriate
sense \,(by analytic continuation) \,since, for large $\,x$,
$\,|\CP^{-\frac{1}{2}+i\sigma}_{00}(x)|=\CO(x^{-\frac{1}{2}})$,
\,see Eqs.\,(\ref{GrR}). Let us split the integration on
the right hand side of (\ref{invtr}) into the one from 1 to 2 and the rest.
\,The first integral gives an entire function
of $\,k$, \,as is easily seen from Eq.\,(\ref{Pm0}), and it contributes
to $\,\CE(k)\,$ a term $\,O(k^2)\,$ or less for small $\,k$.
\,For the other integral, we shall use the expression
\qq
\CP^{-\frac{1}{2}+i\sigma}_{00}(x)\ =\ \frac{\Gamma(-i\sigma)}{\sqrt{\pi}\,
\Gamma(\frac{1}{2}-i\sigma)}\,\zeta^{-\frac{1}{2}-i\sigma}\,{}_2F_1
(\frac{_1}{^2},\frac{_1}{^2}+i\sigma,1+i\sigma;\zeta^{-2})\,+\ c.\,c.
\label{CPexp}
\qqq
for $\,\zeta\equiv x+\sqrt{x^2-1}$. \,Expanding the hypergeometric functions
into a power series in $\,\zeta^{-2}$, \,we see that only the contribution
\qq
\int\limits_2^\infty\zeta^{-\frac{1}{2}\mp iRk}\,\Psi(x)\,{\rm d}x
\label{distint}
\qqq
of the leading term into the integral needs a special treatment. \,Using
the asymptotic expansion for $\,\Psi_{stat}\,$ rewritten in terms
of variable $\,\zeta$, \,we see that the integral (\ref{distint})
contributes few simple poles and a function analytic in $\,k\,$
in a neighborhood of the real axis. The pole at zero appears only
if the power $\,x^{-\frac{1}{2}}\,$ occurs in the large $\,x\,$ expansion
of $\,\Psi_{stat}$. \,The other terms of the expansion of the hypergeometric
functions contribute absolutely convergent integrals
so that
\qq
\int\limits_1^\infty\zeta^{-\frac{1}{2}\mp iRk}\,
\left({}_2F_1(\frac{_1}{^2},\frac{_1}{^2}\pm iRk,1\pm iRk;\zeta^{-2})\,-\,1
\right)
\,\Psi_{stat}(x)\,{\rm d}x\qquad
\qqq
is again analytic in $\,k\,$ in a neighborhood of the real axis.
The prefactors
$\,\pi^{-\frac{1}{2}}\m\Gamma(\mp iRk)/$ $\Gamma(\frac{1}{2}\mp iRk)\,$
in Eq.\,(\ref{CPexp}) are meromorphic in $\,k\,$ in such a region
with a simple pole at $\,k=0$. \,Eq.\,(\ref{invtr}) implies
than that the contribution of $\,\Psi_{stat}\,$ is analytic around
$\,k\geq0$. As for the contribution to $\,\Psi_0$, \,a presence
of the $\,x^{-\frac{2}{2}}\,$ term in its asymptotic expansion
could leave us with a pole at $\,k=0\,$ of up to the second order
(due to up to two derivatives over $\,k\,$ used to produced the
logarithmic terms of $\,\Psi_0$), \,but such pole would be incompatible
with the finiteness of the energy density at finite times. Hence
the contribution of $\,\Psi_0\,$ to the spectrum should also be
analytic around $\,k\geq0$.
\end{appendix}

\end{document}